%% file: acl_latex.tex
\documentclass[11pt]{article}

\usepackage[final]{acl}

\usepackage{times}
\usepackage{latexsym}

\usepackage[T1]{fontenc}

\usepackage[utf8]{inputenc}
\usepackage[table,xcdraw]{xcolor}

\usepackage{microtype}

\usepackage{inconsolata}

\usepackage[table]{xcolor}
\usepackage{multirow}
\usepackage{titletoc}
\usepackage[most]{tcolorbox} 

\newtcolorbox[auto counter, number within=section]{promptbox}[2][]{%
  colback=white,
  colframe=green!50!gray!40!black,
  width=\textwidth,
  arc=1mm,
  boxrule=0.5mm,
  title={\normalsize#2},
  #1
}

\usepackage{algorithm}
\usepackage{algorithmic}
\usepackage{multirow}
\usepackage{booktabs}
\usepackage{amsmath}
\usepackage{amssymb}
\usepackage{graphicx}
\usepackage{bbding}

\usepackage{newfloat}
\usepackage{listings}
\usepackage{pifont}

\definecolor{darkorange}{rgb}{1.0, 0.55, 0.0}
\definecolor{mygreen}{RGB}{0,180,0}
\definecolor{myred}{RGB}{180,0,0}  

\newcommand{\benchmark}{SwanBench-Speech}
\newcommand{\icook}{\textcolor{mygreen}{\ding{51}}} 
\newcommand{\icox}{\textcolor{myred}{\ding{55}}}    
\newcommand{\icohalf}{\textcolor{darkorange}{\ding{51}\kern-0.65em\ding{55}}}

\title{
Comprehensive Benchmarking of Long-Form Speech Generation in Diverse Scenarios
 }

\author{
\hspace*{-0.8cm}
Changhao Pan$^{1}$~\thanks{Equal contribution.}
~Rui Yang~$^{1}$~\footnotemark[1]
~Han Wang~$^{1}$~\footnotemark[1]
~Zhuan Zhou~$^{1}$
~Xuming He~$^{1}$
\\
\textbf{
\hspace*{-0.4cm}
Wenxiang Guo~$^{1}$
~Ziyue Jiang~$^{1}$ 
~Ruiqi Li~$^{2}$ 
~Yu Zhang~$^{2}$ 
~Chenyuhao Wen~$^{1}$ 
}\\
\textbf{
\hspace*{-0.4cm}
~Ke Lei~$^{1}$ 
~Xiang Yin~$^{2}$ 
~Jingyu Lu~$^{1}$
~Zhiyuan Zhu~$^{1}$
~~Zhou Zhao~$^{1}$\thanks{Corresponding author.}
} \\
$^{1}$~Zhejiang University \quad $^{2}$~Bytedance \\
\texttt{\{panch,yangruiii,zhaozhou\}@zju.edu.cn} \\
* Equal contribution. \quad \dag Corresponding author.
}
\begin{document}

\twocolumn[{%
\renewcommand\twocolumn[1][]{#1}%
\maketitle
\vspace{-1em}
\includegraphics[width=\linewidth]{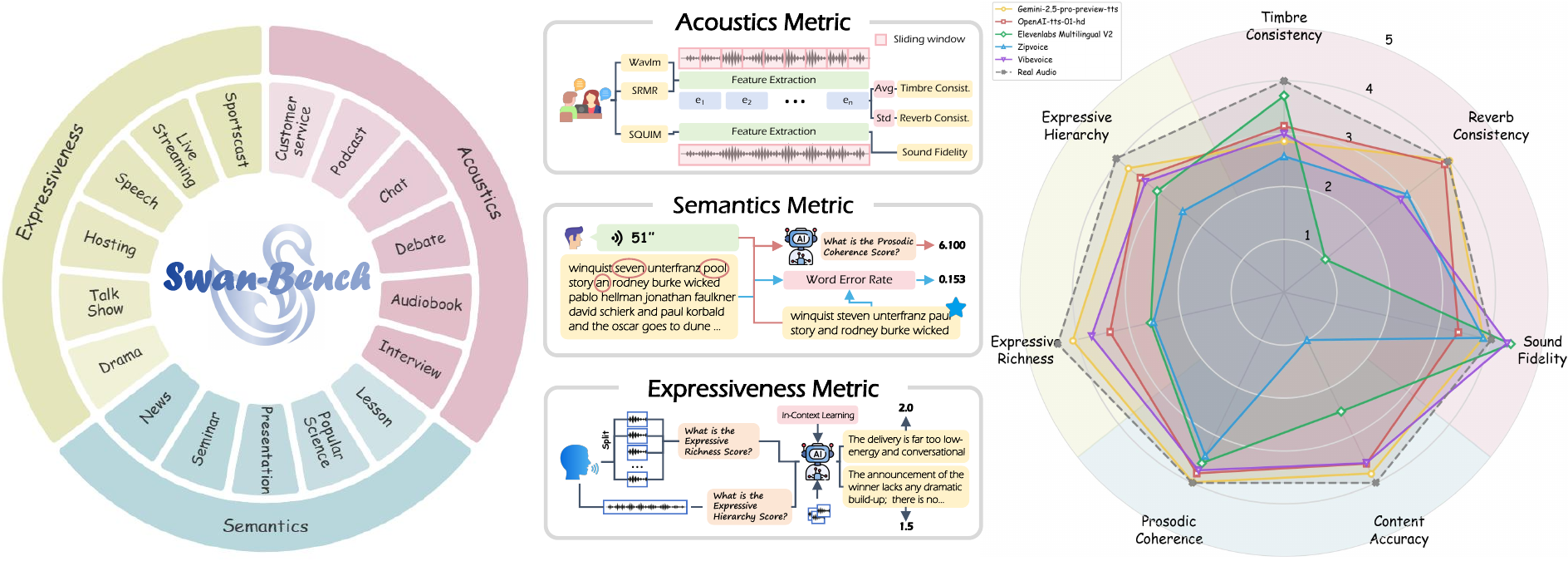}
\label{fig:teaser}
\captionof{figure}{\textbf{Overview of \benchmark.} 
    We propose \benchmark, a comprehensive benchmark designed to evaluate the performance of long-form speech generation models.
    \textbf{\textit{Left:}} We construct test sets across 17 downstream speech scenarios, grounded in three core challenges of long-form generation: \textit{Acoustics}, \textit{Semantics}, and \textit{Expressiveness}.
    %
    \textbf{\textit{Center:}} Along these three challenge axes, we propose seven disentangled metrics to comprehensively assess model performance and validate them through human alignment studies.
    %
    %
    \textbf{\textit{Right:}} Extensive experiments show that existing models still have substantial room for improvement in reverb consistency, prosodic coherence, and expressiveness.
    \vspace{1.5em}
}
\label{fig:teaser}
}]

\begin{abstract}
Recent advances in speech generation have enabled high-fidelity synthesis, yet systematic evaluation of models under long-context conditions remains largely underexplored. A comprehensive evaluation benchmark for long-form speech is indispensable for two reasons: 1) existing test scenarios are often confined to limited domains, creating a significant gap with the diverse downstream applications; 2) existing metrics overlook critical long-text factors such as consistency and coherence, failing to generalize reliably. To this end, we propose \benchmark, a comprehensive benchmark that decomposes “long-form speech quality” into specific, disentangled dimensions. \benchmark ~has three key properties. \textbf{1) Rich speech scenarios}: Focusing on long-form speech generation and dialog generation, \benchmark ~covers acoustics, semantics, and expressiveness challenges, and consists of 1,101 samples spanning 17 common speech scenarios; \textbf{2) Comprehensive evaluation dimensions}: Along the acoustics, semantics, and expressiveness axes, \benchmark ~defines an automated evaluation protocol with seven metrics to provide a comprehensive, accurate, and standardized assessment; \textbf{3) Valuable Insights}: Through extensive experiments, we reveal that current models still struggle in highly expressive scenarios and exhibit a notable gap in consistency and hierarchy compared to real recordings. The code and demo can be found at \url{https://swanaigc.github.io/#bench}.

\end{abstract}

\input{latex/Sections/1_introduction}

\input{latex/Sections/2_related_work}

\input{latex/Sections/3_benchmark}

\input{latex/Sections/4_experiements}

\input{latex/Sections/5_discussions}



\input{latex/Sections/8_conclusions}


\section*{Limitations}
We identify three limitations in this work. First, the linguistic scope of SwanBench-Speech is currently restricted to Chinese and English, leaving low-resource languages and diverse dialects or accents underexplored. Second, our investigation into semantics remains preliminary; while \benchmark's evaluation metrics prioritize acoustic coherence, we lack a robust automated framework to assess emotional and stylistic transitions grounded in deep semantic understanding of long-form text. Finally, the prompt speech utilized in our experiments is derived from only 20 speakers from open-source datasets. This limited speaker diversity may introduce evaluation bias, and we encourage the research community to contribute additional data to facilitate a more comprehensive assessment of model generalization.

\section*{Ethical considerations}

Although this work itself raises no immediate ethical concerns, two potential risks must be addressed when applying our benchmark. 
First, when utilizing our benchmark for evaluation, users must ensure that the prompt speech does not infringe upon the rights of the original voice actors. 
The use of audio from unverified sources or those restricted by regulations is strictly prohibited. 
Second, while our objective is to enhance the holistic performance of long-form synthesis, practitioners must ensure that models trained or evaluated using our methods are not deployed for generating disinformation, such as fabricated news reports or unauthorized political speeches. 
To mitigate these risks, we intend to implement strict usage guidelines upon open-sourcing the benchmark to prevent unethical and unauthorized applications.

\section*{Acknowledgements}

This work was supported by National Natural Science Foundation of China under Grant No.U25B2064.

\bibliography{custom}

\clearpage
\appendix

\input{latex/Sections/9_appendix}

\end{document}

%% file: latex/Sections/1_introduction.tex
\section{Introduction}\label{sec:intro}

Recent advances in generative modeling have revolutionized content creation across modalities~\cite{openai2024sora,esser2024scaling,guo2025deepseek}. 
While Large Language Models (LLMs) have demonstrated impressive capabilities in long-context generation and understanding~\cite{chen2023longlora,xiao2023efficient,bai2024longbench}, the speech community is similarly shifting focus from sentence-level to paragraph-level synthesis~\cite{le2023voicebox,shen2023naturalspeech}. 
Compared to traditional concatenation strategies, end-to-end long-form TTS paradigms promise superior acoustic and semantic consistency, leveraging broader contextual cues~\cite{peng2025vibevoice, park2024long}.

Despite these advancements, the systematic evaluation of long-form speech remains a significant challenge. 
While downstream applications involve complex multi-speaker interactions and rich semantic contexts, existing test scenarios are often confined to limited domains or single-speaker settings~\cite{koizumi2023libritts,zhang2022wenetspeech}.
This discrepancy prevents a thorough assessment of how models handle the rich challenges inherent in long-form generation, leaving their capabilities in complex scenarios largely underexplored.

Furthermore, establishing an effective evaluation protocol that is both scalable and accurate is equally difficult. 
Existing sentence-level metrics like Word Error Rate (WER)\cite{ali2018word} have become saturated~\cite{chen2024vall} and correlate poorly with human perception in long-text contexts~\cite{minixhofer2025ttsds2}.
Although human listening tests are the gold standard, they are non-scalable and costly.
Recently, MLLM-based evaluators have emerged~\cite{chen2024mllm, manku2025emergenttts}, yet they typically provide coarse-grained comparative judgments rather than quantitative metrics, often overlooking the property of consistency~\cite{li2024styletts}. Consequently, the field lacks an automated protocol aligned with the fine-grained nuances of long-form generation.

To this end, we propose \benchmark, a benchmark for long-form TTS models with three core properties: 1) \textbf{rich} scenarios, 2) \textbf{comprehensive} evaluation, and 3) \textbf{valuable} insights.

First, \benchmark ~is defined over two fundamental long-form TTS paradigms: long-form speech generation and dialog generation. Starting from three core dimensions of long-form speech, namely \textit{acoustics}, \textit{semantics}, and \textit{expressiveness}, \benchmark ~constructs 1,101 test samples spanning 17 scenarios, providing broad coverage of long-form TTS applications.

Second, our framework establishes an automatic evaluation protocol that employs a hierarchical approach to decomposing “long-form speech quality”.
Transcending the traditional focus on Fidelity and Accuracy, we introduce novel dimensions tailored for long-form characteristics, specifically Acoustic Consistency, Prosodic Coherence, and Expressive Hierarchy. 
These metrics effectively address the limitations of existing protocols by quantifying temporal stability and expressive dynamics. 
Moreover, we conduct user studies to validate the reliability of these metrics, ensuring they serve as a scalable proxy for perception.

Finally, through extensive experiments on \benchmark, we derive critical insights detailed in Section~\ref{sec:discussion}.
Our empirical results reveal that while current models rival human recordings in fidelity and accuracy, they exhibit substantial gaps in reverb consistency, prosodic coherence, and expressive hierarchy. Notably, performance deteriorates in highly expressive scenarios, underscoring the persisting challenges in modeling long-term dependencies and dynamic stylistic variations.

We’re open-sourcing \benchmark, including test samples, and evaluation scripts with prompts. We’ll also include more models in \benchmark ~to drive forward the field of long-form speech generation.

%% file: latex/Sections/2_related_work.tex
\section{Related Work}\label{sec:related_works}

\input{latex/Tables/related_work}

\paragraph{Long-form TTS}
Generating long-form speech and dialogues presents significant challenges in maintaining prosodic coherence, modeling long sequences, and managing speaker transitions. 
To ensure prosodic consistency, recent studies have explored joint style modeling and cross-sentence memory mechanisms~\cite{guo2024text,li2025long}. 
Concurrently, to enhance long-sequence modeling efficiency, researchers have introduced compact representations via multi-resolution quantization~\cite{nishimura2024hall} or low frame-rate tokenization~\cite{peng2025vibevoice}, as well as state space models to alleviate memory bottlenecks~\cite{park2024long}. 
Regarding speaker transitions, while early works combined autoregressive (AR) and non-autoregressive (NAR) components~\cite{borsos2023soundstorm}, recent advancements have further developed both paradigms: NAR approaches increasingly employ flow-matching techniques, whereas AR models leverage speaker tokens to handle long-context dialogues~\cite{ju2025mooncast,xie2025soulx}. 
Despite these technical strides, existing metrics remain insufficient for evaluating prosodic coherence, emotional richness, and transition quality. 
To bridge this gap, \benchmark ~introduces a unified evaluation framework with targeted test cases and human-aligned metrics designed to quantify these critical properties.

\paragraph{Evaluation for Speech Generation Models} 
Current TTS evaluation mainly relies on four objective metric families: signal-based metrics~\cite{taal2010short}, MOS prediction networks~\cite{saeki2022utmos}, distributional metrics~\cite{minixhofer2024ttsds}, and accuracy metrics~\cite{ali2018word}. 
These metrics are nearly saturated for recent state-of-the-art systems~\cite{ju2024naturalspeech}.
Follow-up benchmarks~\cite{huang2025instructttseval,anastassiou2024seed} increase difficulty via harder texts or controllability, but remain sentence level and are not directly suitable for long-form speech~\cite{clark2019evaluating}.
Long text test sets like MinutesSpeech-~\cite{nishimura2024hall} and LibriSpeech-Long~\cite{park2024long} partially address this gap, yet cover only a narrow range of scenarios, as shown in Table~\ref{tab:related_work}. Benchmarks for dialog models also face similar issues~\cite{ao2024sd}.
Moreover, existing protocols rely heavily on subjective evaluations~\cite{cambre2020choice,zhang2023audiobook}, which do not scale and lack standardized procedures. 
In contrast, \benchmark ~ jointly covers long-form speech and dialog generation, spans 17 scenarios, and provides comprehensive automatic metrics aligned with humans, thereby addressing key limitations of current evaluation practices.

%% file: latex/Tables/related_work.tex
\begin{table}
  \caption{Comparison of speech generation benchmarks and test datasets. \textbf{Pipe.} indicates availability of an automatic evaluation pipeline, and \icohalf ~ marks that only part of the metrics are objectively computable. * denotes non-public data, with results estimated from the paper.}
  \label{tab:related_work}
  \vspace{-2pt}
  \centering
  \footnotesize
  \resizebox{\linewidth}{!}{%
    \begin{tabular}{@{}lcccccc@{}}
      \toprule
      \textbf{Benchmark} & \textbf{Clips} & \textbf{Scenario} &
      \textbf{Spk-Num} & \textbf{Avg-Word} & \textbf{Pipe} & \textbf{Dim.} \\ \midrule
      SeedTTS-Eval       & 6612  & 1  & 1    & 19.57  & \icohalf & 3 \\
      EmergentTTS-Eval   & 1645  & 6  & 1    & 33.93  & \icook   & 5 \\
      TTSDS2             & 60    & 4  & 1    & 24.24  & \icook   & 4 \\
      Choice of Voices   & 1     & 1  & 1    & 988    & \icox    & 5 \\
      MinutesSpeech-test & 1221  & 1  & 1    & 134    & \icohalf & 6 \\
      LibriSpeech-long   & 960   & 1  & 1    & 534.5  & \icohalf & 6 \\
      NeuralTTS-eval     & 250   & 1  & 1    & 260*   & \icox    & 9 \\
      MultiDialog        & 831   & 3  & 2 & 319.8 & \icohalf & 4 \\ \midrule
      \textbf{\benchmark}& 1101  & 17 & 1-4  & 228.6  & \icook   & 7 \\ \bottomrule
    \end{tabular}%
  }
  \vspace{-5pt}
\end{table}

%% file: latex/Sections/3_benchmark.tex
\section{\benchmark}

\subsection{Overview}\label{sec:bench-overview}
Long-form speech generation requires multi-dimensional evaluation to ensure immersion and realism.
For instance, in an online education scenario, a generated lecture must not only preserve timbre and acoustic environment (acoustics) but also deliver accurate content with natural pacing (semantics), while exhibiting dynamic variations to sustain engagement (expressiveness).
Motivated by these requirements, we propose \benchmark, a hierarchical benchmark comprising 1,101 samples across 17 downstream applications.
And as detailed in Section~\ref{sec:bench-eval}, our evaluation protocol is organized around three primary dimensions:

\textbf{Acoustics Challenge} focuses on sound quality, environmental fidelity, and speaker identity.
Hence, we carefully curate samples from six relevant scenarios: customer service, podcast, chat, debate, audiobook, and interview,
and evaluate acoustic performance based on \textit{timbre consistency}, \textit{reverb consistency}, and \textit{sound fidelity}.

\textbf{Semantics Challenge} targets correctness and fluency to probe the upper limits of semantic modeling.
We derive complex test cases from five information-dense scenarios (lesson, popular science, presentation, seminar, and news), evaluating them by \textit{content accuracy} and \textit{prosodic coherence}.

\textbf{Expressiveness Challenge} addresses the issues of flat emotion and low engagement in long-form speech.
We incorporate highly expressive scenarios such as drama, talk show, hosting, speech, live streaming, and sportscast.
Performance is assessed through \textit{expressive richness} (sentence-level emotional impact) and \textit{expressive hierarchy} (paragraph-level expressive dynamics).

\subsection{Data Collection}\label{sec:bench-collection}
To provide a high-quality benchmark, we curate the test samples from three sources: online text corpora, online audio media, and LLM generation.

\begin{figure*}[t]
  \centerline{\includegraphics[width=\textwidth]{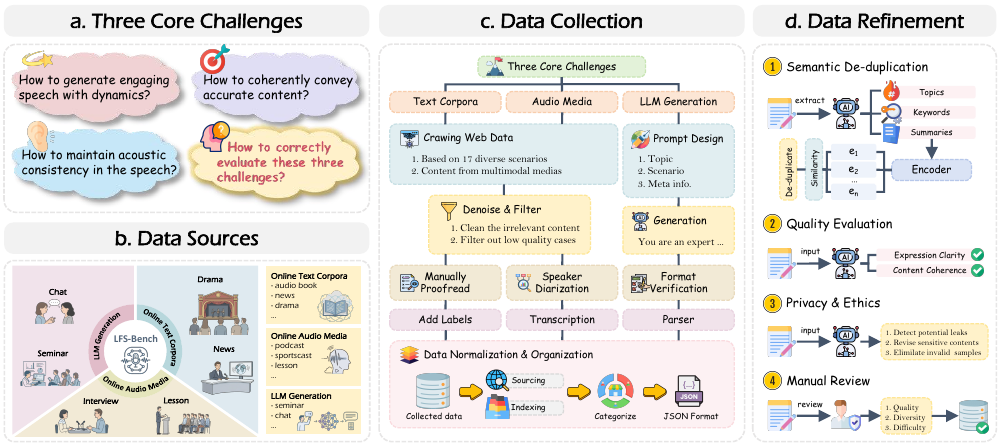}} 
  \caption{ \textbf{Overview of dataset construction and refinement.}
  The process consists of four stages:
    1) Formulating SwanBench-Speech based on three core challenges;
    2) Selecting 17 downstream speech scenarios aligned with these challenges;
    3) Designing a hybrid data collection pipeline;
    4) Performing data refinement on the constructed dataset.
  }
  \label{fig:data}
  \vspace{-5pt}
\end{figure*}

\textbf{Online Text Corpora} 
For scenarios such as audiobooks, drama, and news, where abundant transcripts are available online, we directly construct test sets from the web. 
After crawling the raw data, we clean irrelevant content such as illegal characters, and normalize the text into a clear and readable format. 
We then employ human annotators to proofread the transcripts and add speaker labels, yielding the final curated test samples.

\textbf{Online Audio Media} 
This source constitutes the main component of \benchmark. 
For web audio data, after crawling, we first denoise the raw audio~\cite{wang2025zipenhancer}, and then use DNS-MOS~\cite{reddy2021dnsmos} scores to filter out low-quality cases. 
After that, speaker diarization is conducted~\cite{zheng20233d} to obtain audio segments for each speaker. 
Finally, we use SenseVoice~\cite{an2024funaudiollm} to transcribe audio clips. 
Upon completion of the script processing, we perform manual verification to correct errors from the previous steps and curate the final test samples.

\textbf{LLM generation}
We use GPT-5~\cite{gpt5} to augment our test set and increase the diversity of data sources. 
Specifically, we first design prompts that include scenario, topic, and task information. Then we use them to guide the LLM to generate high-quality test cases. 
All generated samples are then checked and verified by human annotators.

\subsection{Data Refinement}\label{sec:bench-refinement}
To ensure the quality of curated samples, we implement a rigorous refinement pipeline. 
The process begins with semantic de-duplication, where we employ GPT-5 to extract topics, keywords, and summaries for each sample. These fields are concatenated and encoded using SentenceBERT~\cite{reimers2019sentence} to identify and remove highly similar instances based on cosine similarity. 
Subsequently, we filter for content quality by leveraging GPT-5 to evaluate expression clarity and content coherence, discarding any samples that fall below predefined thresholds.
To address privacy and ethical concerns, we utilize DeepSeek V3.2~\cite{liu2024deepseek} with a chain-of-thought~\cite{wei2022chain} procedure to detect potential leaks, revise sensitive content, and eliminate samples posing social or ethical risks. 
Finally, we conduct a manual review to purge remaining low-quality samples and replenish the dataset, ultimately yielding 1,101 samples that cover three core challenges and span 17 downstream scenarios, as shown in the left side of Fig.~\ref{fig:teaser}.

\subsection{Evaluation Metrics}\label{sec:bench-eval}

We disentangle the challenges into seven objective metrics to comprehensively assess the performance of TTS models. More details in Appendix~\ref{app_sec:eval}.

\textbf{Timbre Consistency}.
Compared with prior work that evaluates zero-shot capability using speaker similarity, we directly measure within-utterance timbre consistency to assess a model's ability to maintain or switch speaker identity. 
For single-speaker long-form speech $w$, we apply a sliding window over the waveform and extract a speaker embedding for each window, yielding a sequence $\{\mathbf{e}_i \}_{i=1}^{n}$, where $n$ is the number of windows. 
We then compute the cosine similarity for every pair of distinct embeddings
and take the average of the resulting similarity sequence $\{\mathrm{sim}_{i,j} \}_{i,j=1, i\neq j}^{n}$ as the measure of timbre consistency.
For dialog, we first use forced alignment~\cite{mcauliffe2017montreal} to obtain segments of each speaker.
The final metric is obtained by averaging the consistency scores of individual speakers.

\textbf{Reverb Consistency}.
We assess whether synthesized audio maintains a stable acoustic environment by measuring the consistency of reverberation over time.
For a generated utterance $w$, we apply a sliding window over the waveform and compute the speech-to-reverberation modulation energy ratio~(SRMR) for each window, obtaining a sequence of reverberation scores $\{ r_i \}_{i=1}^{n}$. 
We then compute the standard deviation of this sequence, which serves as our reverb consistency metric; lower variance indicates a more consistent reverberation pattern across the utterance.

\textbf{Sound Fidelity}.
We evaluate the perceptual quality and clarity of the generated speech using the Perceptual Evaluation of Speech Quality (PESQ) metric. 
Given that standard PESQ requires a reference signal unavailable in our setting, we employ \texttt{SQUIM-PESQ} to perform non-intrusive, reference-free evaluation for the synthesized audio.

\textbf{Content Accuracy}.
Faithful content rendering is a cornerstone of robust TTS systems.
To investigate the impact of long-sequence modeling on content fidelity, we employ an ASR-based evaluation, calculating the Word Error Rate (Character Error Rate for Chinese) between the transcripts of the synthesized audio and the ground truth text.

\textbf{Prosodic Coherence}.
While content accuracy ensures lexical correctness, prosodic coherence evaluates the naturalness of delivery.
This metric focuses on pauses, speaking rate, and the consistency of overall prosody to capture the naturalness of generated speech.
\benchmark ~leverages SpeechJudge~\cite{zhang2025speechjudge}, a scoring model fine-tuned from Qwen2.5-Omni-7B~\cite{xu2025qwen2}. 
We refine the input prompt to strengthen the model's sensitivity to prosodic consistency in long-form contexts, utilizing the resulting scalar score (1–5) as our metric for coherence.

\textbf{Expressive Richness}.
In long-form synthesis, expressiveness becomes crucial, as monotonous delivery fails to sustain user engagement or support immersive experiences.
To address this need, \benchmark ~evaluates expressive richness along three dimensions: emotional resonance, character portrayal, and storytelling.
Following EmergentTTS-Eval~\cite{manku2025emergenttts}, we employ LALMs as evaluators using a comprehensive prompt to score audio on a 1–5 scale. 
To ensure fine-grained assessment, we segment inputs into 10-second intervals and calculate the average score across all segments.

\textbf{Expressive Hierarchy}.
Beyond sentence-level expressiveness, paragraph-level expressive hierarchy is a defining characteristic of long-form speech.
We employ LALMs to evaluate this attribute on a scale of 1 to 5,
designing prompts that specifically target emotional variation, vocal dynamics, and scene appropriateness.
Crucially, we evaluate the full utterance rather than via segmentation to preserve the integrity of the narrative flow.

\subsection{Human Perception Alignment Test}\label{sec:bench-human}

To further validate the effectiveness of our evaluation protocol, we conduct a subjective assessment in which human raters score a randomly selected subset of the test data. Additional implementation details and results are provided in the Appendix~\ref{app_sec:human}.

\textbf{Prosody Evaluation}.
We randomly sample 50 pairs of audio clips, each synthesized from identical text by different models, and conduct a subjective preference test with 10 human evaluators. 
For each pair $(A, B)$, raters assess the comparative prosodic coherence on a 5-point scale ranging from -2 to 2. The human preference score is defined as:
\begin{equation}
\mathcal{S}_{\text{pref}}(A, B) = \frac{1}{N} \sum_{i=1}^{N} s_i,
\end{equation}
\noindent where $s_i$ denotes the score assigned by the $i$-th rater, and $N$ represents the total number of raters. We compute the Spearman Rank Correlation Coefficient~(SRCC) between human preference scores and the differential of our metric. The SRCC of 0.82~shows that our metric effectively captures the perceived prosodic coherence of long-form speech.

\textbf{Expressiveness Evaluation}.
We randomly sample 200 audio clips across all models and tasks,
recruiting 10 human evaluators to score each sample, strictly adhering to the same expressiveness prompts used for the LALM evaluation.
In parallel, we benchmark three MOS prediction networks and six LALMs by computing the correlation between their predicted scores and the human Mean Opinion Scores (MOS).
Finally, we select Gemini3-Pro as our primary evaluator, due to its highest alignment with human judgment,
yielding SRCC scores of 0.71 for expressive richness and 0.62 for expressive hierarchy. We also validate the stability of Gemini 3 Pro through independent repeated trials. More results are detailed in Appendix~\ref{app_sec:human_expressiveness}.

%% file: latex/Sections/4_experiements.tex
\section{Experiments} \label{sec:exp}

\subsection{Settings}

\textbf{Model Evaluated} 
\input{latex/Tables/mono_result}
For single-speaker long-form speech, we evaluate ten open-source models: ZipVoice~\cite{zhu2025zipvoice}, SparkTTS~\cite{wang2025spark}, CosyVoice2-0.5B~\cite{du2024cosyvoice2}, CosyVoice3-0.5B~\cite{du2025cosyvoice}, GLM-TTS~\cite{cui2025glm}, MegaTTS3~\cite{jiang2025megatts}, IndexTTS2~\cite{zhou2025indextts2}, FishSpeech-1.5~\cite{liao2024fish}, F5TTS~\cite{chen2024f5}, and VibeVoice~\cite{peng2025vibevoice}. And we evaluate six closed-source flagship systems: Gemini-2.5-pro-preview-tts, OpenAI-tts-1-hd, ElevenLabs Multilingual V2, Minimax-speech-02-hd~\cite{zhang2025minimax}, InWorld-TTS-1-max~\cite{atamanenko2025inworldtts}, and Seed-TTS2~\cite{anastassiou2024seed}.
In the dialogue generation setting, we select six open-source models capable of long-form synthesis—ZipVoice-Dialog~\cite{zhu2025zipvoicedialog}, MoonCast~\cite{ju2025mooncast}, MOSS-TTSD~\cite{zhao2025moss}, FireRedTTS2~\cite{xie2025fireredtts}, VibeVoice, and SoulX-Podcast~\cite{xie2025soulx}—and compare them with four closed-source baselines: Gemini-2.5-pro-preview-tts, OpenAI-tts-1-hd, ElevenLabs Multilingual V2, and SeedTTS-Podcast.

\textbf{Evaluation Models}
For the timbre consistency evaluation, we use WavLM TDCNN\footnote{\url{https://github.com/microsoft/UniSpeech/tree/main/downstreams/speaker_verification}} to extract speaker embeddings, and perform forced alignment with Paraformer\footnote{\url{https://modelscope.cn/models/iic/speech_timestamp_prediction-v1-16k-offline}} on Chinese data and WhisperX~\cite{bain2022whisperx} on English data. 
For WER computation, we adopt FunASR Nano\footnote{\url{https://huggingface.co/FunAudioLLM/Fun-ASR-Nano-2512}} as the transcription model. 
For all expressiveness-related metrics, we use Gemini3-pro~\cite{gemini3} with prompt enhancement as the evaluator.

\subsection{Evaluation from Different Perspectives}
\input{latex/Tables/two_result}
\textbf{Per-Dimension Evaluation}
We demonstrate \benchmark ~scores across all dimensions following the evaluation protocol outlined in Section~\ref{sec:bench-eval}, with results summarized in Tables~\ref{tab:mono_res} and~\ref{tab:two_res}. 
Additionally,  we incorporate two reference baselines:
{\textit{Real Speech}} and {\textit{Real Dialogue}}, which are derived from the source dataset in Section~\ref{sec:bench-collection}, serving as the topological upper bound for audio quality.

\textbf{Per-Scenario Evaluation}
We evaluate the long-form speech and dialog generation models across three core categories spanning 17 different scenarios, and then calculate their performance via the evaluation protocol. Fig.~\ref{fig:exp-scenearios} visualizes the evaluation results of each model in terms of three categories.

\textbf{Evaluations On Generated Length}
We evaluate five representative models~(MegaTTS3, F5TTS, Cosyvoice2, SparkTTS, and VibeVoice) across increasing input lengths among 100 samples in three core scenarios~(Acoustics, Semantics, and Expressiveness). The results are shown in Fig~\ref{fig:exp-time}.

%% file: latex/Tables/mono_result.tex
\begin{table*}[t]
    \caption{
        \textbf{Evaluation results of long-form TTS models across multi-dimensional metrics.} Metrics cover Acoustics (Timbre/Reverb Consistency, Fidelity), Semantics (Content Accuracy, Prosodic Coherence), and Expressiveness (Richness, Hierarchy).
        CER and WER apply to Chinese and English, respectively. 
        Closed-source models and open-source models is separately marked, with the best results in \textbf{bold} and the second best \underline{underlined}.
    }
    \label{tab:mono_res}
    \centering
    \footnotesize
  \resizebox{\linewidth}{!}{
        \begin{tabular}{@{}lccccccc@{}}
        \toprule
        \multicolumn{1}{c|}{}                                 & \multicolumn{3}{c|}{\textbf{Acoustics}}                                    & \multicolumn{2}{c|}{\textbf{Semantics}}                     & \multicolumn{2}{c}{\textbf{Expressiveness}} \\ \cmidrule(l){2-8} 
        \multicolumn{1}{c|}{\multirow{-2}{*}{\textbf{Model}}} & \textbf{Timbre($\uparrow$)} & \textbf{Reverb($\downarrow$)} & \multicolumn{1}{c|}{\textbf{Sound Fidelity($\uparrow$)}}              & \textbf{CER/WER($\downarrow$)} & \multicolumn{1}{c|}{\textbf{Prosody($\uparrow$)}}              & \textbf{Richness($\uparrow$)}          & \textbf{Hierarchy($\uparrow$)}        \\ \midrule
        \rowcolor[HTML]{EFEFEF} 
        \multicolumn{8}{c}{\cellcolor[HTML]{EFEFEF}\textit{\textbf{Open-Source Models}}}                    \\ \midrule
        \multicolumn{1}{c|}{CosyVoice-2}   & {0.92$\pm$0.018} & {2.35$\pm$0.78} & \multicolumn{1}{c|}{3.80$\pm$0.27} & {\textbf{0.032} / 0.168} & \multicolumn{1}{c|}{3.23$\pm$1.01}   & {3.02$\pm$0.68}  & {2.76$\pm$0.88}  \\
        \multicolumn{1}{c|}{CosyVoice-3}   & \textbf{0.94$\pm$0.008}  & {2.26$\pm$0.59}     & \multicolumn{1}{c|}{3.83$\pm$0.10}     & {0.034 / 0.141}    & \multicolumn{1}{c|}{3.31$\pm$0.71}       &      {2.80$\pm$0.70}  & {2.45$\pm$0.75}  \\
        \multicolumn{1}{c|}{FishSpeech}     & {0.93$\pm$0.014} &   {1.79$\pm$0.65}    & \multicolumn{1}{c|}{\textbf{4.10$\pm$0.09}}    &  {0.043 / 0.113}       & \multicolumn{1}{c|}{\underline{3.80$\pm$0.86}}   &  {2.66$\pm$0.78} & {2.90$\pm$0.74}                      \\
        \multicolumn{1}{c|}{F5TTS}       &  {0.90$\pm$0.022}    &  {1.82$\pm$0.77}   & \multicolumn{1}{c|}{3.39$\pm$0.33}   &    {0.072 / 0.113}    & \multicolumn{1}{c|}{3.41$\pm$0.99}        & {3.07$\pm$0.63}   &  {2.77$\pm$0.84}    \\
        \multicolumn{1}{c|}{GLM-TTS}   &  \underline{0.94$\pm$0.010}   &  \textbf{1.62$\pm$0.61}    & \multicolumn{1}{c|}{\underline{3.95$\pm$0.13}}  &   {0.035 / 0.118}  & \multicolumn{1}{c|}{3.64$\pm$0.87} & {2.68$\pm$0.71} & {2.54$\pm$0.88}                     \\
        \multicolumn{1}{c|}{IndexTTS-2}    &  \textbf{0.94$\pm$0.008}   & \underline{1.72$\pm$0.53} & \multicolumn{1}{c|}{2.77$\pm$0.41}   &  {\underline{0.033} / 0.135}  & \multicolumn{1}{c|}{3.64$\pm$0.52}    &  \underline{3.59$\pm$0.72}      &  \underline{2.96$\pm$0.81}           \\
        \multicolumn{1}{c|}{MegaTTS-3}   &  {0.93$\pm$0.008}  &   {1.81$\pm$0.45}   & \multicolumn{1}{c|}{3.55$\pm$0.19} &   {0.035 / \textbf{0.108}}    & \multicolumn{1}{c|}{3.61$\pm$0.84}   &  {2.81$\pm$0.55}    & {2.53$\pm$0.63}   \\
        \multicolumn{1}{c|}{SparkTTS}   & {0.93$\pm$0.033} & {1.79$\pm$1.70} & \multicolumn{1}{c|}{3.59$\pm$0.40}   &  {0.329 / 0.240}  & \multicolumn{1}{c|}{2.58$\pm$1.24}      & {3.47$\pm$0.58}   & {2.38$\pm$0.83}         \\
        \multicolumn{1}{c|}{VibeVoice}     &   {0.93$\pm$0.024}  &  {2.15$\pm$0.88}  & \multicolumn{1}{c|}{3.82$\pm$0.42}     &    {0.047 / \underline{0.111}}   & \multicolumn{1}{c|}{\textbf{3.90$\pm$0.79}} & \textbf{3.71$\pm$0.58}     & \textbf{3.34$\pm$0.88}  \\
        \multicolumn{1}{c|}{ZipVoice}      &  {0.90$\pm$0.011}   & {2.06$\pm$1.08}     & \multicolumn{1}{c|}{3.51$\pm$0.19}        & {0.072 / 0.396}   & \multicolumn{1}{c|}{3.19$\pm$1.11}   & {2.44$\pm$0.85}   & {2.11$\pm$1.05}       \\
        \rowcolor[HTML]{FFFC9E} 
        \multicolumn{1}{c|}{Average}  &  0.93  & 1.95   & \multicolumn{1}{c|}{3.63} &    {0.073 / 0.164}   & \multicolumn{1}{c|}{3.43} &   3.03    &   2.67          \\ \midrule
        \rowcolor[HTML]{EFEFEF} 
        \multicolumn{8}{c}{\cellcolor[HTML]{EFEFEF}\textbf{\textit{Closed-Source models}}}                                                                                                                                                                     \\ \midrule
        \multicolumn{1}{c|}{Elevenlabs Multilingual V2}     & \textbf{0.96$\pm$0.008}   &   {3.05$\pm$0.59}  & \multicolumn{1}{c|}{\textbf{4.02$\pm$0.11}}    &   {0.100 / \underline{0.115}}    & \multicolumn{1}{c|}{3.50$\pm$0.73}    & {2.33$\pm$0.74}  &  {2.68$\pm$0.81}         \\
        \multicolumn{1}{c|}{Gemini-2.5-pro-preview-tts}      & {0.91$\pm$0.018} &   {\underline{1.44$\pm$0.50}}   & \multicolumn{1}{c|}{3.16$\pm$0.36}          &  {0.058 / 0.169}    & \multicolumn{1}{c|}{3.91$\pm$0.72}    &   {\textbf{4.14$\pm$0.65}}  & {\textbf{3.51$\pm$0.84}}        \\
        \multicolumn{1}{c|}{Inworld-TTS-1-max}       &  {0.93$\pm$0.025}   &  {2.19$\pm$0.64}    & \multicolumn{1}{c|}{3.73$\pm$0.17}   &    {0.053 / \textbf{0.113}}  & \multicolumn{1}{c|}{3.71$\pm$0.51}      &  {3.68$\pm$0.86}      & {3.03$\pm$0.92}       \\
        \multicolumn{1}{c|}{Minimax-Speech-02-hd}    & {0.93$\pm$0.010}   &  {\textbf{1.38$\pm$0.35}}   & \multicolumn{1}{c|}{3.82$\pm$0.09}   &   {\textbf{0.032} / 0.119}   & \multicolumn{1}{c|}{\textbf{3.95$\pm$0.73}}    &  {\underline{3.80$\pm$0.44}}      & {\underline{3.26$\pm$0.79}}      \\
         \multicolumn{1}{c|}{OpenAI-tts-01-hd}       &  {0.92$\pm$0.011}  &  {1.74$\pm$0.42}    & \multicolumn{1}{c|}{2.68$\pm$0.12}  &   {\underline{0.043} / 0.119}  & \multicolumn{1}{c|}{\underline{3.91$\pm$0.52}}    &   {3.46$\pm$0.62}  &   {3.25$\pm$0.81}      \\
        \multicolumn{1}{c|}{SeedTTS-2}   & \underline{0.94$\pm$0.022} &   {1.95$\pm$0.74}  & \multicolumn{1}{c|}{\underline{3.88$\pm$0.18}}   &  {0.106 / 0.193}   & \multicolumn{1}{c|}{3.74$\pm$0.44}   &  {3.10$\pm$0.80}  &  {2.34$\pm$0.65}   \\
        \rowcolor[HTML]{FFFC9E} 
        \multicolumn{1}{c|}{Average}  & 0.93  &  1.96    & \multicolumn{1}{c|}{3.55} &    {0.065 / 0.138}  & \multicolumn{1}{c|}{3.79} &  3.42   &   3.01        \\  \midrule
        \multicolumn{1}{c|}{\textbf{Real Speech}}                                           &   0.96    &  1.91   & \multicolumn{1}{c|}{3.62}  &    {0.070 / 0.074}   & \multicolumn{1}{c|}{4.04}       &   4.35    &  3.94       \\
        \bottomrule
        \end{tabular}
    }
    \vspace{-5pt}
\end{table*}

%% file: latex/Tables/two_result.tex
\begin{table*}[t]
    \caption{
        \textbf{Results of dialogue generation models across LFS-Bench's metrics}.
        The performance of closed-source models and open-source models is separately marked, with the best results in \textbf{bold} and the second best \underline{underlined}.
    }
    \label{tab:two_res}
    \centering
    \footnotesize
  \resizebox{\linewidth}{!}{
        \begin{tabular}{@{}lccccccc@{}}
        \toprule
        \multicolumn{1}{c|}{}                                 & \multicolumn{3}{c|}{\textbf{Acoustics}}                                    & \multicolumn{2}{c|}{\textbf{Semantics}}                     & \multicolumn{2}{c}{\textbf{Expressiveness}} \\ \cmidrule(l){2-8} 
        \multicolumn{1}{c|}{\multirow{-2}{*}{\textbf{Model}}} & \textbf{Timbre($\uparrow$)} & \textbf{Reverb($\downarrow$)} & \multicolumn{1}{c|}{\textbf{Sound Fidelity($\uparrow$)}}              & \textbf{CER/WER($\downarrow$)} & \multicolumn{1}{c|}{\textbf{Prosody($\uparrow$)}}              & \textbf{Richness($\uparrow$)}          & \textbf{Hierarchy($\uparrow$)}          \\ \midrule
        \rowcolor[HTML]{EFEFEF} 
        \multicolumn{8}{c}{\cellcolor[HTML]{EFEFEF}\textit{\textbf{Open-Source Models}}}                                                                                                                                                              \\ \midrule
        \multicolumn{1}{c|}{FireRedTTS-2}   & {\underline{0.93$\pm$0.017}} & \underline{3.48$\pm$1.06} & \multicolumn{1}{c|}{2.62$\pm$0.69}      & {0.075 / 0.131} & \multicolumn{1}{c|}{3.24$\pm$1.04}   & {2.72$\pm$0.75} & {2.81$\pm$0.97} \\
        \multicolumn{1}{c|}{MoonCast}       & {0.90$\pm$0.022} & \textbf{3.06$\pm$1.84}  & \multicolumn{1}{c|}{2.62$\pm$0.37}    &  {0.313 / 0.125}  & \multicolumn{1}{c|}{3.16$\pm$1.18}    & {2.68$\pm$0.68} & {2.70$\pm$0.99}       \\
        \multicolumn{1}{c|}{MOSS-TTSD}      &  {0.91$\pm$0.028} & {3.55$\pm$1.16}  & \multicolumn{1}{c|}{2.89$\pm$0.55}    & {0.148 / 0.239}   & \multicolumn{1}{c|}{2.79$\pm$1.14}    & {3.21$\pm$0.79}     & {2.99$\pm$1.06}     \\
        \multicolumn{1}{c|}{SoulX-Podcast}  &  {\textbf{0.93$\pm$0.016}} & {3.51$\pm$0.80}  & \multicolumn{1}{c|}{\textbf{3.96$\pm$0.09}}    & {\textbf{0.061} / \textbf{0.090}}  & \multicolumn{1}{c|}{\textbf{4.01$\pm$0.78}}  & \underline{3.44$\pm$0.69}  & \textbf{3.71$\pm$0.81}        \\
        \multicolumn{1}{c|}{VibeVoice}      & {0.91$\pm$0.028} & {3.59$\pm$0.85}  & \multicolumn{1}{c|}{\underline{3.35$\pm$0.72}}    &  {\underline{0.106} / 0.125}   & \multicolumn{1}{c|}{3.57$\pm$1.05}    & \textbf{3.76$\pm$0.63}  & \underline{3.37$\pm$0.83}                     \\
            \multicolumn{1}{c|}{ZipVoice-Dialog}     & {0.91$\pm$0.021} & {3.53$\pm$0.85} & \multicolumn{1}{c|}{2.66$\pm$0.24}   & {0.069 / \underline{0.114}}   & \multicolumn{1}{c|}{\underline{3.67$\pm$0.89}}   & {2.62$\pm$0.60}  & {2.80$\pm$0.88}      \\
        \rowcolor[HTML]{FFFC9E} 
        \multicolumn{1}{c|}{Average}  &  0.92  &   3.45   & \multicolumn{1}{c|}{3.02} &    {0.129 / 0.137}  & \multicolumn{1}{c|}{3.41} &  3.07   & 3.06              \\ \midrule
        \rowcolor[HTML]{EFEFEF} 
        \multicolumn{8}{c}{\cellcolor[HTML]{EFEFEF}\textbf{\textit{Closed-Source models}}}                                                  \\ \midrule
        
        \multicolumn{1}{c|}{Elevenlabs Multilingual V2} & {\underline{0.93$\pm$0.016}}  & {4.43$\pm$1.01}  & \multicolumn{1}{c|}{\underline{3.48$\pm$0.44}}  & {0.127 / 0.109} & \multicolumn{1}{c|}{3.67$\pm$0.78}   &  {2.84$\pm$0.79}  &  {3.46$\pm$0.87}   \\
        \multicolumn{1}{c|}{Gemini-2.5-pro-preview-tts}        &   {0.92$\pm$0.017} &  {3.17$\pm$0.68}     & \multicolumn{1}{c|}{3.01$\pm$0.24}    &  {\underline{0.086} / \textbf{0.092}} & \multicolumn{1}{c|}{\textbf{4.06$\pm$0.39}}    &   {\textbf{4.06$\pm$0.48}}  &  {\textbf{4.02$\pm$0.68}}                      \\
        \multicolumn{1}{c|}{OpenAI-tts-1-hd}     &  {\textbf{0.93$\pm$0.013}} &  \underline{{2.98$\pm$0.63}}  & \multicolumn{1}{c|}{2.28$\pm$0.17}        & {0.104 / \underline{0.103}}   & \multicolumn{1}{c|}{3.69$\pm$0.62}    &    {3.29$\pm$0.75}      & {3.70$\pm$0.88}   \\
        \multicolumn{1}{c|}{SeedTTS-Podcast}     & {0.91$\pm$0.017}  &   \textbf{{2.85$\pm$0.78}}    & \multicolumn{1}{c|}{\textbf{3.89$\pm$0.17}}      &  {\textbf{0.063} / 0.108} & \multicolumn{1}{c|}{\underline{3.93$\pm$0.46}}    &    {\underline{3.84$\pm$0.72}} & {\underline{3.84$\pm$0.88}}  \\
        \rowcolor[HTML]{FFFC9E} 
        \multicolumn{1}{c|}{Average}  & 0.92  &  3.36  & \multicolumn{1}{c|}{3.17} &   {0.095 / 0.103}   & \multicolumn{1}{c|}{3.83} &    3.51      &   3.76          \\  \midrule
        \multicolumn{1}{c|}{\textbf{Real Dialogue}}        &    0.95  &  2.73   & \multicolumn{1}{c|}{2.94}     &  {0.050 / 0.137} & \multicolumn{1}{c|}{3.95}   &  4.42    &    4.17 \\
        \bottomrule
        \end{tabular}
    }
    \vspace{-5pt}
\end{table*}

%% file: latex/Sections/5_discussions.tex
\section{Insights and Discussions}\label{sec:discussion}


\subsection{Observations}
\textbf{Gap to Ground-Truth Audio}
As shown in Tables~\ref{tab:mono_res} and~\ref{tab:two_res}, among the evaluated systems, \textit{\textbf{VibeVoice}} and \textit{\textbf{SoulX-Podcast}} emerge as the strongest open-source models, 
while \textit{\textbf{Minimax-Speech-02-hd}} and \textit{\textbf{Gemini-2.5-pro-preview-tts}} lead their proprietary counterparts. 
We also observe that, although SOTA open-source models already match or even surpass the best proprietary systems on several evaluation dimensions, 
Proprietary models still exhibit consistently stronger overall performance than open-source models for long-form speech generation.
However, benchmarking against real recordings reveals persistent and systematic gaps.
For long-form synthesized speech, even the best-performing models remain below human speech in overall expressiveness: the closed-source average lags behind real speech by nearly one MOS point in richness and over half a point in hierarchy. 
A similar pattern holds in dialog scenarios, where closed-source systems obtain higher expressiveness, but still fall short of the natural expressivity implied by real dialogue.
In acoustic metrics, synthesized speech approaches real recordings in Fidelity, but long-form outputs show a deficit in Timbre Consistency.
For dialog generation, the marked gap in Reverb Consistency (3.36 vs. 2.73) underscores a core challenge: sustaining global acoustic consistency across multiple speakers.
In terms of Semantics, current models achieve Content Accuracy comparable to real speech, demonstrating strong capability in pronunciation. 
Nevertheless, deficiencies in prosodic coherence persist, limiting the naturalness of the synthesized audio.

\begin{figure*}[t]
  \centerline{\includegraphics[width=\textwidth]{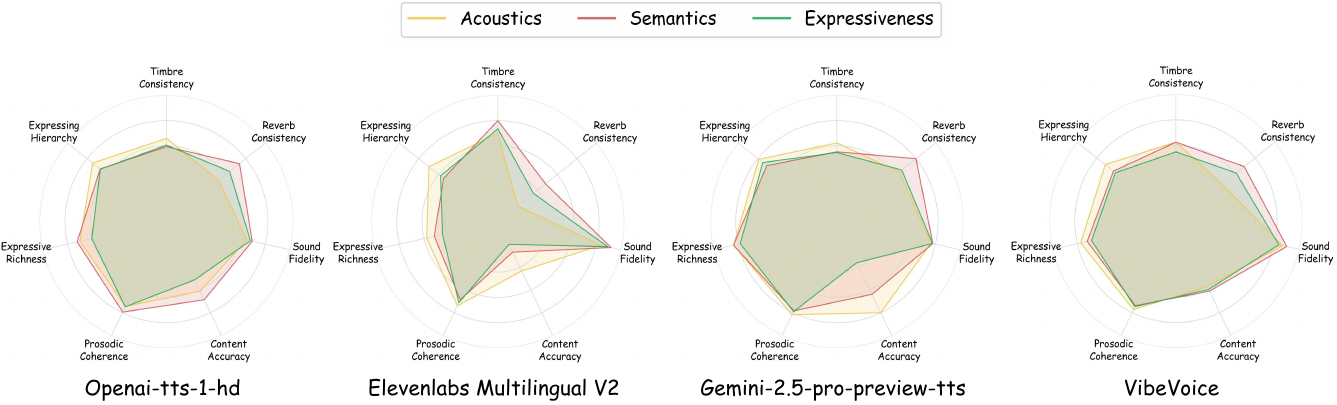}} 
  \caption{
    \textbf{LFS-Bench Results across Three Core Challenges.} For each chart, we plot the evaluation results across three core challenges. The results are normalized between 1 and 5~(larger is better) for visibility across challenges. 
  }
    \vspace{-5pt}
  \label{fig:exp-scenearios}
\end{figure*}

\textbf{Impact of Scenarios.} 
As illustrated in Figure~\ref{fig:exp-scenearios}, downstream scenarios significantly impact generation performance.
\textit{Acoustic challenge scenarios} present distinct difficulties, particularly in maintaining acoustic field consistency. This struggle likely stems from frequent speaker transitions that disrupt reverberation unity, also causing minor fidelity degradation. Notably, however, timbre consistency remains stable, demonstrating the robustness of current models in this dimension.
For \textit{semantic-dominated scenarios}, linguistic complexity in semantic-dominated scenarios does not compromise content accuracy, thanks to robust text normalization. 
However, it poses substantial challenges to prosody modeling, indicating a need for improved comprehension of intricate syntactic structures.
An intriguing finding emerges \textit{in expressiveness settings}.
Here, all models exhibit performance degradation across nearly all metrics, particularly in Expressive Richness.
Theoretically, these scenarios should represent a higher upper bound for expressiveness.
Consequently, this counter-intuitive performance suggests that models may lack effective training on expressive data.
Furthermore, it highlights the substantial gap remaining in achieving immersive and expressive generation. More data support, experimental results, and detailed analysis can be found in Appendix~\ref{app_sec:scenarios}.

\subsection{Discussions}

\begin{figure}[t]
  \centerline{\includegraphics[width=\linewidth]{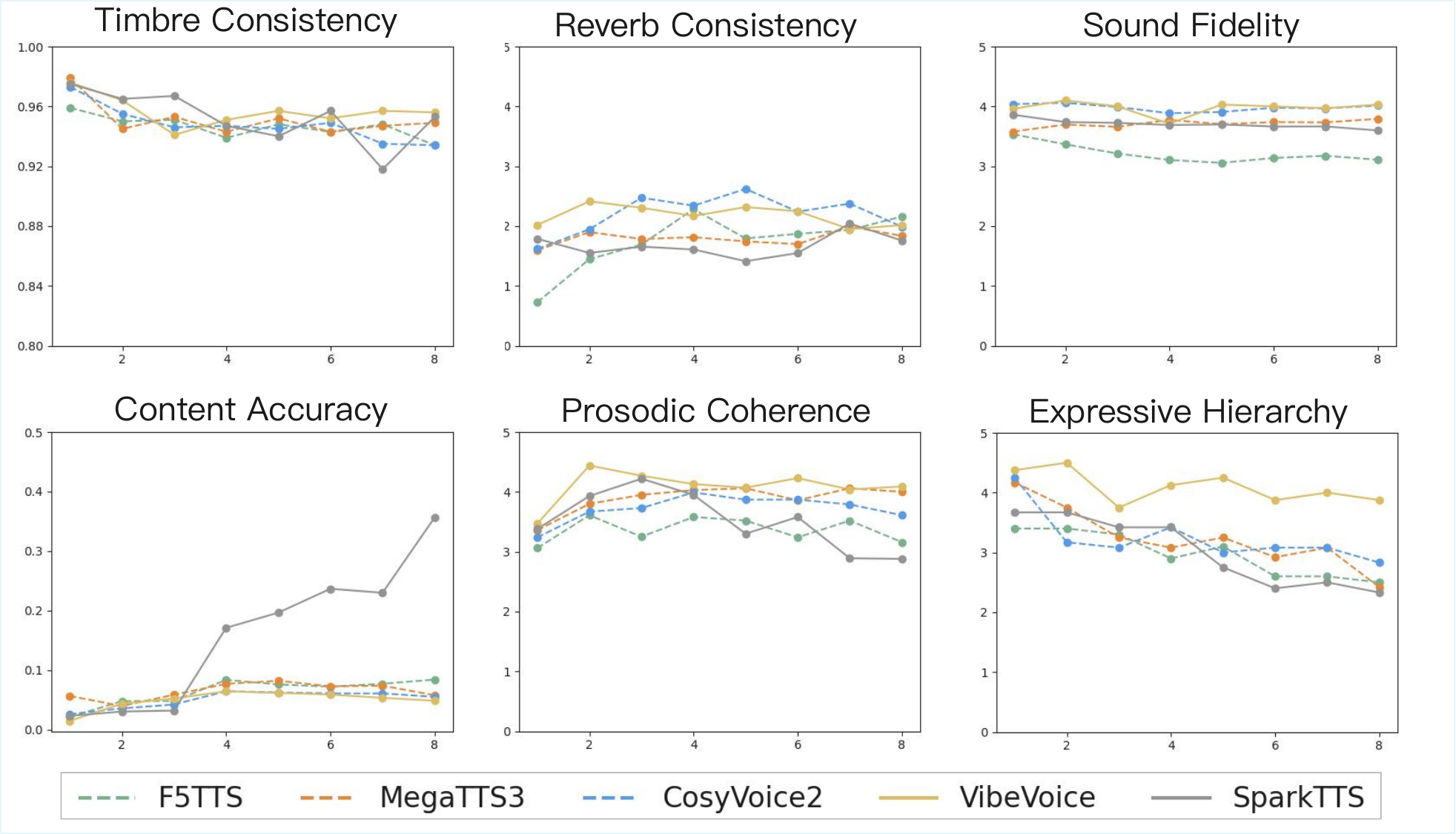}} 
  \caption{
    \textbf{Results on Sequence Length}. 
    The horizontal axis represents the number of sentences in the text.
  }
  \label{fig:exp-time}
  \vspace{-5pt}
\end{figure}

\textbf{AR v.s. NAR} 
In long-form TTS, the choice between AR and NAR paradigms centers on the trade-off between expressiveness and robustness.
NAR models, leveraging parallel generation mechanisms, demonstrate superior robustness and efficiency in long-text synthesis~\cite{ren2020fastspeech}.
However, they tend to produce over-smoothed rhythms, often failing to capture the vocal dynamics and emotional nuances required for extended narration. 
As observed in Table~\ref{tab:mono_res} and~\ref{tab:two_res}, F5TTS, despite being the top-performing NAR model, lags significantly behind most AR counterparts in expressive hierarchy. 
Similarly, ZipVoice-Dialog ranks among the lowest in expressiveness within the dialogue category.
Conversely, AR models, typically built upon language model backbones, excel in prosody modeling but suffer from error propagation in long-form scenarios. While they achieve superior expressiveness, they exhibit a lower bound on Content Accuracy; 
for instance, both SparkTTS and MoonCast show suboptimal performance in this dimension. 
Furthermore, as illustrated in Figure~\ref{fig:exp-time}, SparkTTS suffers from a substantial decline in content accuracy as sequence length increases, whereas NAR models maintain stability without significant degradation.
Consequently, we propose that future long-form TTS architectures should evolve beyond this binary choice toward a Coarse-to-Fine Architecture~\cite{kharitonov2023speak,ju2024naturalspeech}, thereby effectively reconciling long-range semantic coherence with local generation stability.


\textbf{Data Quality v.s. Data Quantity} 
While scaling laws have advanced speech synthesis by leveraging more data and bigger parameters~\cite{du2025cosyvoice}, our analysis suggests that relying solely on mainstream datasets presents three critical impediments to long-form audio generation:
1) \textbf{Fragmentation in open-source data}~\cite{chen2021gigaspeech} induces a short-form bias that compromises discourse coherence. For instance, SparkTTS is trained on VoxBox, a dataset characterized by an average segment duration of less than 10 seconds. Consequently, the model exhibits significant degradation in both content accuracy and prosodic coherence as the generation length extends, as illustrated in Figure~\ref{fig:exp-time};
2) \textbf{Acoustic instability} in web-crawled data~\cite{he2024emilia}, such as variable noise and recording conditions, triggers acoustic drift. For example, CosyVoice3 utilizes extensive in-the-wild data for training. As a result, it significantly lags behind other models in reverb consistency, as shown in Table~\ref{tab:mono_res};
and 
3) The \textbf{averaging effect} of scaling enhances generalization but homogenizes expressiveness. As shown in Table~\ref{tab:mono_res}, flagship models such as GLM-TTS and FishSpeech excel in acoustic metrics. However, they underperform in the expressiveness dimension despite their large scale. Consequently, they fail to capture the dynamic nuances required for narration.
Therefore, the path forward requires a strategic shift towards prioritizing data quality and temporal continuity over raw quantity.
We advocate for the adoption of curriculum-learning strategies~\cite{wang2021survey} that progressively transition from sentence-level to paragraph-level training. 
By leveraging high-fidelity, long-context recordings, future models can more effectively capture the long-range dependencies essential for coherent and expressive narration.

%% file: latex/Sections/8_conclusions.tex
\section{Conclusion}\label{sec:con}
In this work, we present \benchmark, a holistic benchmark tailored for evaluating long-form TTS models.
\benchmark ~addresses three core challenges in long-form generation, encompassing 1,101 carefully curated instances across 17 downstream scenarios.
To facilitate precise and automatic assessment, we propose a disentangled, human-aligned evaluation protocol featuring seven complementary metric dimensions.
Through extensive benchmarking of over 20 models, we provide an in-depth analysis of current capabilities and limitations from the perspectives of model architectures as well as training data and strategy.
We envision \benchmark ~as a standardized testbed for future research, propelling the development of more robust and immersive long-form speech synthesis.

%% file: latex/Sections/9_appendix.tex
\section*{Appendix Contents}
The Appendix is structured as follows:
\begin{itemize}
\setlength{\leftskip}{0.5em}
    \item Section~\ref{app_sec:dataset}: Details of dataset construction, including the detailed explanation of scenarios as well as the complete process of data collection and refinement.
    \item Section~\ref{app_sec:stat}: Statistics of LFS-Bench.
    \item Section~\ref{app_sec:eval}: Details of Evaluation Protocols.
    \item Section~\ref{app_sec:human}: Details of the validation of human alignment and the user study.
    \item Section~\ref{app_sec:detail}: The details of the experiment's setting. 
    \item Section~\ref{app_sec:exp}: Ablation studies and experiments related to multi-speaker dialogue evaluation.
    \item Section~\ref{app_sec:analysis}: More reuslts and analysis of the experiments.
    \item Section~\ref{app_sec:future}: Limitations and future works.
    \item Section~\ref{app_sec:social}: Potential social impact of \benchmark.
\end{itemize}

\input{latex/app_sec/01_datasets}

\input{latex/app_sec/02_stat}
\input{latex/app_sec/03_eval}

\input{latex/app_sec/04_human}

\input{latex/app_sec/05_implement_detail}

\input{latex/app_sec/06_more_exp}
\input{latex/app_sec/07_detailed_analysis}
\input{latex/app_sec/future_work_social_impact}
\input{latex/app_figs/prompt_prosody_coherence}
\input{latex/app_figs/prompt_expressive_hierarchy}
\input{latex/app_figs/prompt_expressive_richness}

%% file: latex/app_sec/01_datasets.tex
\section{Details of \benchmark's Construction}\label{app_sec:dataset}

\subsection{Explanation of Scenarios}

\benchmark ~systematically categorizes the challenges inherent in current long-form speech generation into three primary dimensions:~\textbf{Acoustics, Semantics, and Expressiveness}.
To facilitate a more fine-grained and precise assessment, we curate a dataset of 1,101 audio samples aligned with these dimensions, encompassing 17 downstream scenarios such as audiobooks, podcasts, talk shows, and news broadcasting. 
In the following section, we comprehensively detail the audio scenarios and data selection criteria associated with each challenge category.

\subsubsection{Scenarios for Acoustics Challenges}
In the context of long-form TTS and dialogue generation tasks, the primary user concerns regarding acoustic performance are categorized as follows:
\begin{itemize}
\setlength{\leftskip}{0.1em}
    \item \textbf{Audio Quality:} As a fundamental requirement, the generated audio must be devoid of background noise and electronic artifacts, ensuring high fidelity and clear auditory perception for the user.
    \item \textbf{Timbre Consistency:} In single-speaker settings, the speaker's timbre must remain perceptually consistent throughout the sequence, analogous to identity preservation in video generation tasks. In multi-speaker dialog scenarios, accurate speaker transitions are critical, requiring precise alignment between the dialogue script and the corresponding speaker identities.
    \item \textbf{Acoustic Environment Consistency:} The ability to maintain a stable sound field is a core capability in long-form speech generation. This requires unity across acoustic dimensions, such as the recording environment and sound imaging. Furthermore, in multi-speaker contexts, ensuring that different speakers appear to share a unified acoustic scene is a crucial objective.
\end{itemize}
Based on the above basic requirements, we select six audio downstream scenarios to construct test cases related to the acoustic dimension, which are specifically introduced as follows.

\textbf{Customer Service} Widely deployed in e-commerce, AI agents frequently deliver lengthy responses detailing policies and products. This scenario demands high-fidelity, artifact-free audio to maintain professional credibility and ensure a trustworthy user experience.

\textbf{Audiobooks} As a quintessential long-form scenario, audiobooks demand rigorous acoustic consistency. The synthesis must maintain timbre stability to mitigate "speaker drift," preserve a stationary acoustic environment to ensure immersion, and guarantee high-fidelity quality for prolonged listening comfort.

\textbf{Podcasts} This scenario focuses on multi-turn dialogue generation and natural interaction. Characterized by an informal or semi-formal conversational style, this domain places relatively lower demands on dramatic expressiveness; however, it imposes strict requirements on turn-taking transitions. Consequently, this scenario necessitates that TTS models not only execute accurate speaker switching but also synthesize appropriate and stable reverberation to reconstruct an authentic and vivid conversational atmosphere.

\textbf{Chat, Debate, and Interview} While lacking direct commercial applications, these real-world scenarios serve as benchmarks for acoustic modeling limits. The frequent speaker transitions inherent in these domains pose significant challenges to synthesis systems. Furthermore, the associated complex acoustic environments introduce additional layers of difficulty regarding background noise and channel variability.

\subsubsection{Scenarios for Semantics Challenges}

In the semantic dimension, long-form speech generation is categorized into two sub-dimensions: accuracy and naturalness.
\begin{itemize}
\setlength{\leftskip}{0.1em}
    \item \textbf{Content Accuracy}: Evaluates the alignment between the generated speech and the input text. 
    In long-sequence generation, this metric primarily assesses the model's robustness against omissions, repetitions, and hallucinations, ensuring high content fidelity.
    \item \textbf{Prosodic Coherence}: Evaluates the consistency between prosodic structure and semantic logic. 
    Beyond natural pausing, this includes the appropriate handling of stress and intonation, ensuring a fluent rhythm at the paragraph level and avoiding mechanical or disjointed delivery.
\end{itemize}
To rigorously evaluate model performance regarding semantic challenges, we construct test cases across five downstream scenarios, specifically targeting the two aforementioned dimensions.

\textbf{News and Popular Science} 
In these scenarios, content correctness is paramount, as users exhibit minimal tolerance for semantic deviations. 
Consequently, we curate instances featuring linguistic complexity, challenging pronunciations, and domain-specific knowledge to comprehensively assess model robustness.

\textbf{Lesson, Seminar, and Presentation}
Beyond basic accuracy, these scenarios impose higher demands on naturalness. Speakers are expected to enhance auditory perception through appropriate stress and rhythmic cadence. 
Therefore, in addition to content complexity, we incorporated colloquial expressions and diverse prosodic structures to further evaluate the model's prosodic coherence.

\subsubsection{Scenarios for Expressiveness Challenges}

Immersion and high expressiveness are the ultimate goals of audio synthesis. For long-form generation, given its temporal complexity, we decompose expressiveness into Richness and Hierarchy.
\begin{itemize}
\setlength{\leftskip}{0.1em}
    \item \textbf{Expressive Richness}: Evaluates the overall expressive quality through the lenses of emotional resonance, character portrayal, and storytelling. Similar to sentence-level synthesis, this metric primarily focuses on the **average magnitude** of expressiveness maintained throughout the entire audio sequence.
    \item \textbf{Expressive Hierarchy}: Represents the fundamental distinction between paragraph-level and sentence-level generation. The extended context necessitates a focus on dynamic variations (e.g., shifts in emotion and volume) and the alignment between the acoustic evolution and the semantic scenario.
\end{itemize}

Guided by these evaluation dimensions, we curate test cases across six highly expressive downstream scenarios to rigorously probe the upper boundaries of model capabilities within SwanBench-Speech.

\textbf{Sportcast and Live Streaming}: These scenarios predominantly challenge Expressive Richness. Characterized by sustained high-intensity delivery and emotional saturation, they demand that the model maintain a consistently elevated energy level to match the fast-paced nature of the content.

\textbf{Speech, Host, Talkshow, and Drama}: These domains necessitate a synergy of both Richness and Hierarchy. Beyond high emotional fidelity, they require sophisticated control over dynamic evolution, such as tension building in drama or rhythmic variation in hosting, to ensure the acoustic delivery aligns seamlessly with the narrative arc.

\subsection{Details of Data Collection}
In this section, we provide further elaboration on the data sources and processing pipeline of SwanBench-Speech.

\subsubsection{Online Text Corpora} 
For the Audiobook, News, Drama, and Host scenarios, we harvest long-form texts from diverse online resources, spanning classic literature, web novels, and TouTiao\footnote{\url{https://app.toutiao.com/news_article/}}. 
Following data acquisition via OCR or web crawling, we employ the \texttt{clean-text}\footnote{\url{https://pypi.org/project/clean-text/}} library to sanitize the raw corpus by removing artifacts such as URLs, emojis, and garbled characters.
Subsequently, human annotators conduct rigorous quality assurance and enrich the dataset with metadata labels for scenario, topic, and speaker identity.

\subsubsection{Online Audio Media}

We extensively utilize online audio materials across various scenarios, with data sources including YouTube\footnote{\url{https://www.youtube.com}}, Bilibili\footnote{\url{https://www.bilibili.com}}, Spotify\footnote{\url{https://open.spotify.com/}}, RedNote\footnote{\url{https://www.xiaohongshu.com/}}, and Apple Podcasts\footnote{\url{https://podcasts.apple.com/}}. 
First, we crawl audio materials tailored to our target scenarios from these platforms~\cite{chen2025wavrag}. Subsequently, we denoise the raw audio using Zipenhancer~\cite{wang2025zipenhancer} to ensure processing accuracy. 
After obtaining cleaner data, we filter out samples with low expressiveness and quality based on a DNS-MOS~\cite{reddy2021dnsmos} threshold of 3.5. 
We then perform speaker diarization using 3D-Speaker~\cite{zheng20233d} and transcribed the resulting audio segments via SenseVoice-Small\footnote{\url{https://huggingface.co/FunAudioLLM/SenseVoiceSmall}}. 
Finally, human annotators are employed to proofread the machine-generated transcripts against the ground truth and update the metadata labels.

\subsubsection{LLM Generation}

\input{latex/app_figs/llm_text_generation}
In scenarios such as chat, presentations, and customer service, we leverage GPT-5~\cite{gpt5} to facilitate the generation of high-quality test cases. 
Specifically, we designe sophisticated prompts to guide the LLM in producing structured content that aligns with specific scenarios and topics while maintaining a certain level of generation complexity. 
Figure~\ref{app_fig:llm_generation} illustrates a set of prompts used for generating presentation topics for computer science students. These structured prompts serve as customizable templates, allowing users to adapt them for generating diverse long-form data.
After LLM generation, the generated content is mutually proofread by annotators.

We recruit three undergraduate students for data annotation and verification, compensated at a rate of \$0.20 per instance. To ensure quality, all data samples are double-checked. The total expenditure for the data collection process amount to \$220.

Using LLM to generate data is to better achieve data scaling and update test cases. As introduced in Section 3.2 and Section 3.3, we conduct comprehensive checks on LLM-generated cases through multiple dimensions including repetition detection, quality checks, privacy checks, and social and ethical reviews. This multi-faceted approach aims to mitigate issues associated with LLM-generated data, such as data quality degradation and privacy infringement.

\subsection{Details of Data Refinement}
\input{latex/app_figs/llm_text_eval}

\subsubsection{Semantic De-duplication}
To ensure data diversity, we perform topic-level deduplication on both crawled and generated test instances. Specifically, we utilized GPT-5 to extract topics, keywords, and summaries from each long-text instance. These elements are concatenated and encoded into embeddings using Sentence-BERT\footnote{\url{https://huggingface.co/sentence-transformers/all-MiniLM-L6-v2}}~\cite{reimers2019sentence}. We then filter out semantically redundant samples based on a cosine similarity threshold of 0.8 and replenish the dataset via LLM-based generation.

\subsubsection{Quality Evaluation}
We further employ GPT-5 to assess the quality of the de-duplicated samples. Specifically, we design prompts to evaluate textual expressiveness and content consistency, guiding the LLM to rate the suitability of each instance for long-form speech generation on a scale of 1 to 5. Only samples with recommendation scores exceeding 2 are retained. The specific prompt used for this quality assessment is in Figure~\ref{app_fig:llm_eval}.

\subsubsection{Privacy and Ethical Filtering}

To ensure the safety and anonymity of our dataset, we employ DeepSeek V3.2~\cite{liu2024deepseek} to conduct a rigorous privacy and ethical assessment. Specifically, we design a prompt incorporating Chain-of-Thought (CoT)~\cite{wei2022chain} reasoning to guide the model through a two-step analysis:

\begin{enumerate}
    \item \textbf{Selective PII Anonymization}: The model is instructed to specifically identify and anonymize the names of \textbf{private individuals} (non-public figures). While the names of celebrities or public entities are retained to preserve contextual integrity, the names of ordinary citizens are replaced with generic placeholders or synthetic alternatives.
    \item \textbf{Ethical Risk Assessment}: The model then scrutinizes the content for social and ethical risks, including hate speech, violence, sexual explicitness, and bias. 
\end{enumerate}

Based on this analysis, samples containing toxic content are discarded, while those with minor sensitivity issues are revised. The specific prompt used for this filtering is presented in Figure~\ref{app_fig:safety_prompt}.

\input{latex/app_figs/llm_safety}

\subsubsection{Manual Review}

Following the automated filtering pipelines, we implement a three-stage human-in-the-loop review process to finalize the dataset. Expert annotators execute the following operations:

\begin{enumerate}
    \item \textbf{Harmless Placeholder Infilling}: For samples that underwent privacy anonymization, the automated generic tags (e.g., \texttt{[NAME]}, \texttt{[LOC]}) are replaced with specific but fictitious entities. This step ensures the text remains natural and grammatically fluid while strictly maintaining the harmlessness and anonymity.
    \item \textbf{Residual Error Purging}: Annotators then scrutinize the dataset to identify subtle logical inconsistencies, formatting errors, or context mismatches that might have evaded the automated filters. Samples deemed substandard or unnatural are strictly discarded.
    \item \textbf{Dataset Replenishment}: To compensate for the discarded samples and maintain the volume, new instances are constructed. These replenished samples undergo the same process before being added to the final pool.
\end{enumerate}

Five undergraduate students are enlisted for this manual review, receiving a compensation of \$0.30 per instance. The cumulative expenditure for the data collection process totaled \$330.

\subsection{Instructions for Use}

The test set will be released on Hugging Face under the \textbf{CC BY-NC-SA 4.0} license, allowing for free non-commercial use. For evaluations involving additional voice profiles on our benchmark, users must strictly adhere to the specific licenses associated with those assets. Furthermore, the complete codebase for data processing and evaluation will be made publicly available on our GitHub repository.

%% file: latex/app_figs/llm_text_generation.tex
\begin{figure*}[t]
\centering
\begin{promptbox}{Prompt for generating structured presentation data}
\small
You are an expert computer science professor and content creator. Your task is to generate a high-quality, long-form presentation script on the topic: \textbf{[Insert Topic Here]}.

\textbf{Generation Requirements:}
1. \textbf{Complexity}: The content must be academically rigorous, suitable for computer science students. Include technical terminology and logical reasoning.
2. \textbf{Structure}: The speech should be coherent but segmented into logical paragraphs.
3. \textbf{Format}: You must strictly output a valid JSON object without any Markdown formatting.

\textbf{JSON Schema:} \\
\{ \\
\hspace*{1em} "content": [ \\
\hspace*{2em} \{ \\
\hspace*{3em} "speaker": "Speaker1", \\
\hspace*{3em} "text": "The first paragraph of the speech..." \\
\hspace*{2em} \}, \\
\hspace*{2em} \{ \\
\hspace*{3em} "speaker": "Speaker1", \\
\hspace*{3em} "text": "The second paragraph of the speech..." \\
\hspace*{2em} \} \\
\hspace*{1em} ], \\
\hspace*{1em} "num\_speakers": 1, \\
\hspace*{1em} "theme": "[Insert Topic Here]", \\
\hspace*{1em} "source": "LLM Generation", \\
\hspace*{1em} "TLDR": "A one-sentence summary of the presentation." \\
\}
\end{promptbox}
\vspace{-2pt}
\caption{Prompt template used for generating presentation topics for computer science students.}
\vspace{-5pt}
\label{app_fig:llm_generation}
\end{figure*}

%% file: latex/app_figs/llm_text_eval.tex
\begin{figure*}[t]
\centering
\begin{promptbox}{Prompt for the evaluation of long-form instances}
\small
You are an expert linguist and data quality evaluator. Your task is to assess the suitability of the following text sample for \textbf{long-form speech generation}.

Please evaluate the text based on the following two criteria:

1. \textbf{Textual Expressiveness}: Assess the fluency, naturalness, and rhetorical quality of the text. Is the language vivid and rhythmically suitable for long-duration speech synthesis?

2. \textbf{Content Consistency}: Assess the logical coherence and semantic stability of the text. Is the narrative or argument consistent throughout without contradictions or abrupt topic shifts?

Rate each criterion on a scale of 1 to 5 (1 = Poor, 5 = Excellent). Based on these, provide an \textbf{Overall Score} (1-5) representing your recommendation for retaining this sample.

\textbf{Output Requirement:}

You must output the result strictly in the following JSON format:

\{ \\
\quad \quad \quad \quad \quad \quad "reasoning": "Provide a brief analysis explaining the scores, highlighting pros and cons.", \\
\quad \quad \quad \quad \quad \quad "textual\_expressiveness\_score": <integer between 1 and 5>, \\
\quad \quad \quad \quad \quad \quad "content\_consistency\_score": <integer between 1 and 5>, \\
\quad \quad \quad \quad \quad \quad "overall\_score": <integer between 1 and 5> \\
\}

\textbf{Text to Evaluate:}

[Insert Text Here]
\end{promptbox}
\vspace{-2pt}
\caption{Prompt template for the quality evaluation of test instances.}
\label{app_fig:llm_eval}
\end{figure*}

%% file: latex/app_figs/llm_safety.tex
\begin{figure*}[t]
\centering
\begin{promptbox}{Prompt for Privacy and Ethical Filtering}
\small
\textbf{Role:} You are an expert data safety and privacy compliance assistant. Your task is to review the input text for privacy leaks and ethical risks.

\textbf{Instructions:}
Please analyze the input text following these steps (Chain-of-Thought):
\begin{enumerate}
    \item \textbf{PII Detection (Selective):} Identify all person names. 
    \begin{itemize}
        \item If the name belongs to a \textbf{public figure} (celebrity, politician, historical figure), \underline{retain it} to preserve context.
        \item If the name belongs to a \textbf{private individual} (ordinary citizen), \underline{anonymize it} using a placeholder (e.g., [NAME]).
    \end{itemize}
    \item \textbf{Ethical Risk Assessment:} Check for hate speech, explicit violence, sexual content, or severe bias.
    \begin{itemize}
        \item If the risk is severe and cannot be mitigated, mark as invalid.
        \item If the risk is minor or related to PII, provide a revised version.
    \end{itemize}
\end{enumerate}

\textbf{Output Format:}
Output the result in a strict JSON format with the following keys:
\begin{itemize}
    \item \texttt{"reasoning"}: A brief explanation of your analysis regarding PII and safety risks.
    \item \texttt{"valid"}: Boolean (true/false). Set to \texttt{false} only if the content contains unmitigable toxic content. Set to \texttt{true} if it is safe or has been successfully anonymized/revised.
    \item \texttt{"revised\_text"}: The clean version of the text after anonymizing private names and removing minor risks. If invalid, return an empty string.
\end{itemize}

\textbf{Input Text:} [INPUT\_TEXT]
\end{promptbox}
\caption{The prompt template used for privacy and ethical filtering. It guides the LLM to selectively anonymize private individuals' names while retaining public figures, and outputs the decision in a structured JSON format.}
\label{app_fig:safety_prompt}
\end{figure*}

%% file: latex/app_sec/02_stat.tex
\section{Statistics of SwanBench-Speech}\label{app_sec:stat}

\subsection{Categorical Statistics}
\begin{figure*}[t]
  \centerline{\includegraphics[width=\textwidth]{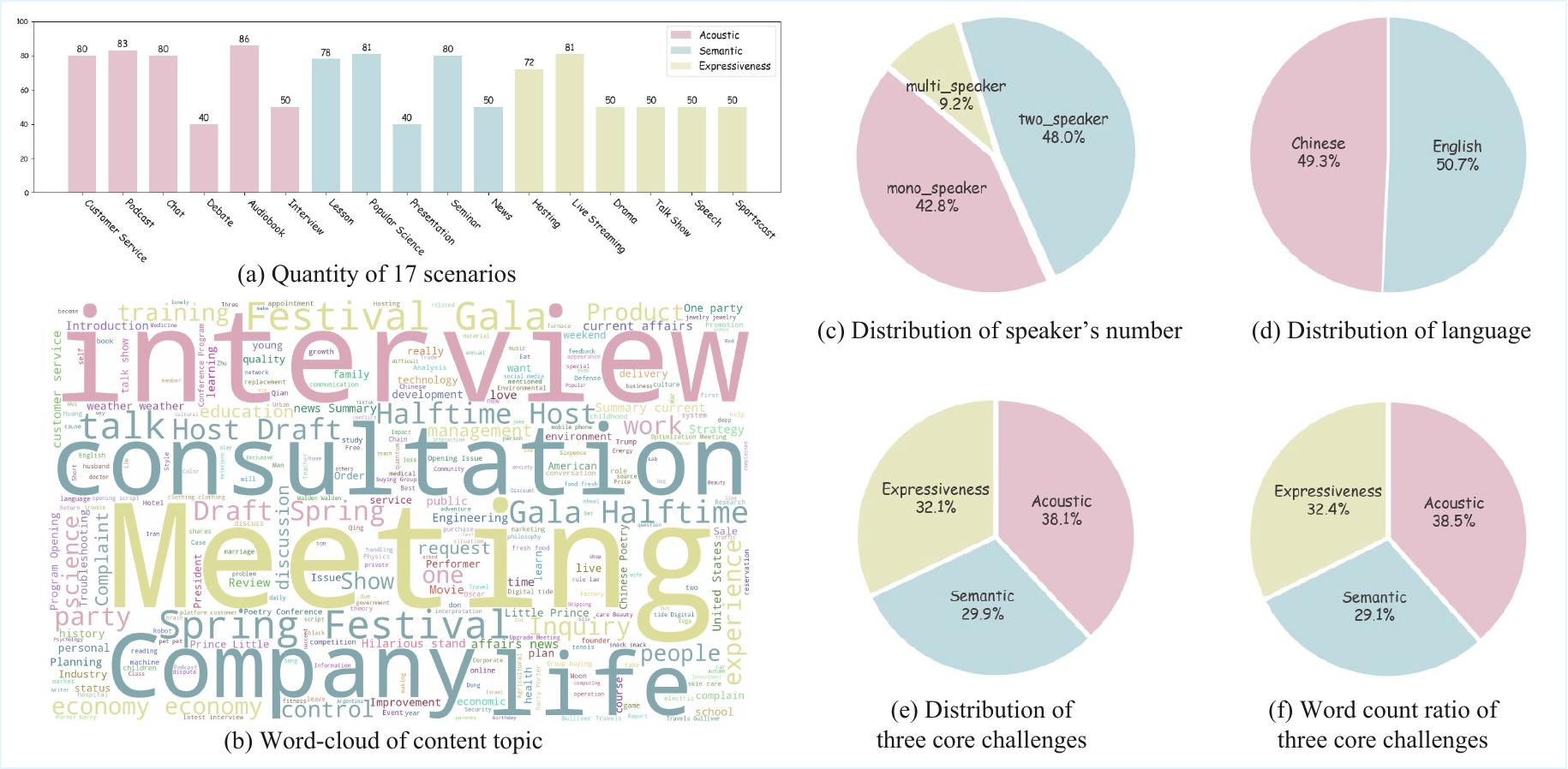}} 
  \caption{
    The categorical statistics of SwanBench-Speech across five key dimensions: language, speaker numbers, core challenges, content topics and scenarios.
  }
  \label{app_fig:cate_stats}
\end{figure*}

We present a comprehensive statistical analysis of the 1,101 samples in SwanBench-Speech across five key dimensions: language (Chinese/English), speaker configuration (single/dual/multi-speaker), core challenges (Acoustics, Semantics, Expressiveness), scenarios, and content topics, as illustrated in Figure~\ref{app_fig:cate_stats}.
As observed, SwanBench-Speech maintains a strictly balanced language ratio, comprising 49.3\% Chinese and 50.7\% English samples.Regarding language selection, given that the application ecosystems for both Chinese and English in long-form speech generation tasks are already relatively mature, we have decided to focus solely on these two languages at this stage in order to include as many baseline models as possible and to validate the effectiveness and necessity of SwanBench-Speech. 

Regarding speaker configuration, while the dataset primarily focuses on single-speaker long-form speech and dual-speaker dialogue, we explicitly include 101 multi-speaker samples (involving 3 or 4 speakers) to facilitate the evaluation of multi-talker generation capabilities. Furthermore, the dataset exhibits a relatively even distribution across the three core challenges, with the Acoustics challenge accounting for the largest proportion at 34.5\%. We also quantify the sample distribution across 17 specific downstream scenarios and generate a word cloud to visualize the topic diversity. This balanced scenario distribution, combined with a rich variety of content topics, minimizes potential bias during the evaluation process.

\subsection{Distributional Statistics}
\begin{figure*}[t]
  \centerline{\includegraphics[width=\textwidth]{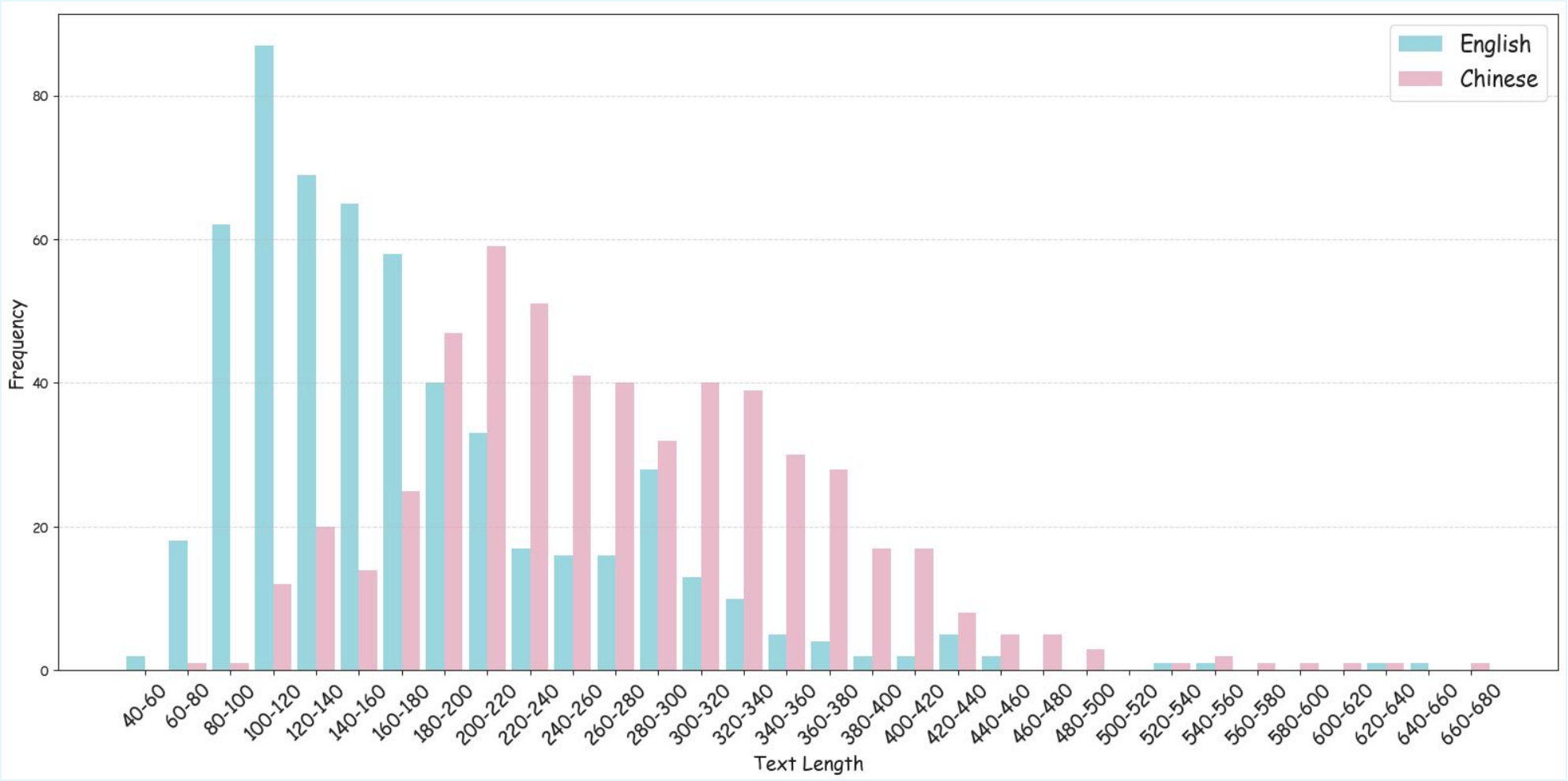}} 
  \caption{
    The statistics of the text length distribution within SwanBench-Speech. 
    The red dashed line indicates the average text length of English, and the green dashed line indicates the average text length of Chinese.
  }
  \label{app_fig:distribution_stats}
\end{figure*}

We also conduct a detailed analysis of the text length distribution within SwanBench-Speech, as illustrated in Figure~\ref{app_fig:distribution_stats}.
Specifically, text length is quantified by the number of characters for Chinese data and the number of words for English data, excluding non-phonetic elements such as punctuation. The results indicate that text lengths for both languages follow an approximate normal distribution, primarily concentrate within the interval $[80, 500]$, with mean lengths of $271.8$ for Chinese and $174.6$ for English. 

Although application scenarios like audiobooks may require speech synthesis lasting over 10 minutes or even an hour, for the vast majority of application scenarios demanding extended speech—such as live streaming, customer service, and talk shows—minute-level synthesis quality remains the primary concern for users. Therefore, we selected the word count corresponding to minute-level speech from the perspective of downstream scenarios, specifically the range of 200 to 400 words. Previous studies ~\cite{guo2024fireredtts, zhang2025isdrama,xie2025soulx}also indicate that such duration is sufficient to reveal long-term dependency issues during synthesis. Through experiments on generated length in Section 4.2 and Appendix F.3, we found that when synthesis exceeds 100 words, most models exhibit varying degrees of performance degradation across multiple dimensions,including Timbre Consistency, Reverb Consistency, and Expressive Hierarchy. This indicates that this length range can already reveal inherent dependency issues in long-form speech generation. This distribution effectively supports the rigorous and realistic evaluation of long-form speech generation capabilities.

%% file: latex/app_sec/03_eval.tex
\section{Details of Evaluation Protocol}\label{app_sec:eval}

\subsection{Timbre Consistency}
To evaluate timbre consistency, we adopt a segment-based speaker similarity approach following prior zero-shot vocal generation studies~\cite{du2024cosyvoice2,ji2024wavtokenizer,zhang2024stylesinger,zhang2024tcsinger}. 

Specifically, for a \textbf{single-speaker} long-form speech sample $w$, we apply a sliding window over the waveform to extract a sequence of speaker embeddings $\{\mathbf{e}_i \}_{i=1}^{n}$ by WavLM TDCNN~\footnote{\url{https://huggingface.co/docs/transformers/en/model_doc/unispeech-sat}}, where $n$ denotes the number of windows.
Given that speaker embeddings are sensitive to segment duration and verification models are typically optimized for 2--4s segments, \textbf{we employ a window length of 3s with a stride of 2s.} We then compute the pairwise cosine similarity between all distinct embeddings:
\begin{equation}
  \mathrm{sim}_{i,j} = \cos\!\left( \frac{\mathbf{e}_i}{\lVert \mathbf{e}_i \rVert}, \frac{\mathbf{e}_j}{\lVert \mathbf{e}_j \rVert} \right), \quad \forall i \neq j.
\end{equation}
Finally, we utilize the average score of the resulting similarity sequence $\{\mathrm{sim}_{i,j} \}$ as the quantitative metric for timbre consistency.

Evaluating dual and multi-speaker scenarios is inherently more complex due to the involvement of speaker transitions. To ensure validity, we first utilize 3D-Speaker~\cite{zheng20233d} to verify the number of speakers, confirming that at least one successful speaker turn occurs.
Subsequently, let $K$ denote the number of distinct speakers in the generated audio. We employ forced alignment to obtain sentence-level timestamps and concatenate speech segments belonging to each speaker $k \in \{1, \dots, K\}$, yielding a speaker-specific audio stream $\tilde{w}_k$. We utilize a Paraformer-based Align Model\footnote{\url{https://modelscope.cn/models/iic/speech_timestamp_prediction-v1-16k-offline}}~\cite{gao2022paraformer} for Chinese data and WhisperX\footnote{\url{https://github.com/m-bain/whisperX}}~\cite{bain2022whisperx} for English data. Both models demonstrate alignment errors of less than 100ms on minute-level recordings, minimizing error accumulation.
Finally, for each speaker-specific stream $\tilde{w}_k$, we compute its similarity average $a_k$ following the single-speaker protocol defined above. The final metric is calculated as the average across all speakers:
\begin{equation}
  \mathrm{Score}_{\text{multi}} = \frac{1}{K} \sum_{k=1}^{K} a_k.
\end{equation}

\subsection{Reverb Consistency}

We employ the Speech-to-Reverberation Modulation Energy Ratio (SRMR) to quantify reverberation intensity, analyzing its temporal fluctuations to evaluate the model's ability to maintain a consistent acoustic environment.

Specifically, for a generated utterance $w$, we apply a sliding window to compute the SRMR for each segment using the SRMRpy toolkit\footnote{\url{https://github.com/jfsantos/SRMRpy}}. To balance estimation reliability with the temporal resolution required to detect ``reverberation drift'', we adopt a window size of 3s and a stride of 2s, consistent with our timbre consistency evaluation.

Furthermore, to mitigate the impact of non-speech segments (e.g., silence or noise) on the statistical analysis, we pre-filter each window using a Voice Activity Detection (VAD) model\footnote{\url{https://modelscope.cn/models/iic/speech_fsmn_vad_zh-cn-16k-common-pytorch}, \\ \url{https://github.com/snakers4/silero-vad}}. Any window containing more than 60\% non-speech frames is discarded. 
This process yields a sequence of valid reverberation scores $\{r_i\}_{i=1}^n$, where $n$ denotes the number of effective windows. 

Finally, we compute the standard deviation of this sequence as our Reverb Consistency metric; a lower value indicates a more stable reverberation pattern throughout the utterance.

It is important to note that this metric is predicated on the assumption that the \textbf{acoustic environment within a single long-form segment should remain stable}. We acknowledge that specific scenarios, such as \textit{Outdoor Live Streaming}, may inherently require dynamic acoustic shifts for semantic correctness. However, for the majority of standard long-form synthesis tasks, acoustic stability serves as a critical indicator of generation robustness; therefore, we treat high variance as a penalty in this evaluation framework.

\subsection{Sound Fidelity}

To achieve a non-intrusive, reference-free assessment of audio fidelity, we directly utilize the SQUIM-PESQ metric via the official \texttt{Torchaudio} interface\footnote{\url{https://docs.pytorch.org/audio/main/tutorials/squim_tutorial.html}}. This metric yields scores ranging from -0.5 to 4.5, with values typically exceeding 1.0 for speech audio.

\subsection{Content Accuarcy}

To quantify content accuracy, we employ Character Error Rate (CER) for Chinese and Word Error Rate (WER) for English. The evaluation pipeline proceeds as follows: First, we obtain the transcribed text $T_{\text{pred}}$ from the generated audio using FunASR-Nano\footnote{\url{https://github.com/FunAudioLLM/Fun-ASR}}. 
Subsequently, we perform rigorous normalization on both the ground truth $T_{\text{gt}}$ and the prediction $T_{\text{pred}}$. This process includes: 
1) \textbf{Punctuation Removal}: eliminating punctuation via \texttt{string.punctuation} and \texttt{zhon.hanzi.punctuation}\footnote{\url{https://pypi.org/project/zhon/}}; 
2) \textbf{Whitespace Standardization}: trimming leading/trailing spaces and collapsing multiple spaces;
and 3) \textbf{Character Normalization}: converting Traditional Chinese to Simplified using \texttt{zhconv}\footnote{\url{https://pypi.org/project/zhconv/}} while filtering non-ASCII characters in English text via \texttt{clean-text}\footnote{\url{https://pypi.org/project/clean-text/}}.
Finally, following the methodology of F5-TTS~\cite{chen2024f5}, we calculate the WER and CER using the \texttt{JiWER} library\footnote{\url{https://pypi.org/project/jiwer/}}.

It is worth noting that our selected transcription system, FunASR-Nano, demonstrates exceptional performance on clean speech benchmarks, achieving a WER of 1.76\% on Librispeech-clean (EN) and a CER of 2.56\% on Fleurs-zh. These results are competitive with state-of-the-art models of similar parameter scale~\cite{srivastav2025open}. Utilizing such a high-performance ASR model minimizes transcription-induced errors, ensuring that the reported metrics accurately reflect the content fidelity of the generated audio.

\subsection{Prosodic Coherence}
For prosody evaluation, we utilize SpeechJudge~\cite{zhang2025speechjudge}, a fine-tuned Qwen2.5-Omni model specialized for audio assessment. 
To specifically target long-form modeling capabilities, we refine the original prompt design, decomposing the evaluation criteria into three granular dimensions: \textit{Prosodic Coherence \& Flow}, \textit{Rhythmic Hierarchy \& Layering}, and \textit{Overall Naturalness}. 
Ratings are assigned on a scale from 1.0 to 5.0, as detailed in Figure~\ref{app_fig:eval_prosody}. 
Furthermore, to mitigate the inherent variance of LALMs, we conduct 10 independent evaluations for each generated audio sample and calculate the mean to derive the final prosody score.

\subsection{Expressive Richness}
This dimension assesses the global expressive quality of the generated speech, representing the average level of expressiveness~\cite{chen2026dualaxis}. Formally, we segment the audio waveform into a sequence of non-overlapping 10-second chunks $\{c_i\}_{i=1}^M$. An LALM is then employed to assign an expressiveness score $s_i$ to each chunk $c_i$. The final Expressive Richness metric is defined as the arithmetic mean of these segment scores:
\begin{equation}
  \text{Score}_{\text{rich}} = \frac{1}{M} \sum_{i=1}^{M} s_i.
\end{equation}
The 10-second segmentation window is selected to align with the typical generation duration of chunk-based long-form synthesis pipelines. This strategy effectively mitigates the confounding effects of inter-chunk inconsistencies, allowing for a more focused evaluation of intrinsic expressiveness. The prompt template used for this assessment is illustrated in Figure~\ref{app_fig:expressive_richness}.

\subsection{Expressive Hierarchy}

Complementing the local expressiveness defined above, paragraph-level expressive hierarchy is equally critical in long-form settings. \cite{E200080,E210196}
Unlike the segment-based approach for \textbf{Expressive Richness}, we leverage the long-context understanding capabilities of modern LALMs to conduct a holistic assessment. 
Specifically, the entire audio sequence is fed into the model, which is instructed to evaluate the speech based on three dimensions: \textbf{Emotional Variation}, \textbf{Vocal Dynamics}, and \textbf{Scene Appropriateness}.

The prompt template used for this assessment is illustrated in Figure~\ref{app_fig:expressive_hierarchy}.

%% file: latex/app_sec/04_human.tex
\section{User Study}\label{app_sec:human}

For the subjective evaluation, we recruit a balanced cohort of 10 expert listeners (5 male, 5 female) with diverse professional backgrounds, including audio engineers from the internet industry, live streaming specialists, and academic researchers (professors and PhD candidates) in signal processing. 
All participants possess extensive experience in audio quality assessment. 
In all subjective tests, we conduct Mean Opinion Score~(MOS) evaluation~\cite{zhang2024gtsinger,zhang2025tcsinger,pan2025synthetic}.
They are compensated at a rate of \$1.00 per evaluation instance (either a single sample or a paired comparison), with the total expenditure for the user study amounting to \$2,000.

\subsection{Validation of Timbre Consistency}

In this experiment, we randomly select 50 samples from the test set for subjective evaluation. Listeners are instructed to rate the ``Timbre Maintenance'' capability using a Mean Opinion Score (MOS). They are explicitly required to focus exclusively on timbre stability, disregarding other acoustic factors (e.g., sound field, audio quality) and semantic dimensions (e.g., pronunciation, prosody). If the expressiveness of the audio does not affect the timbre, it can also be ignored.

We concurrently compute the objective Timbre Consistency score for each sample. The correlation analysis between the subjective MOS and our objective metric yields the following results: SRCC=$0.75$, PLCC=$0.77$, and KRCC=$0.59$. These results demonstrate that our proposed timbre consistency evaluation aligns closely with human perception.

Furthermore, the user study reveals several statistical thresholds regarding our objective metric:
\begin{enumerate}
    \item \textbf{Score $<$ 0.85}: Indicates significant timbre drift. In multi-speaker scenarios, this may also suggest inaccurate speaker transitions.
    \item \textbf{Score $<$ 0.93}: Demonstrates superior timbre maintenance, with performance comparable to ground truth recordings.
    \item \textbf{Score $\in$ [0.85, 0.90]}: Represents generally acceptable performance, typically characterized by minor local timbre mutations or artifacts.
\end{enumerate}

Besides, the robustness of this metric presents room for improvement. Potential misclassifications may arise in specific edge cases, such as audio exhibiting periodic timbre variations (e.g., looping patterns). Since our metric relies on global averages, it may fail to penalize such rhythmic fluctuations, yielding a favorable score despite perceptual inconsistency. Future work will aim to incorporate temporal modeling to address these dynamic artifacts.

\subsection{Validation of Sound Fidelity}

Considering that \texttt{SQUIM-PESQ} is trained on English sentence-level data, we select 50 samples from the test set to verify its generalization to Chinese and long-form scenarios. 
Listeners are instructed to rate ``Clarity and Fidelity'' using MOS~\cite{zhang2024tcsinger,chen2026wavalign}. Specifically, they are required to focus exclusively on factors such as background noise, artifacts, and articulation, while disregarding prosody and expressiveness. 
We concurrently compute the \texttt{SQUIM-PESQ} scores for these samples. 
The correlation analysis between subjective MOS and \texttt{SQUIM-PESQ} yield an SRCC of 0.72, a PLCC of 0.47, and a KRCC of 0.53.
These results demonstrate that the metric aligns closely with human perception.

\subsection{Validation of Prosodic Coherence}

To validate the Prosodic Coherence metric, we adopt the methodology of SpeechJudge~\cite{zhang2025speechjudge}, conducting a human preference test to assess the model's evaluation performance. In addition to the robust correlation reported in Section~\ref{sec:bench-human}, our analysis yields the following statistical insights:

\begin{enumerate}
    \item \textbf{Score Divergence $> 1$}: A difference of more than 1 points indicates a substantial and perceptually obvious gap in prosodic quality between audio samples.
    \item \textbf{Score $\ge 4$}: Audio samples achieving this threshold demonstrate competent basic prosody and rhythmic structure.
    \item \textbf{Score $\ge 4.5$}: Performance at this level is considered virtually indistinguishable from ground truth recordings.
\end{enumerate}

\subsection{Validation of Expressiveness}\label{app_sec:human_expressiveness}

\input{latex/Tables/expressive_richness}

In this experiment, we curate a diverse set of 200 samples spanning all models and tasks for subjective evaluation. Listeners are tasked with rating the audio strictly adhering to the same prompt criteria provided to the LALMs.

Concurrently, we benchmark this 200-sample test set against 4 specialized MOS prediction models (UTMOS~\cite{saeki2022utmos}, UTMOSv2~\cite{baba2024t05}, SQUIM-MOS~\cite{kumar2023torchaudio}, DNS-MOS~\cite{reddy2021dnsmos}) and 8 flagship LALMs (GPT-4o, Qwen3Omni-Instruct-30B-A3B~\cite{xu2025qwen3}, Qwen3Omni-Flash, StepFun-Audio-R1~\cite{tian2025step}, Gemini-2.5-flash, Gemini-2.5-pro, Gemini-3-flash, Gemini-3-pro). Notably, due to context length constraints, only a subset of these LALMs is employed for the Expressive Hierarchy evaluation.

We examine the correlation between the mean listener ratings and the model-predicted scores, with results summarized in Table~\ref{tab:human_er} and Table~\ref{tab:human_eh}. Notably, \textbf{Gemini3-Pro} demonstrates superior performance, significantly outperforming other models across both metrics. From the tables, we can also observe that open-source models such as Qwen3-Omni-Flash and Qwen3-Omni-Instruct demonstrated superior performance compared to GPT-4o, with a relatively small gap to Gemini-2.5-Pro, indicating that open-source models also have the potential to become excellent evaluators. In the future, as the testing scale continues to increase, we will also distill better and more stable open-source evaluators based on open-source models to further enhance reproducibility. It is also worth noting that all traditional MOS prediction networks exhibited poor correlation with human perception regarding expressiveness. This suggests that standard MOS training datasets likely lack a specific focus on expressive qualities.

Moreover, we conduct independent repeated trials on this test set to validate the stability of our selected evaluator, Gemini 3 Pro.
Specifically, we perform five independent scoring iterations for each audio sample, where Gemini 3 Pro yields inconsistent scores for only 11 instances, demonstrating a level of robustness comparable to human evaluators. Consequently, we adopt \textbf{a single-pass evaluation strategy} for this metric.

\input{latex/Tables/expressive_hierarchy}

Furthermore, to ensure consistency in the rating scales adopted by our recruited listeners, we computed the correlation between each individual rater and the mean score of the remaining raters. As shown in Table~\ref{app_tab:raters}, the high inter-rater correlation confirms the reliability and validity of our evaluation protocol.
\input{latex/app_tables/human_score_check}

%% file: latex/Tables/expressive_richness.tex
\begin{table}
    \caption{Human alignment comparison across different LALMs on Expressive Richness.}
    \label{tab:human_er}
    \centering
    \footnotesize
  \resizebox{\linewidth}{!}{
        \begin{tabular}{@{}c c c c c @{}}
        \toprule
        { \textbf{Models}}  & { \textbf{PLCC}} & { \textbf{SRCC}} & { \textbf{QWK}} & { \textbf{MAE}} \\ \midrule
        { \textbf{UTMOS}}   & { -0.0203} & { -0.0433} & { -0.0313}& { 1.043}  \\
        { \textbf{UTMOSv2}} & { -0.0745} & { -0.0789} & { -0.0827}& { 0.9012} \\
        { \textbf{SQUIM-MOS}} & { -0.3145} & { -0.2767}  & { -0.0825}& { 1.3177} \\
        { \textbf{DNS-MOS}} & { -0.0243} & { -0.0189} & { -0.0034}& { 0.8537} \\
        \midrule
        { \textbf{GPT-4o}}  & {0.1549}   & { 0.2002}  & { 0.1435}  & { 0.7982}  \\
        { \textbf{Qwen3Omni-Flash}}    & { 0.1464}  & { 0.1696}    & { 0.0812} & { 1.0401} \\
        { \textbf{Qwen3Omni-Instruct}} & { 0.2245}  & { 0.2488}   & { 0.1172} & { 1.0809} \\
        { \textbf{Gemini2.5-flash}}    & { 0.4166}  & { 0.4079} & { 0.2623} & { 0.8123} \\
        { \textbf{Gemini2.5-pro}}& { 0.5085}  & { 0.5160}  & { 0.4242} & { 0.7635} \\
        { \textbf{Gemini3-flash}}& \underline{0.5224}  & \underline{0.5266}  & \underline{0.5066} & \underline{0.6562} \\
        { \textbf{Gemini3-Pro}}  & { \textbf{0.7061}}  & { \textbf{0.7080}}  & { \textbf{0.6772}} & { \textbf{0.5879}} \\ \bottomrule
        \end{tabular}
    }
\end{table}

%% file: latex/Tables/expressive_hierarchy.tex
\begin{table}
    \caption{Human alignment comparison across different LALMs on Expressive Hierarchy.}
    \label{tab:human_eh}
    \centering
    \footnotesize
  \resizebox{\linewidth}{!}{
    \begin{tabular}{@{}c c c c c @{}}
\toprule
{ \textbf{Models}} & { \textbf{PLCC}}   & { \textbf{SRCC}}   & { \textbf{QWK}}    & { \textbf{MAE}}    \\ \midrule
{ \textbf{GPT-4o}} & { 0.1328}  & { 0.1171}  & { 0.0803}  & { 0.7604}  \\
{ \textbf{Qwen3Omni-Flash}}    & { 0.3263}  & { 0.2496}  & { 0.2193}  & { 0.8426}  \\
{ \textbf{Qwen3Omni-Instruct}} & { 0.1641}  & { 0.1181}  & { 0.0869}  & { 0.9130}  \\
{ \textbf{Gemini2.5-flash}}    & { 0.0421}  & { 0.0005}  & { 0.0256}  & { 0.8673}  \\
{ \textbf{Gemini2.5-pro}}      & { 0.3732}  & { 0.3744}  & \underline{0.2871}  & \underline{0.800}   \\
{ \textbf{Gemini3-flash}}      & {\underline{0.406}}   & { \underline{0.3924}}  & {0.2032}  & { 1.1837}  \\
{ \textbf{Gemini3-Pro}}& { \textbf{0.6041}} & { \textbf{0.6234}} & { \textbf{0.5452}} & { \textbf{0.7204}} \\ \bottomrule
    \end{tabular}
    }
\end{table}

%% file: latex/app_tables/human_score_check.tex
\begin{table*}[t]
\centering
\renewcommand{\arraystretch}{1.2}
\caption{Correlation analysis among different evaluators (A denotes Annotator).}
\label{app_tab:raters}
\resizebox{\linewidth}{!}{
\begin{tabular}{lcccccccccc}
\toprule
     & \textbf{A1} & \textbf{A2} & \textbf{A3} & \textbf{A4} & \textbf{A5} & \textbf{A6} & \textbf{A7} & \textbf{A8} & \textbf{A9} & \textbf{A10} \\
\midrule
\textbf{PLCC($\uparrow$)} & 0.8696      & 0.8426      & 0.9014      & 0.9035      & 0.9163      & 0.8766      & 0.9022      & 0.7080      & 0.8830      & 0.7623       \\
\textbf{SRCC($\uparrow$)} & 0.8711      & 0.8296      & 0.9028      & 0.9025      & 0.9143      & 0.8635      & 0.8945      & 0.7010      & 0.8820      & 0.7585       \\
\textbf{KRCC($\uparrow$)} & 0.7255      & 0.6804      & 0.7726      & 0.7678      & 0.7872      & 0.7238      & 0.7575      & 0.5399      & 0.7405      & 0.6011       \\
\textbf{QWK($\uparrow$)}  & 0.8732      & 0.8330      & 0.9030      & 0.8984      & 0.9079      & 0.8544      & 0.8938      & 0.7002      & 0.8740      & 0.7596       \\
\textbf{MAE($\downarrow$)}  & 0.3713      & 0.4398      & 0.3336      & 0.3452      & 0.3336      & 0.3994      & 0.3541      & 0.5800      & 0.3892      & 0.5402       \\
\bottomrule
\end{tabular}
}
\end{table*}

%% file: latex/app_sec/05_implement_detail.tex
\section{Implementation Detail}\label{app_sec:detail}

\subsection{Computational Resources and Environments}

All inference and evaluation experiments for open-source models are conducted on a server equipped with 8 NVIDIA GeForce RTX 4090 GPUs and an Intel Xeon Gold 6530 CPU, running Ubuntu 22.04. For model inference, we strictly adhere to the environment specifications provided in the respective official repositories. The core dependencies for our evaluation pipeline include Python 3.10, PyTorch 2.8.0, Torchaudio 2.8.0, and Transformers 4.57.3.

\subsection{Selected Voice}
\input{latex/app_tables/voice_source}

For open-source models, we curate a set of 25 reference audio prompts from diverse datasets, including Emilia~\cite{he2024emilia}, AISHELL-3~\cite{shi2020aishell}, NCSSD~\cite{liu2024generative}, LibriSpeech~\cite{panayotov2015librispeech}, MSPPodcast~\cite{martinez2020msp}, and ChildMandarin~\cite{zhou2025childmandarin}, as well as reference voices provided in specific model repositories (see Table~\ref{app_tab:open-voice}). Over 20 representative timbres from multiple open-source datasets cover various dimensions including language, gender, and age, to evaluate model generation capabilities as comprehensively as possible. We conduct extensive evaluations across these prompts and reported the results of the best-performing voice for each model. This strategy aims to minimize the impact of biases arising from training data discrepancies and inherent voice preferences. We acknowledge that the current timbre coverage may still have limitations. However, our evaluation pipeline imposes no constraints on reference timbres, and users can freely select a wider range of timbre categories to perform evaluations based on our provided evaluation dataset and pipeline.

For closed-source models, we selected official voices characterized by high fidelity, superior prosody, and rich expressiveness. Detailed specifications are provided in Table~\ref{app_tab:close_voice}.

\input{latex/app_tables/close_voice_source}

\subsection{Synthesis Strategy}

For open-source models, we strictly adhere to the default configurations provided in their official repositories. Specific adjustments for MegaTTS3, CosyVoice3, and IndexTTS2 are detailed below:

\textbf{MegaTTS3} As the official VAE Encoder~\cite{kingma2013auto} is not publicly available, we obtain the VAE latents for our reference prompt speech by contacting the model maintainers.

\textbf{IndexTTS2} To ensure a fair and objective comparison, we disabled the text sentiment analysis module by setting \texttt{use\_emo\_text} to \texttt{false}.

\textbf{CosyVoice3} We utilized the system text prompt ``You are a helpful assistant'' during generation, consistent with the official implementation.

For closed-source models, we similarly followed the default synthesis strategies without manually adjusting attributes such as emotion, pitch, or speaking rate.

All open-source models are evaluated in a zero-shot setting for long-form and dialogue generation, whereas closed-source models generated speech using designated voice profiles. Finally, all generated audio is resampled to 24kHz for consistent evaluation.

%% file: latex/app_tables/voice_source.tex
\begin{table}[ht]
    \centering
    \caption{Sources and related information of the voice used in LFS-Bench for open-source models' inference.}
    \label{app_tab:open-voice}
    \resizebox{\linewidth}{!}{
    \begin{tabular}{clcll}
        \toprule
        \textbf{No.} & \textbf{Gender} & \textbf{Age Group} & \textbf{Language} & \textbf{Data Source} \\
        \midrule
        1  & Female & \multirow{4}{*}{Children}    & English & Emilia \\
        2  & Male   &                              & English & Emilia \\
        3  & Female &                              & Chinese & ChildMandarin \\
        4  & Male   &                              & Chinese & ChildMandarin \\
        \cmidrule{1-5} 
        5  & Female & \multirow{4}{*}{Teenager}    & English & NCSSD\_R\_EN  \\
        6  & Male   &                              & English & NCSSD\_R\_EN  \\
        7  & Female &                              & Chinese & AISHELL-3     \\
        8  & Male   &                              & Chinese & NCSSD\_R\_ZH  \\
        \cmidrule{1-5}
        9  & Female & \multirow{8}{*}{Youth-Adult} & English & msppodcast    \\
        10 & Male   &                              & English & NCSSD\_R\_EN  \\
        11 & Female &                              & Chinese & AISHELL-3     \\
        12 & Male   &                              & Chinese & NCSSD\_R\_ZH  \\
        13 & Male   &                              & Chinese & VibeVoice Github  \\
        14 & Female  &                             & Chinese & VibeVoice Github  \\
        15 & Male   &                              & English & VibeVoice Github  \\
        16 & Female  &                             & English & VibeVoice Github  \\
        \cmidrule{1-5}
        17 & Female & \multirow{4}{*}{Middle-Aged} & English & LibriSpeech   \\
        18 & Male   &                              & English & Emilia        \\
        19 & Female &                              & Chinese & NCSSD\_C\_ZH  \\
        20 & Male   &                              & Chinese & NCSSD\_C\_ZH  \\
        21 & Male   &                              & Chinese & SparkTTS Github  \\
        \cmidrule{1-5}
        22 & Female & \multirow{4}{*}{Elderly}     & English & msppodcast    \\
        23 & Male   &                              & English & msppodcast    \\
        24 & Female &                              & Chinese & NCSSD\_C\_ZH  \\
        25 & Male   &                              & Chinese & NCSSD\_C\_ZH  \\
        \bottomrule
    \end{tabular}
    }
\end{table}

%% file: latex/app_tables/close_voice_source.tex
\begin{table*}[t]
    \centering
    \caption{the information of the voices selected in the evaluation for closed-source models.}
    \label{app_tab:close_voice}
    \renewcommand{\arraystretch}{1.5} 
    \small 
    \resizebox{\linewidth}{!}{
    \begin{tabular}{ccccc}
        \toprule
        \textbf{Provider} & \textbf{Language} & \textbf{Single Speaker} & \textbf{Two Speakers} & \textbf{Multi Speakers} \\
        \midrule
        \textbf{OpenAI} & General & Alloy & Onyx, Nova & Round-robin: [``alloy'', ``echo'', ``fable'', ``onyx'', ``nova'', ``shimmer''] \\
        \midrule
        \textbf{Gemini} & General & Puck & Puck, Aoede & Round-robin: [``Puck'', ``Aoede'', ``Charon'', ``Kore'', ``Fenrir''] \\
        \midrule
        \textbf{ElevenLabs} & General & Rachel & Charlie, Rachel & Charlie, Rachel, George, Bella, Antoni \\
        \midrule
        \multirow{2}{*}{\textbf{Minimax}}
        & English & male-qn-qingse & -- & -- \\
        & Chinese & Chinese (Mandarin)\_ Male\_Announcer & -- & -- \\
        \midrule
        \multirow{2}{*}{\textbf{Seed-TTS}}
        & English & BV503\_streaming & -- & -- \\
        & Chinese & BV005\_streaming & -- & -- \\
        \midrule
        \multirow{2}{*}{\textbf{Seed-TTS-Podcast}} 
        & \multirow{2}{*}{General} 
        & \multirow{2}{*}{--} 
        & zh\_male\_dayixiansheng\_v2\_saturn\_bigtts,
        & \multirow{2}{*}{--} \\
        & & & zh\_female\_mizaitongxue\_v2\_saturn\_bigtts & \\
        \midrule
        \multirow{2}{*}{\textbf{Inworld}}
        & English & Deborah, Alex & -- & -- \\
        & Chinese & Jing, Yichen & -- & -- \\
        \bottomrule
    \end{tabular}
    }
\end{table*}

%% file: latex/app_sec/06_more_exp.tex
\section{Supplementary Experiment}\label{app_sec:exp}

\subsection{Inference Speed}

The capability to efficiently generate long-form speech is a pivotal performance criterion, garnering widespread attention across both academia and industry. To assess this, we evaluate the computational efficiency of various open-source models using the Real Time Factor (RTF) metric. The RTF is defined as:
\begin{equation}
  \text{RTF} = \frac{T_{\text{inference}}}{T_{\text{audio}}},
\end{equation}
where $T_{\text{inference}}$ denotes the time required for generation and $T_{\text{audio}}$ represents the duration of the generated audio. The computational efficiency results for each model are summarized in Table~\ref{app_tab:single_rtf} and Table~\ref{app_tab:two_rtf}. 
We observe that non-autoregressive models exhibit a significant advantage in generation speed compared to their autoregressive counterparts. This finding is consistent with the inherent parallel decoding mechanism of non-autoregressive architectures.

\input{latex/app_tables/single_rtf}

\input{latex/app_tables/two_rtf}

\subsection{Ablation on Window Size}

The computation of both Timbre Consistency and Reverb Consistency may be sensitive to the sliding window configuration. 
To validate the rationality of our selected window size and stride, we conduct an ablation study across these two dimensions. 
The experimental results are in Table~\ref{app_tab:timbre_window} and Table~\ref{app_tab:reverb_window}.

\input{latex/app_tables/timbre_window}
\input{latex/app_tables/reverb_window}

In the ablation study for \textbf{timbre consistency}, we observe that a window size of $\le$ 2s results in real data exhibiting lower consistency than CosyVoice3, suggesting a misalignment with human perception. 
Conversely, window sizes of $\ge$ 4s gradually reduce the discrepancy between real and synthetic data, indicating that larger windows tend to average out transient timbre mutations. 
Regarding the stride, comparative experiments reveal no significant impact on the results. 
Consequently, to enhance evaluation efficiency and reduce computational overhead, we opt for a larger stride. Based on these findings, we select a window size of 3s and a stride of 2s.

In the ablation study for \textbf{reverb consistency}, a window size of 1s provides sufficient differentiation but proved unstable. 
Specifically, VibeVoice exhibit an excessively high standard deviation relative to its mean reverb score of 9.25, indicating hypersensitivity at this scale. 
Conversely, window sizes of $\ge$ 4s reduce the inter-model differences, implying that overly large windows overlook small-scale acoustic field mutations. 
Balancing computational efficiency and resource overhead, we similarly select a window size of 3s and a stride of 2s. 
Notably, our evaluation method demonstrates overall stability, as the relative rankings of the models remain consistent.

\subsection{Ablation on Generated Length}

\begin{figure*}[t]
  \centerline{\includegraphics[width=\linewidth]{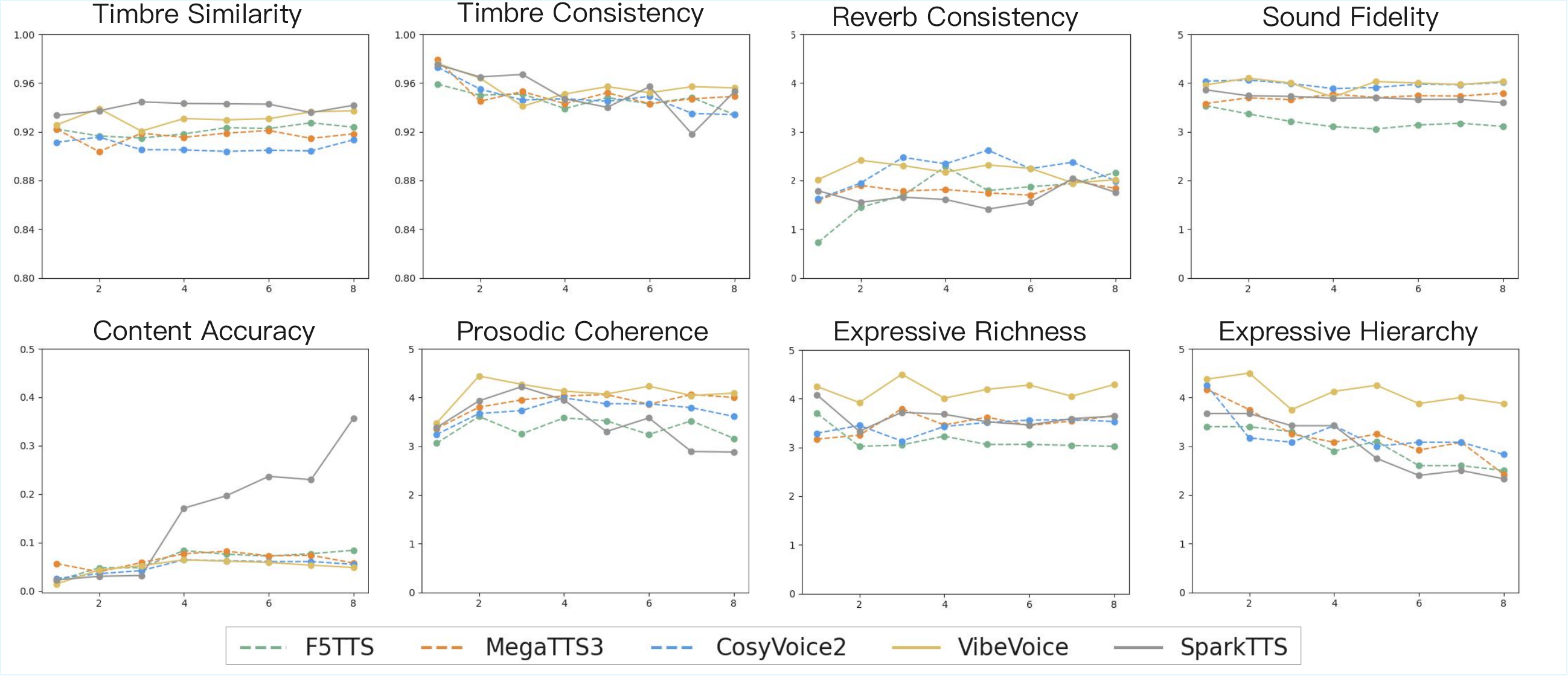}} 
  \caption{
    \textbf{Results on Sequence Length}. 
    The horizontal axis represents the number of sentences in the text.
    Solid lines denote models using the End-to-End strategy, while dashed lines represent the chunked synthesis.
  }
  \label{app_fig:time}
\end{figure*}

To further verify the impact of long-sequence modeling on acoustic, semantic, and expressive performance, we extend the analysis presented in Figure~\ref{fig:exp-time}. 
Beyond the original six dimensions, we additionally track the evolution of Timbre Consistency and Timbre Similarity with respect to increasing generation length, as shown in Figure~\ref{app_fig:time}.

Regarding the Timbre Similarity metric, we adopt the methodology from prior works~\cite{huynh2025ozspeech}. 
Specifically, the generated audio $w$ is segmented into a sequence $\{w_i\}_{i=1}^n$ using a window size of 3s and a stride of 2s.
We then utilize WavLM TDCNN\footnote{\url{https://huggingface.co/docs/transformers/en/model_doc/unispeech-sat}} to extract and normalize speaker embeddings for each segment $w_i$ and the reference audio $w_{ref}$, yielding the embedding sequence $\{e_i\}_{i=1}^n$ and the reference embedding $e_{ref}$. 
Finally, we calculate the average cosine similarity between the generated segment embeddings and the reference embedding to serve as the quantitative indicator of Timbre Similarity.

Overall, we observe a general performance decay across nearly all metrics as the generation duration increases. Specifically, Reverb Consistency, Prosodic Coherence, and Expressive Hierarchy exhibits the most significant degradation.
These findings suggest that current models struggle to maintain acoustic field stability and effectively capture long-term dependencies in long-form settings. 
Conversely, Timbre Similarity and Timbre Consistency remained relatively stable compared to other acoustic dimensions. This stability highlights the effectiveness of ``in-context learning'' paradigms~\cite{du2024cosyvoice2,jiang2025megatts} in preserving speaker identity. 
Additionally, with the exception of SparkTTS, most models demonstrate robust Content Accuracy. This can be attributed to the strong text understanding and alignment capabilities inherent in modern TTS architectures.

\subsection{Multi-Speaker Dialogue Generation}

To facilitate future research in multi-speaker long-form speech synthesis, SwanBench-Speech incorporates 101 test cases specifically designed for 3- and 4-speaker dialog scenarios. Using this subset, we evaluate three closed-source models capable of multi-speaker generation: ElevenLabs Multilingual V2, Gemini-2.5-pro-preview-tts, and OpenAI-tts-1-hd. The experimental results are shown in Table~\ref{app_tab:multi_res}.

\input{latex/app_tables/multi_spk}

%% file: latex/app_tables/single_rtf.tex
\begin{table}[t]
\caption{The Real Time Factor of mono-speaker long form speech generation models.}
\centering
\begin{tabular}{@{}cc@{}}
\toprule
             Models      & RTF                        \\ \midrule
\multicolumn{2}{c}{\textit{Autoregressive Models}}     \\ \midrule
CosyVoice-2~(0.5B)        &    1.061 $\pm$ 0.031      \\
CosyVoice-3~(0.5B)        &    0.747  $\pm$ 0.048                    \\
FishSpeech~(0.5B)         &    1.351 $\pm$ 0.131               \\
GLM-TTS~(1.5B)            &   2.400 $\pm$ 0.158       \\
IndexTTS-2~(0.1B)        &    1.065 $\pm$ 0.037                        \\
SparkTTS~(0.5B)           &    2.046 $\pm$ 0.212       \\
VibeVoice~(1.5B)          &    3.801 $\pm$ 0.317       \\ \midrule
\multicolumn{2}{c}{\textit{Non-Autoregressive Models}} \\ \midrule
F5TTS~(0.3B)           &    0.198 $\pm$ 0.006            \\
MegaTTS3~(0.45B)          &    0.172 $\pm$ 0.002                 \\
ZipVoice~(0.12B)    &    0.338 $\pm$ 0.013             \\ \bottomrule
\end{tabular}
\label{app_tab:single_rtf}
\end{table}

%% file: latex/app_tables/two_rtf.tex
\begin{table}[t]
\caption{The Real Time Factor of two-speaker dialogue generation models. MOSS-TTSD supports batch inference, thus we directly report the RTF of batch process( batchsize = 32)}
\centering
\begin{tabular}{@{}cc@{}}
\toprule
             Models      & RTF                        \\ \midrule
FireRedTTS2               &   4.717 $\pm$ 0.963        \\
MoonCast~(2.6B)           &   5.219 $\pm$ 0.048        \\
MOSS-TTSD~(1.7B)          &   0.219 $\pm$ 0.019     \\
SoulX-PodCast~(1.7B)      &   2.143 $\pm$ 0.169      \\
VibeVoice~(1.5B)          &   4.092 $\pm$ 0.305      \\ 

ZipVoice-Dialog~(0.12B)    &  0.305 $\pm$ 0.030      \\ \bottomrule
\end{tabular}
\label{app_tab:two_rtf}
\end{table}

%% file: latex/app_tables/timbre_window.tex
\begin{table}[t]
\caption{The Ablation study of window setting for timbre consistency. We select the representative models, CosyVoice3 and OpenAI-tts-1-hd, to conduct this ablation in single-spaeker settings.}
\resizebox{\linewidth}{!}{
\begin{tabular}{@{}ccccc@{}}
\toprule
\multicolumn{2}{c}{Window Setting} & \multirow{2}{*}{CosyVoice3} & \multirow{2}{*}{OpenAI} & \multirow{2}{*}{Real-Speech} \\ \cmidrule(r){1-2}
Size~(s)            & Stride~(s)           &                            &                        &                              \\ \midrule
1               & 0.5              &0.868       &0.824          &0.844                    \\
2               & 1                &0.911       &0.887          &0.901                    \\
3               & 1                &0.930       &0.916          &0.956               \\
3               & 2                &0.929       &0.915          &0.955               \\
4               & 2                &0.941       &0.931          &0.963              \\
5               & 2                &0.942       &0.949          &0.967              \\
10              & 4                &0.968       &0.971          &0.971                              \\ \bottomrule
\end{tabular}
}
\label{app_tab:timbre_window}
\end{table}

%% file: latex/app_tables/reverb_window.tex
\begin{table}[t]
\caption{The Ablation study of window setting for reverb consistency. We select the representative models, VibeVoice and Gemini-2.5-pro-preview-tts, to conduct this ablation in two-spaeker settings.}
\resizebox{\linewidth}{!}{
\begin{tabular}{@{}ccccc@{}}

    \toprule
    \multicolumn{2}{c}{Window Setting} & \multirow{2}{*}{VibeVoice} & \multirow{2}{*}{Gemini} & \multirow{2}{*}{Real-Dialog} \\ \cmidrule(r){1-2}
    Size~(s)            & Stride~(s)          &        &       &                              \\ \midrule
    1               & 0.5              &6.40        &4.99     &3.87                              \\
    2               & 1                &4.27        &3.62     &3.20                              \\
    3               & 1                &3.58        &3.17     &2.67                              \\
    3               & 2                &3.59        &3.17     &2.74                              \\
    4               & 2                &3.20        &2.85     &2.51                              \\
    5               & 2                &2.95        &2.61     &2.41                              \\
    10              & 4                &2.23        &1.88     &1.60                            \\ \bottomrule
    \end{tabular}
}
\label{app_tab:reverb_window}
\end{table}

%% file: latex/app_tables/multi_spk.tex
\begin{table*}[t]
    \caption{
        \textbf{Results of multi-speaker dialogue generation models across LFS-Bench's metrics}.
        The best results are in \textbf{bold} and the second best are \underline{underlined}.
    }
    \label{app_tab:multi_res}
    \centering
    \footnotesize
  \resizebox{\linewidth}{!}{
        \begin{tabular}{@{}lccccccc@{}}
        \toprule
        \multicolumn{1}{c|}{}                                 & \multicolumn{3}{c|}{\textbf{Acoustics}}                                    & \multicolumn{2}{c|}{\textbf{Semantics}}                     & \multicolumn{2}{c}{\textbf{Expressiveness}} \\ \cmidrule(l){2-8} 
        \multicolumn{1}{c|}{\multirow{-2}{*}{\textbf{Model}}} & \textbf{Timbre($\uparrow$)} & \textbf{Reverb($\downarrow$)} & \multicolumn{1}{c|}{\textbf{Sound Fidelity($\uparrow$)}}              & \textbf{CER/WER($\downarrow$)} & \multicolumn{1}{c|}{\textbf{Prosody($\uparrow$)}}              & \textbf{Richness($\uparrow$)}          & \textbf{Hierarchy($\uparrow$)}          \\ \midrule

        \multicolumn{1}{c|}{Elevenlabs Multilingual V2} &  \textbf{0.93$\pm$0.030}  & {4.72$\pm$0.69}  & \multicolumn{1}{c|}{\textbf{3.19$\pm$0.37}}  & {0.183 / 0.12} & \multicolumn{1}{c|}{3.28$\pm$0.87}   &  \underline{3.23$\pm$0.54}  &  {3.52$\pm$0.82}   \\
        \multicolumn{1}{c|}{Gemini-2.5-pro-preview-tts}        &  {0.92$\pm$0.012}   &  \underline{3.28$\pm$0.75}     & \multicolumn{1}{c|}{\underline{3.04$\pm$0.17}}    &  {\textbf{0.077} / \textbf{0.102}} & \multicolumn{1}{c|}{\textbf{3.92$\pm$0.36}}    &   {\textbf{3.86$\pm$0.46}}  &  {\textbf{4.05$\pm$0.62}}                      \\
        \multicolumn{1}{c|}{OpenAI-tts-1-hd}     & \underline{0.92$\pm$0.011}   &  \textbf{{1.91$\pm$0.38}}  & \multicolumn{1}{c|}{2.29$\pm$0.17}        & {\underline{0.106} / \underline{0.104}}   & \multicolumn{1}{c|}{\underline{3.78$\pm$0.63}}    &    {2.93$\pm$0.60}      & \underline{3.77$\pm$0.84}   \\
        \rowcolor[HTML]{FFFC9E} 
        \multicolumn{1}{c|}{Average}  &  0.92 &  3.30  & \multicolumn{1}{c|}{2.84} &   { 0.122 / 0.109 }   & \multicolumn{1}{c|}{3.66} &    3.34      &   3.78          \\ 
        \bottomrule
        \end{tabular}
    }
\end{table*}

%% file: latex/app_sec/07_detailed_analysis.tex
\section{More Analysis Based on SwanBench-Speech}\label{app_sec:analysis}

\subsection{Detailed analysis on each metric}

\paragraph{Timbre Consistency} Although experimental results indicate that real data generally outperforms synthetic data in timbre consistency (single speaker: 0.96 vs. 0.93; two-speaker: 0.95 vs. 0.92), this gap is not significant. 
This suggests that the consistency performance of current models is acceptable. 
However, we offer two deeper insights. 
First, open-source models exhibit a relatively larger standard deviation compared to closed-source models, indicating that their stability still lags behind commercial solutions. 
Second, dialogue models demonstrate greater variance in timbre consistency than single-speaker long-form speech. 
Given that we have minimized error accumulation from forced alignment, this increased variance likely reflects that models are still hindered by speaker transitions.

\paragraph{Reverb Consistency}
In this dimension, single-speaker performance is comparable to human recordings. 
Apart from the CosyVoice series and ElevenLabs models, which underperform on this metric,
other models maintain robust reverb consistency, demonstrating strong acoustic field preservation over extended durations. 
Conversely, in dialogue scenarios, all open-source models and the majority of closed-source models show a significant performance gap compared to real data (Open average: 3.45; Closed average: 3.36). 
Feedback from our user study further reveals inconsistencies in sound fields and volume between speakers in generated dialogues. 
This indicates a need to enhance the models' ability to disentangle prompt speech attributes. Consequently, future work should prioritize maintaining acoustic unity during speaker transitions.

\paragraph{Sound Fidelity}
Regarding this metric, the performance of generated speech aligns closely with that of real data. 
Notably, models such as FishSpeech and ElevenLabs achieve scores significantly surpassing the mean of real data. 
This suggests that contemporary models have largely resolved sound quality constraints. 
The fact that generated speech outperforms human recordings likely stems from the composition of the real data. 
Since the majority of real data is web-crawled rather than studio-recorded, it is susceptible to device and environmental noise, which compromises its fidelity.

\paragraph{Content Accuracy}
Prior studies indicate that metrics such as WER have reached saturation in sentence-level speech generation~\cite{chen2024vall}. 
This finding extends to chunk-based in-context learning approaches, where models like CosyVoice2 and MegaTTS3 demonstrate exceptional performance.
However, the metric remains relevant for autoregressive end-to-end architectures.
For instance, SparkTTS exhibits suboptimal Content Accuracy in long-form generation.
As in Figure~\ref{app_fig:time}, deeper ablation studies confirm that the character accuracy of such models declines as the text length increases.

\paragraph{Prosodic Coherence}

Regarding prosodic coherence, we observe a distinct gap between real and synthetic speech, suggesting that current models require further improvement in prosody modeling. 
Notably, closed-source models significantly outperform their open-source counterparts in this dimension. 
This indicates that while open-source models achieve parity with state-of-the-art systems in fidelity and content accuracy, they still lag in perceptual metrics such as prosodic naturalness.

\paragraph{Expressive Richness}
Experimental results identify expressiveness as the primary differentiator between real and synthetic audio. 
Specifically, open-source models trail real data by approximately 1.5 points in Expressive Richness. 
While closed-source models demonstrate marked improvement, they still exhibit a gap of nearly 1.0 point. 
Furthermore, our scenario-based analysis confirms that models underperform in high-expressiveness settings. 
These findings consistently underscore that generating realistic, highly expressive speech remains a pivotal challenge for achieving immersive audio generation.

\paragraph{Expressive Hierarchy}

Similar to Expressive Richness, real data outperforms synthetic speech in this metric, with closed-source models surpassing their open-source counterparts. 
Notably, in single-speaker tasks, models consistently achieve lower scores on Expressive Hierarchy compared to Expressive Richness. 
This indicates that capturing and modeling paragraph-level hierarchical structure remains a significant challenge. 
Furthermore, dialog models generally exhibit superior hierarchical performance compared to single-speaker models. 
We attribute this to the inherent semantic logic of dialog interactions, which likely provides stronger contextual cues that facilitate the learning of hierarchical patterns.

\subsection{Analysis based on the scenarios}\label{app_sec:scenarios}

\begin{figure*}[t]
  \centerline{\includegraphics[width=\linewidth]{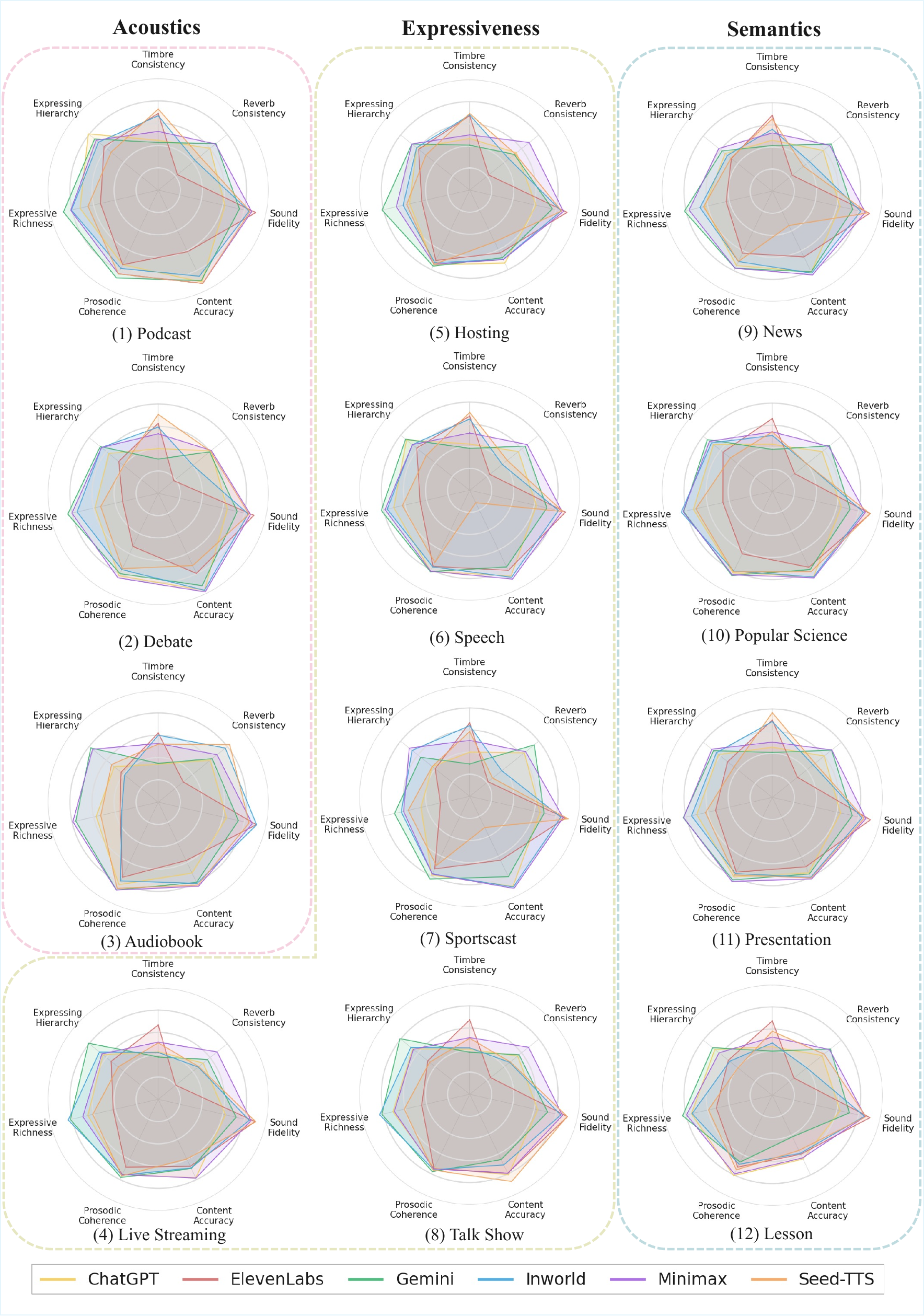}} 
  \caption{
  We visualize the performance of closed-source models in single-speaker long-form generation across various downstream scenarios using a radar chart. 
  To ensure consistency, we normalize the metrics for Timbre Consistency, Reverb Consistency, and Content Accuracy within their respective minimum and maximum ranges. 
  As a result, all metrics are presented such that higher values indicate better performance.
}
\label{app_fig:radar_scenarios}
\end{figure*}

We extend our analysis by providing scenario-based performance results, visualizing the metrics of closed-source models via a radar chart in Figure~\ref{app_fig:radar_scenarios}.
These detailed findings corroborate our primary conclusion: most metrics exhibit varying degrees of degradation in high-expressiveness scenarios. 
A granular visualization reveals that challenging scenarios such as sportscast, host, and talk-show suffer the most severe performance decline. 
This further indicates that current models lack the capacity to effectively model highly dynamic prosody and intense emotional variations.

We provide a detailed explanation of the normalization procedures applied to the radar charts in Figure~\ref{app_fig:radar_scenarios}. 
For LALM-based metrics (Expressive Richness, Expressive Hierarchy, Prosodic Coherence), we directly utilize the original values as its definition is consistent with that of MOS scores.
For Fidelity, quantified by \texttt{SQUIM-PESQ} (range $[-0.5, 4.5]$), we apply a linear shift of $+0.5$ for alignment. 
Regarding Timbre Consistency, Reverb Consistency, and Content Accuracy, we first identify the global maximum $s_{\max}$ and minimum $s_{\min}$ across all models in all scenarios. 
Then, we employ a mapping function $f$ that projects the range $[s_{\min}, s_{\max}]$ onto the interval $[1, 5]$. 
This transformation ensures that for all dimensions in the radar chart, a larger value consistently represents superior performance. The radar charts in Figure~\ref{fig:exp-scenearios} and Figure~\ref{fig:teaser} follow this identical normalization protocol.

\subsection{Analysis based on the Languages}

We also present the experimental results for the evaluated models across the two covered languages, Chinese and English, as shown in Table~\ref{app_tab:mono_res_zh_en} and Table~\ref{app_tab:two_res_zh_en}. 

We observe that although all evaluated models claim bilingual capabilities, the target language significantly impacts performance for the majority. 
For instance, despite utilizing identical voice profiles, ElevenLabs Multilingual V2 exhibits a marked disparity in Expressive Richness between Chinese and English (1.79 vs. 2.87). 
A similar divergence is evident in Seed-TTS-Podcast (Chinese: 4.19 vs. English: 3.49). 
In contrast, Gemini-2.5-pro-preview-tts stands out by not only delivering exceptional performance in prosody and expressiveness but also maintaining a consistent balance across both languages.

\input{latex/app_tables/mono_zh_en_res}
\input{latex/app_tables/two_zh_en_res}

%% file: latex/app_tables/mono_zh_en_res.tex
\begin{table*}[t]
    \caption{
        \textbf{Evaluation results of long-form TTS models across two languages.} Metrics cover Acoustics (Timbre/Reverb Consistency, Fidelity), Semantics (Content Accuracy, Prosodic Coherence), and Expressiveness (Richness, Hierarchy).
        Closed-source models and open-source models are separately marked, with the best results in \textbf{bold} and the second best \textit{italic}.
        Chinese results and English results are separately marked as well, with Chinese in black and English in red.    
    }
    \label{app_tab:mono_res_zh_en}
    \centering
    \footnotesize
  \resizebox{\linewidth}{!}{
\begin{tabular}{@{}ccccccccc@{}}
\toprule
          &            & \multicolumn{3}{c}{\textbf{Acoustics}}         & \multicolumn{2}{c}{\textbf{Semantics}}     & \multicolumn{2}{c}{\textbf{Expressiveness}}     \\ \cmidrule(l){3-5} \cmidrule(l){6-7} \cmidrule(l){8-9} 
\multirow{-2}{*}{\textbf{Models}} & \multirow{-2}{*}{\textbf{Languages}} & \textbf{Timbre($\uparrow$)}      & \textbf{Reverb($\downarrow$)}     & \textbf{Fidelity($\uparrow$)}   & \textbf{Content($\downarrow$)}     & \textbf{Prosody($\uparrow$)}           & \textbf{Richness($\uparrow$)}   & \textbf{Hierarchy($\uparrow$)}     \\ \midrule
\multicolumn{9}{c}{\cellcolor[HTML]{FAF1D1}\textit{Open-Source Models}}              \\ \midrule
          & ZH & {0.90} & {\textbf{0.79}} & {3.47} & {0.329} & {2.37} & {3.29} & {2.11} \\
\multirow{-2}{*}{SparkTTS}    & EN & {\color{red}\textbf{0.95}} & {2.96} & {3.70} & {0.240} & {2.78} & {\color{red}\emph{3.64}} & {2.65} \\ \midrule
          & ZH & {0.90} & {1.65} & {3.55} & {0.072} & {3.24} & {3.16} & {2.87}\\
\multirow{-2}{*}{ZipVoice}        & EN & {0.89} & {2.47} & {3.47} & {0.396} & {3.13} & {1.71} & {1.34}         \\ \midrule
          & ZH & {0.93} & {1.52} & \emph{3.99} & {0.035} & {\textbf{4.07}} & {3.17} & {3.12}        \\
\multirow{-2}{*}{GLM-TTS}         & EN & {\color{red}\emph{0.94}} & {\color{red}\emph{1.70}} & {3.90} & {0.118} & {3.21} & {2.19} & {1.96}        \\ \midrule
          & ZH & {0.90} & {1.74} & {3.57} & {\textbf{0.032}} & {3.62} & {3.47} & {3.13}        \\
\multirow{-2}{*}{CosyVoice2}      & EN & {0.93} & {2.95} & {\color{red}\emph{4.02}} & {0.168} & {2.84} & {2.56} & {2.39}         \\ \midrule
          & ZH & {\emph{0.94}} & {1.83} & {3.83} & {\emph{0.034}} & {3.92} & {3.36} & {2.83}\\
\multirow{-2}{*}{CosyVoice3} & EN & {0.93} & {2.68} & {3.82} & {0.141} & {2.83} & {2.23} & {2.07}\\ \midrule
          & ZH & {0.93} & {2.12} & {3.67} & {0.035} & {3.92} & {3.02} & {2.88} \\
\multirow{-2}{*}{MegaTTS3}        & EN & {0.93} & {\color{red}\textbf{1.50}} & {3.43} & {\color{red}\textbf{0.108}} & {3.30} & {2.60} & {2.17} \\ \midrule
          & ZH & {\textbf{0.95}} & {1.28} & {2.39} & {\emph{0.033}} & {3.96} & {\textbf{4.02}} & {\emph{3.30}}\\
\multirow{-2}{*}{IndexTTS2}       & EN & {0.92} & {2.15} & {3.15} & {0.135} & {3.33} & {3.15} & {2.62}\\ \midrule
          & ZH & {0.92} & {1.76} & {\textbf{4.06}} & {0.043} & {\emph{4.03}} & {3.25} & {3.16}     \\
\multirow{-2}{*}{FishSpeech}      & EN & {0.93} & {1.81} & {\color{red}\textbf{4.13}} & {0.113} & {\color{red}\emph{3.56}} & {2.06} & {2.63}         \\ \midrule
          & ZH & {0.91} & {1.54} & {3.88} & {0.047} & {3.91} & {3.47} & {\textbf{3.34}}        \\
\multirow{-2}{*}{VibeVoice}       & EN & {\color{red}\textbf{0.95}} & {2.75} & {3.75} & {\color{red}\emph{0.111}} & {\color{red}\textbf{3.88}} & {\color{red}\textbf{3.95}} & {\color{red}\textbf{3.34}}      \\ \midrule
          & ZH         & {0.88} & {\emph{1.13}} & {3.12} & {0.072}           & {3.28}           & {\emph{3.50}} & 2.73\\
\multirow{-2}{*}{F5TTS} & EN         & {0.92} & {2.51} & {3.65} & {0.113}           & {3.54}           & {2.64} & {\color{red}\emph{2.81}}\\ \midrule
          & ZH         & {0.92}               & {1.54}               & {3.55}               & {0.073}                & {3.63}    & {3.37}          & 2.95         \\
\multirow{-2}{*}{Average}         & EN         & {0.93}               & {2.35}               & {3.70}               & {0.164}                & {3.24}    & {2.67}              & 2.40           \\ \midrule
\multicolumn{9}{c}{\cellcolor[HTML]{FAF1D1}\textit{Closed-Source Models}}             \\ \midrule
          & ZH         & {0.90}    & {\textbf{1.38}}   & {3.13}   & {0.059}         & {\emph{4.13}}          & {\textbf{4.20}} & {\emph{3.53}}   \\
\multirow{-2}{*}{gemini-2.5-pro-preview-tts}      & EN         & {0.92}   & {\color{red}\emph{1.49}}   & {3.19}   & {0.169}         & {3.69}          & {\color{red}\textbf{4.07}} & {\color{red}\textbf{3.48}}   \\ \midrule
          & ZH         & {0.91}   & {1.65}   & {2.69}   & {0.043}         & {4.00}          & {3.20} & {3.07}   \\
\multirow{-2}{*}{OpanAI-tts-1-hd} & EN         & {0.92}   & {1.82}   & {2.60}   & {0.119}         & {\color{red}\textbf{3.82}} & {3.71} & {\color{red}\emph{3.43}}   \\ \midrule
          & ZH         & {0.93}   & {\emph{1.43}}   & {3.83}   & {\textbf{0.030}} & {\textbf{4.14}} & {\emph{4.00}} & {\textbf{3.56}}   \\
\multirow{-2}{*}{MiniMax-Speech-2.6-hd}           & EN         & {0.92}   & {\color{red}\textbf{1.32}}   & {3.81}   & {0.119}         & {\color{red}\emph{3.77}}          & {3.60} & {2.95}   \\ \midrule
          & ZH         & {\textbf{0.95}}   & {3.04}   & \textbf{4.00}   & {0.100}         & {3.26}          & {1.79} & {2.38}   \\
\multirow{-2}{*}{Elevenlabs Multilingual V2}      & EN         & {\color{red}\textbf{0.96}}  & {3.05}   & \color{red}\textbf{4.04}   & {\color{red}\emph{0.115}}   & {3.73}          & {2.87} & {2.97}   \\ \midrule
          & ZH         & {\emph{0.94}}   & {2.19}   & {3.72}   & {0.053}         & {3.73}          & {3.41} & {2.92}   \\
\multirow{-2}{*}{Inworld-tts-1-max}               & EN         & {0.92}   & {2.19}   & {3.74}   & {\color{red}\textbf{0.114}}   & {3.69}          & {\color{red}\emph{3.95}} & {3.13}   \\ \midrule
          & ZH         & {\emph{0.94}}   & {1.99}   & \emph{3.86}   & {0.106}         & {3.86}          & {3.06} & {2.46}   \\
\multirow{-2}{*}{Seed-TTS2}        & EN         & {\color{red}\emph{0.94}}   & {1.91}   & \color{red}\emph{3.89}   & {0.193}         & {3.62}           & {3.14} & {2.21}   \\ \midrule
          & ZH         &  {0.93}           & {1.95}           & {3.54}           & {0.065}             &  {3.85}  &  {3.28}          &  2.99          \\
\multirow{-2}{*}{Average}          & EN         &  {0.93}           & {1.96}           &  {3.55}           &   {0.138}           & {3.72}   &  {3.56}          & 3.03       \\ \bottomrule
\end{tabular}
    }
\end{table*}

%% file: latex/app_tables/two_zh_en_res.tex

\begin{table*}[t]
\caption{
    \textbf{Evaluation results of dialog generation models across two languages.} Metrics cover Acoustics (Timbre/Reverb Consistency, Fidelity), Semantics (Content Accuracy, Prosodic Coherence), and Expressiveness (Richness, Hierarchy).
    Closed-source models and open-source models are separately marked, with the best results in \textbf{bold} and the second best \textit{italic}.
    Chinese results and English results are separately marked as well, with Chinese in black and English in red.    
}
\label{app_tab:two_res_zh_en}
\centering
\footnotesize
\resizebox{\linewidth}{!}{
\begin{tabular}{@{}ccccccccc@{}}
\toprule
      &        & \multicolumn{3}{c}{\textbf{Acoustics}}          & \multicolumn{2}{c}{\textbf{Semantics}}              & \multicolumn{2}{c}{\textbf{Expressiveness}} \\ \cmidrule(l){3-5} \cmidrule(l){6-7} \cmidrule(l){8-9}
\multirow{-2}{*}{\textbf{Models}}      & \multirow{-2}{*}{\textbf{Languages}} & \textbf{Timbre($\uparrow$)}      & \textbf{Reverb($\downarrow$)}     & \textbf{Fidelity($\uparrow$)}   & \textbf{Content($\downarrow$)}     & \textbf{Prosody($\uparrow$)}               & \textbf{Richness($\uparrow$)}   & \textbf{Hierarchy($\uparrow$)}    \\ \midrule
\multicolumn{9}{c}{\cellcolor[HTML]{FAF1D1}\textit{Open-Source Models}}       \\ \midrule
      & ZH     & {0.90} & 3.15        & 2.65        & \textit{0.069}  & \textbf{4.01}   & 3.01      & 2.87            \\
\multirow{-2}{*}{ZipVoice}             & EN     & 0.91 & 3.91        & 2.67       & \textit{\color{red} 0.114}  & 3.34 & 2.24      & 2.72      \\ \midrule
      & ZH     & 0.89 & \textit{3.11}       & 2.56      & 0.313  & 3.25   & 2.58       & 2.60        \\
\multirow{-2}{*}{MoonCast}             & EN     & 0.91 & \textbf{\color{red} 3.01} & 2.68     & 0.125  & 3.08   & 2.78        & 2.79          \\ \midrule
      & ZH     & \textbf{0.92} & \textit{3.32}     & 3.16      & 0.075  & \textit{3.57} & 3.16        & 3.03            \\
\multirow{-2}{*}{FireRedTTS2}          & EN     & \textit{\color{red} 0.93} & \textit{\color{red} 3.64}    &  2.08 & 0.131  & 2.91 & 2.29      & 2.58            \\ \midrule
      & ZH     & {0.90} & \textbf{3.02}        & 3.13       & 0.148  & 3.10  & \textit{3.66}      & \textit{3.26}        \\
\multirow{-2}{*}{MOSS-TTSD}            & EN     & 0.91 & 4.07   &  2.64        & 0.239  & 2.47  & 2.75    & 2.71        \\ \midrule
      & ZH     & \textit{0.90} & 3.26       & \textit{3.32}      & 0.106  & 3.48  & \textbf{3.74}      & 3.34           \\
\multirow{-2}{*}{VibeVoice}            & EN     & 0.91 & 3.91       & \color{red}\textit{3.38}       & 0.125  & \textit{\color{red} 3.66}  & \textbf{\color{red} 3.78}     & \textit{\color{red} 3.39}      \\ \midrule
      & ZH     & \textbf{0.92} & 3.31        & \textbf{3.94}      & \textbf{0.061}        & \textbf{4.01}  & 3.69        & \textbf{3.82}         \\
\multirow{-2}{*}{SoulXPodcast}         & EN     & \textbf{\color{red} 0.94} & 3.70        & \color{red}\textbf{3.98}       & \textbf{\color{red} 0.090} & \textbf{\color{red} 4.00} & \textit{\color{red} 3.18}       & \textbf{\color{red} 3.59}         \\ \midrule
      & ZH     & 0.91           &  3.20     & 3.13      &   0.129     & 3.42     &  3.31   & 3.15      \\
\multirow{-2}{*}{Average}& EN     & 0.92   &   3.71   &   3.07    &  0.154      & 3.24    &  2.84   &  2.96     \\ \midrule
\multicolumn{9}{c}{\cellcolor[HTML]{FAF1D1}\textit{Closed-Source Models}}     \\ \midrule
      & ZH     & 0.91   & 3.07   & {3.05}   & \textit{0.086}          & \textit{4.12}   & \textit{4.10}          & \textit{4.11}        \\
\multirow{-2}{*}{Gemini-2.5-pro-preview-tts}   & EN     & \textbf{\color{red} 0.93}   & 3.26   & 2.96  & \textbf{\color{red} 0.092}          & \textbf{\color{red} 4.00}  & \textbf{\color{red} 4.02}          & \textbf{\color{red} 3.93}        \\ \midrule
      & ZH     & \textit{0.92}   & \textit{2.97}          & 2.26         & 0.104  & 3.52  & 3.17          & 3.56           \\
\multirow{-2}{*}{OpenAI-tts-1-hd)}    & EN     & \textbf{\color{red} 0.93}   & \textbf{\color{red} 2.99}          & 2.29          & \textit{\color{red} 0.103}  & 3.86           & 3.41          & \textit{\color{red} 3.84}        \\ \midrule
      & ZH     & \textbf{0.93}   & 4.55          & \textit{3.38}       & 0.127          & 3.44           & 2.32        & 3.11           \\
\multirow{-2}{*}{Elevenlabs Multilingual V2)} & EN     & \textbf{\color{red} 0.93}   & 4.31           & \color{red}\textit{3.58}         & 0.109          & \textit{\color{red} 3.89}           & 3.36       & 3.81            \\ \midrule
      & ZH     & \textit{0.92}   & \textbf{2.48} & \textbf{3.90}    & \textbf{0.063}  & \textbf{4.16}        & \textbf{4.19}       & \textbf{4.26}          \\
\multirow{-2}{*}{Seed-TTS-Podcast}     & EN     & \textit{\color{red} 0.91}   & \textit{\color{red} 3.22}           & \color{red}\textbf{3.88}           & 0.108  & 3.70  & \textit{\color{red} 3.49}        & 3.42           \\ \midrule
      & ZH     &  0.92     & 3.27      &  3.15     & 0.095  & 3.81       &  3.45   & 3.76   \\
\multirow{-2}{*}{Average}& EN     &     0.93   &  3.45     &   3.18    & 0.103  & 3.86       &   3.57  & 3.75 \\ \bottomrule
\end{tabular}
}
\end{table*}

%% file: latex/app_sec/future_work_social_impact.tex
\section{Future Works}\label{app_sec:future}

While SwanBench-Speech provides a comprehensive evaluation framework for long-form speech generation, several challenges warrant further exploration:

\textbf{Dependency on Closed-source Models}: The evaluation of Expressiveness in SwanBench-Speech currently relies on closed-source models such as Gemini 3 Pro. The absence of open-source alternatives poses a risk to reproducibility due to potential updates in closed-source APIs. Future work will focus on distilling high-performance open-source evaluators using data derived from both human assessments and closed-source model outputs~\cite{ji2024wavchat}.

\textbf{Limited Language Coverage}: Our current dataset focuses exclusively on English and Chinese, omitting other languages, particularly low-resource ones. Future efforts should aim to expand the linguistic breadth of long-form speech generation evaluation.

\textbf{Timbre Sensitivity}: To ensure diversity, SwanBench-Speech utilizes over 20 reference voices spanning various genders and ages. However, as noted in prior work~\cite{manku2025emergenttts}, model performance in expressiveness and prosody is highly sensitive to the reference voice. Our current selection may not be sufficiently diverse. Future research should investigate the impact of input voice characteristics on long-form synthesis more deeply.

\textbf{Instruction Following Capabilities}: SwanBench-Speech primarily evaluates models in zero-shot settings. However, recent advancements have introduced models capable of Instruct-based speech generation~\cite{huang2025instructttseval,wang2025spark,zhou2025indextts2,xu2025qwen3}. Developing long-form InstructTTS systems and evaluating their instruction-following capabilities in long-context settings represent significant avenues for future research.

\section{Social Impacts}\label{app_sec:social}

This work aims to advance immersive and robust long-form speech generation, facilitating superior downstream applications. 
However, enhanced generative capabilities inevitably heighten the risk of misuse, potentially violating ethical norms and legal regulations. 
These risks highlight the critical need for ethically aligned practices and sufficient oversight. 
To mitigate these concerns, we subjected our text data to rigorous ethical review and anonymization. 
We also verified that the accompanying audio samples are free of Personally Identifiable Information (PII). 
Additionally, we mandate that all researchers utilizing this benchmark strictly adhere to the CC BY-NC-SA 4.0 license. 
We hope that the progress in speech generation technology will benefit society through responsible and ethical deployment.

%% file: latex/app_figs/prompt_prosody_coherence.tex


\begin{figure*}[t]

\centering

\begin{promptbox}{Prompt for Prosody Coherence}

\small

\textbf{Role:} Senior Linguistic Expert \& Prosody Analyst. You are an expert in assessing speech naturalness, with a hyper-sensitivity to prosodic coherence, rhythmic hierarchy, and robotic artifacts.

\textbf{Input Data:}
\begin{itemize}
    \item \textbf{Target Text:} The reference text script that needs to be synthesized.
    \item \textbf{Audio Output:} The speech audio generated by the TTS model (labeled as Output A).
\end{itemize}

\textbf{Generation Requirements:}
\begin{enumerate}
    \item \textbf{Core Task}: Evaluate the audio's naturalness by analyzing its \textbf{prosodic structure} and \textbf{coherence} against the target text, rather than just audio quality.
    \item \textbf{Dimension 1 - Prosody Coherence \& Flow}: Assess the smoothness of the speech stream. Check for unnatural pauses, abrupt disjoints between words/phrases, and the logical flow of intonation across sentence boundaries.
    \item \textbf{Dimension 2 - Rhythmic Hierarchy \& Layering}: Evaluate the structural stress patterns. Does the speaker correctly emphasize content words while de-emphasizing function words? Is there a natural "melody" (intonation contour) rather than a flat or repetitive beat?
    \item \textbf{Dimension 3 - Overall Naturalness}: Check for presence of human-like micro-prosody (e.g., breathiness, slight pitch variations).
    \item \textbf{Format}: Strictly output a valid JSON object. No other text.

\end{enumerate}

\textbf{Scoring Guidelines (1.0--5.0, step of 0.5):}

\begin{itemize}

    \item \textbf{5.0 (Human-Parity):} Indistinguishable from a professional human speaker; perfect coherence and rich prosodic hierarchy.
    \item \textbf{4.0 (Natural):} Very smooth and pleasant; minor prosodic flaws only noticeable to experts; good structural layering.
    \item \textbf{3.0 (Acceptable):} Intelligible and decent flow; but lacks depth (flat hierarchy) or contains audible TTS artifacts.
    \item \textbf{2.0 (Mechanical):} Disjointed flow; unnatural pauses; wrong stress placement (e.g., stressing every word equally).
    \item \textbf{1.0 (Robotic):} Completely lifeless; broken prosody; difficult to listen to.

\end{itemize}

\textbf{JSON Schema:} \\

\{ \\

\hspace*{1em} "Overall\_Impression": "[Brief summary of naturalness and flaws]", \\
\hspace*{1em} "Detailed\_Analysis": \{ \\
\hspace*{2em} "Coherence\_and\_Flow": "[Critique the smoothness and connection...]", \\
\hspace*{2em} "Hierarchy\_and\_Layering": "[Analyze stress patterns and intonation curves...]", \\
\hspace*{2em} "Naturalness": "[Comments on naturalness]" \\
\hspace*{1em} \}, \\
\hspace*{1em} "Score": [Number 1.0-5.0], \\

\}

\end{promptbox}

\vspace{-2pt}

\caption{Structured prompt for evaluating long-form audio's performance in Prosody Coherence.}
\label{app_fig:eval_prosody}

\end{figure*}

%% file: latex/app_figs/prompt_expressive_hierarchy.tex
\begin{figure*}[t]
\centering
\begin{promptbox}{Prompt for Expressive Hierarchy}
\small
\textbf{Role:} Senior Voice Director \& Audio Engineer (Long-Form Specialist). You are an expert in long-form narration (audiobooks, documentaries), hyper-sensitive to monotony, repetitive patterns, and lack of structural progression.

\textbf{Generation Requirements:}
\begin{enumerate}
    \item \textbf{Core Task}: Analyze how the performance \textbf{evolves over time}, focusing on "Layering and Hierarchy".
    \item \textbf{Dimension 1 - Emotional Variation \& Arc}: Evaluate progression from beginning to end, distinction between climax and exposition, and avoidance of "one-note" acting.
    \item \textbf{Dimension 2 - Vocal Dynamics}: Check for macro/micro dynamics (volume/tempo shifts).
    \item \textbf{Dimension 3 - Scene Appropriateness \& Structural Fit}: Assess contextual adaptation to content structure and long-term engagement.
    \item \textbf{Format}: Strictly output a valid JSON object. No other text.
\end{enumerate}

\textbf{Scoring Guidelines (1.0--5.0, step of 0.5):}
\begin{itemize}
    \item \textbf{5.0 (Masterful):} A journey with rich variety; no repetitive patterns; perfect for long listening.
    \item \textbf{4.0 (Strong):} Good dynamics and clear emotional shifts; avoids obvious monotony.
    \item \textbf{3.0 (Acceptable but Static):} Pleasant but lacks progression; risks boring the listener over time.
    \item \textbf{2.0 (Repetitive):} Clear signs of "looping prosody"; same intonation for every sentence.
    \item \textbf{1.0 (Robotic):} Lifeless; no dynamic range or emotional change; raw TTS-like.
\end{itemize}

\textbf{JSON Schema:} \\
\{ \\
\hspace*{1em} "Overall\_Impression": "[A brief summary of the long-form experience]", \\
\hspace*{1em} "Hierarchy\_Analysis": \{ \\
\hspace*{2em} "Emotional\_Arc": "[Describe the emotional progression...]", \\
\hspace*{2em} "Dynamics\_and\_Rhythm": "[Critique the pacing and prosody...]", \\
\hspace*{2em} "Scene\_Fit": "[How well does it adapt to the structure?]" \\
\hspace*{1em} \}, \\
\hspace*{1em} "Score": [Number 1.0-5.0], \\
\hspace*{1em} "Final\_Recommendation": "[Highly Recommended / Recommended with Reservations / Not Recommended]" \\
\}
\end{promptbox}
\vspace{-2pt}
\caption{Structured prompt for evaluating long-form audio performance, focusing on expressive hierarchy.}
\label{app_fig:expressive_hierarchy}
\end{figure*}

%% file: latex/app_figs/prompt_expressive_richness.tex
\begin{figure*}[t]
\centering
\begin{promptbox}{Prompt for Expressive Richness}
\small
\textbf{Role:} You are a Senior Voice Director and Audio Engineer with standards equivalent to a top-tier animation studio. Your task is to meticulously evaluate a voice recording and determine if it meets professional standards.

\textbf{Evaluation Dimension: Performance \& Expressiveness}
\begin{itemize}
    \item \textbf{Emotional Resonance:} Genuine, layered emotion vs. flat/forced.
    \item \textbf{Character Portrayal:} Believable, consistent character; tone/age/personality coherence.
    \item \textbf{Storytelling \& Immersion:} Narrative flow, atmosphere, and engagement.
\end{itemize}

\textbf{Exclusions:} Ignore sudden stop, audio quality, timbre consistency, and pronunciation accuracy.

\textbf{Scoring Guidelines (1.0--5.0):}
\begin{itemize}
    \item \textbf{5.0 (Outstanding):} Richly expressive, immersive, and artistically elevated.
    \item \textbf{4.0 (Strong):} High expressiveness, close to professional but lacks fine nuance.
    \item \textbf{3.0 (Adequate):} Meets basic requirements; emotions may be somewhat generic.
    \item \textbf{2.0 (Flat):} Unconvincing, weak emotional expression, clearly subpar.
    \item \textbf{1.0 (Mechanical):} Synthetic/lifeless, no emotional color or dynamics.
\end{itemize}

\textbf{JSON Schema:} \\
\{ \\
\hspace*{1em} "Overall\_Impression": "A brief, one-sentence summary of the audio.", \\
\hspace*{1em} "Expressiveness": "Detailed professional analysis of the performance dimension.", \\
\hspace*{1em} "Expressiveness\_Score": [Number between 1.0 and 5.0 in 0.5 increments], \\
\hspace*{1em} "Final\_Recommendation": "[Highly Recommended / Recommended with Reservations / Not Recommended]" \\
\}
\end{promptbox}
\vspace{-2pt}
\caption{The structured prompt used for professional voice performance and expressiveness assessment.}
\label{app_fig:expressive_richness}
\end{figure*}

%% file: custom.bib
@Article{E210196,
title = {Improved Cross-Corpus Speech Emotion Recognition Using Deep Local Domain Adaptation},
journal = {Chinese Journal of Electronics},
volume = {32},
number = {3},
pages = {640-646},
year = {2023},
issn = {},
doi = {10.23919/cje.2021.00.196},	
url = {https://cje.ejournal.org.cn/en/article/doi/10.23919/cje.2021.00.196},
author = {ZHAO Huijuan and YE Ning and WANG Ruchuan}
}

@Article{E200080,
title = {Multi-Distributed Speech Emotion Recognition Based on Mel Frequency Cepstogram and Parameter Transfer},
journal = {Chinese Journal of Electronics},
volume = {31},
number = {1},
pages = {155-167},
year = {2022},
issn = {},
doi = {10.1049/cje.2020.00.080},	
url = {https://cje.ejournal.org.cn/en/article/doi/10.1049/cje.2020.00.080},
author = {LIN Long and TAN Liang}
}

@article{chen2026wavalign,
  title={WavAlign: Enhancing Intelligence and Expressiveness in Spoken Dialogue Models via Adaptive Hybrid Post-Training},
  author={Chen, Yifu and Ji, Shengpeng and Chen, Qian and Liang, Tianle and Li, Yangzhuo and Wang, Ziqing and Wang, Wen and Lu, Jingyu and Wang, Haoxiao and Pu, Xueyi and Zhuo, Fan and Zhao, Zhou},
  journal={arXiv preprint arXiv:2604.14932},
  year={2026}
}

@article{chen2026dualaxis,
  title={Dual-Axis Generative Reward Model Toward Semantic and Turn-taking Robustness in Interactive Spoken Dialogue Models},
  author={Chen, Yifu and Ji, Shengpeng and Liu, Zhengqing and Chen, Qian and Wang, Wen and Wang, Ziqing and Li, Yangzhuo and Liang, Tianle and Zhao, Zhou},
  journal={arXiv preprint arXiv:2604.14920},
  year={2026}
}

@inproceedings{wang2025zipenhancer,
  title={ZipEnhancer: Dual-Path Down-Up Sampling-based Zipformer for Monaural Speech Enhancement},
  author={Wang, Haoxu and Tian, Biao},
  booktitle={ICASSP 2025-2025 IEEE International Conference on Acoustics, Speech and Signal Processing (ICASSP)},
  pages={1--5},
  year={2025},
}

@article{an2024funaudiollm,
  title={Funaudiollm: Voice understanding and generation foundation models for natural interaction between humans and llms},
  author={An, Keyu and Chen, Qian and Deng, Chong and Du, Zhihao and Gao, Changfeng and Gao, Zhifu and Gu, Yue and He, Ting and Hu, Hangrui and Hu, Kai and others},
  journal={arXiv preprint arXiv:2407.04051},
  year={2024}
}

@inproceedings{zhang2022wenetspeech,
  title={Wenetspeech: A 10000+ hours multi-domain mandarin corpus for speech recognition},
  author={Zhang, Binbin and Lv, Hang and Guo, Pengcheng and Shao, Qijie and Yang, Chao and Xie, Lei and Xu, Xin and Bu, Hui and Chen, Xiaoyu and Zeng, Chenchen and others},
  booktitle={ICASSP 2022-2022 IEEE International Conference on Acoustics, Speech and Signal Processing (ICASSP)},
  pages={6182--6186},
  year={2022},
}

@article{kharitonov2023speak,
  title={Speak, read and prompt: High-fidelity text-to-speech with minimal supervision},
  author={Kharitonov, Eugene and Vincent, Damien and Borsos, Zal{\'a}n and Marinier, Rapha{\"e}l and Girgin, Sertan and Pietquin, Olivier and Sharifi, Matt and Tagliasacchi, Marco and Zeghidour, Neil},
  journal={Transactions of the Association for Computational Linguistics},
  volume={11},
  pages={1703--1718},
  year={2023},
}

@article{wang2021survey,
  title={A survey on curriculum learning},
  author={Wang, Xin and Chen, Yudong and Zhu, Wenwu},
  journal={IEEE transactions on pattern analysis and machine intelligence},
  pages={4555--4576},
  year={2021},
}

@article{gao2022paraformer,
  title={Paraformer: Fast and accurate parallel transformer for non-autoregressive end-to-end speech recognition},
  author={Gao, Zhifu and Zhang, Shiliang and McLoughlin, Ian and Yan, Zhijie},
  journal={arXiv preprint arXiv:2206.08317},
  year={2022}
}

@inproceedings{esser2024scaling,
  title={Scaling rectified flow transformers for high-resolution image synthesis},
  author={Esser, Patrick and Kulal, Sumith and Blattmann, Andreas and Entezari, Rahim and M{\"u}ller, Jonas and Saini, Harry and Levi, Yam and Lorenz, Dominik and Sauer, Axel and Boesel, Frederic and others},
  booktitle={Forty-first international conference on machine learning},
  year={2024}
}

@article{chen2024vall,
  title={Vall-e 2: Neural codec language models are human parity zero-shot text to speech synthesizers},
  author={Chen, Sanyuan and Liu, Shujie and Zhou, Long and Liu, Yanqing and Tan, Xu and Li, Jinyu and Zhao, Sheng and Qian, Yao and Wei, Furu},
  journal={arXiv preprint arXiv:2406.05370},
  year={2024}
}

@inproceedings{bai2024longbench,
  title={Longbench: A bilingual, multitask benchmark for long context understanding},
  author={Bai, Yushi and Lv, Xin and Zhang, Jiajie and Lyu, Hongchang and Tang, Jiankai and Huang, Zhidian and Du, Zhengxiao and Liu, Xiao and Zeng, Aohan and Hou, Lei and others},
  booktitle={Proceedings of the 62nd annual meeting of the association for computational linguistics (volume 1: Long papers)},
  pages={3119--3137},
  year={2024}
}

@article{le2023voicebox,
  title={Voicebox: Text-guided multilingual universal speech generation at scale},
  author={Le, Matthew and Vyas, Apoorv and Shi, Bowen and Karrer, Brian and Sari, Leda and Moritz, Rashel and Williamson, Mary and Manohar, Vimal and Adi, Yossi and Mahadeokar, Jay and others},
  journal={Advances in neural information processing systems},
  volume={36},
  pages={14005--14034},
  year={2023}
}

@article{chen2023longlora,
  title={Longlora: Efficient fine-tuning of long-context large language models},
  author={Chen, Yukang and Qian, Shengju and Tang, Haotian and Lai, Xin and Liu, Zhijian and Han, Song and Jia, Jiaya},
  journal={arXiv preprint arXiv:2309.12307},
  year={2023}
}

@article{xiao2023efficient,
  title={Efficient streaming language models with attention sinks},
  author={Xiao, Guangxuan and Tian, Yuandong and Chen, Beidi and Han, Song and Lewis, Mike},
  journal={arXiv preprint arXiv:2309.17453},
  year={2023}
}

@inproceedings{he2024emilia,
  title={Emilia: An extensive, multilingual, and diverse speech dataset for large-scale speech generation},
  author={He, Haorui and Shang, Zengqiang and Wang, Chaoren and Li, Xuyuan and Gu, Yicheng and Hua, Hua and Liu, Liwei and Yang, Chen and Li, Jiaqi and Shi, Peiyang and others},
  booktitle={2024 IEEE Spoken Language Technology Workshop (SLT)},
  pages={885--890},
  year={2024},
}

@article{chen2021gigaspeech,
  title={Gigaspeech: An evolving, multi-domain asr corpus with 10,000 hours of transcribed audio},
  author={Chen, Guoguo and Chai, Shuzhou and Wang, Guanbo and Du, Jiayu and Zhang, Wei-Qiang and Weng, Chao and Su, Dan and Povey, Daniel and Trmal, Jan and Zhang, Junbo and others},
  journal={arXiv preprint arXiv:2106.06909},
  year={2021}
}

@article{zhang2025speechjudge,
  title={SpeechJudge: Towards Human-Level Judgment for Speech Naturalness},
  author={Zhang, Xueyao and Wang, Chaoren and Liao, Huan and Li, Ziniu and Wang, Yuancheng and Wang, Li and Jia, Dongya and Chen, Yuanzhe and Li, Xiulin and Chen, Zhuo and others},
  journal={arXiv preprint arXiv:2511.07931},
  year={2025}
}

@article{reimers2019sentence,
  title={Sentence-bert: Sentence embeddings using siamese bert-networks},
  author={Reimers, Nils and Gurevych, Iryna},
  journal={arXiv preprint arXiv:1908.10084},
  year={2019}
}

@article{xu2025qwen2,
  title={Qwen2. 5-omni technical report},
  author={Xu, Jin and Guo, Zhifang and He, Jinzheng and Hu, Hangrui and He, Ting and Bai, Shuai and Chen, Keqin and Wang, Jialin and Fan, Yang and Dang, Kai and others},
  journal={arXiv preprint arXiv:2503.20215},
  year={2025}
}

@article{zheng20233d,
  title={3d-speaker: A large-scale multi-device, multi-distance, and multi-dialect corpus for speech representation disentanglement},
  author={Zheng, Siqi and Cheng, Luyao and Chen, Yafeng and Wang, Hui and Chen, Qian},
  journal={arXiv preprint arXiv:2306.15354},
  year={2023}
}

@inproceedings{reddy2021dnsmos,
  title={DNSMOS: A non-intrusive perceptual objective speech quality metric to evaluate noise suppressors},
  author={Reddy, Chandan KA and Gopal, Vishak and Cutler, Ross},
  booktitle={ICASSP 2021-2021 IEEE International Conference on Acoustics, Speech and Signal Processing (ICASSP)},
  pages={6493--6497},
  year={2021},
}

@article{park2024long,
  title={Long-form speech generation with spoken language models},
  author={Park, Se Jin and Salazar, Julian and Jansen, Aren and Kinoshita, Keisuke and Ro, Yong Man and Skerry-Ryan, RJ},
  journal={arXiv preprint arXiv:2412.18603},
  year={2024}
}

@article{minixhofer2025ttsds2,
  title={TTSDS2: Resources and Benchmark for Evaluating Human-Quality Text to Speech Systems},
  author={Minixhofer, Christoph and Klejch, Ondrej and Bell, Peter},
  journal={arXiv preprint arXiv:2506.19441},
  year={2025}
}

@inproceedings{chen2024mllm,
  title={Mllm-as-a-judge: Assessing multimodal llm-as-a-judge with vision-language benchmark},
  author={Chen, Dongping and Chen, Ruoxi and Zhang, Shilin and Wang, Yaochen and Liu, Yinuo and Zhou, Huichi and Zhang, Qihui and Wan, Yao and Zhou, Pan and Sun, Lichao},
  booktitle={Forty-first International Conference on Machine Learning},
  year={2024}
}

@article{guo2024text,
  title={Text-aware and Context-aware Expressive Audiobook Speech Synthesis},
  author={Guo, Dake and Zhu, Xinfa and Xue, Liumeng and Zhang, Yongmao and Tian, Wenjie and Xie, Lei},
  journal={arXiv preprint arXiv:2406.05672},
  year={2024}
}

@inproceedings{taal2010short,
  title={A short-time objective intelligibility measure for time-frequency weighted noisy speech},
  author={Taal, Cees H and Hendriks, Richard C and Heusdens, Richard and Jensen, Jesper},
  booktitle={IEEE international conference on acoustics, speech and signal processing},
  year={2010},
}

@inproceedings{minixhofer2024ttsds,
  title={TTSDS-Text-to-Speech Distribution Score},
  author={Minixhofer, Christoph and Klejch, Ond{\v{r}}ej and Bell, Peter},
  booktitle={2024 IEEE Spoken Language Technology Workshop (SLT)},
  pages={766--773},
  year={2024},
}

@article{wei2022chain,
  title={Chain-of-thought prompting elicits reasoning in large language models},
  author={Wei, Jason and Wang, Xuezhi and Schuurmans, Dale and Bosma, Maarten and Xia, Fei and Chi, Ed and Le, Quoc V and Zhou, Denny and others},
  journal={Advances in neural information processing systems},
  year={2022}
}

@article{clark2019evaluating,
  title={Evaluating long-form text-to-speech: Comparing the ratings of sentences and paragraphs},
  author={Clark, Rob and Silen, Hanna and Kenter, Tom and Leith, Ralph},
  journal={arXiv preprint arXiv:1909.03965},
  year={2019}
}

@inproceedings{ali2018word,
  title={Word error rate estimation for speech recognition: e-WER},
  author={Ali, Ahmed and Renals, Steve},
  booktitle={Proceedings of the 56th Annual Meeting of the Association for Computational Linguistics (Volume 2: Short Papers)},
  pages={20--24},
  year={2018},
}

@inproceedings{zhang2023audiobook,
  title={Audiobook synthesis with long-form neural text-to-speech},
  author={Zhang, Weicheng and Yeh, Cheng-Chieh and Beckman, Will and Raitio, Tuomo and Rasipuram, Ramya and Golipour, Ladan and Winarsky, David},
  booktitle={12th Speech Synthesis Workshop (SSW) 2023},
  year={2023}
}

@article{nishimura2024hall,
  title={HALL-E: hierarchical neural codec language model for minute-long zero-shot text-to-speech synthesis},
  author={Nishimura, Yuto and Hirose, Takumi and Ohi, Masanari and Nakayama, Hideki and Inoue, Nakamasa},
  journal={arXiv preprint arXiv:2410.04380},
  year={2024}
}

@inproceedings{cambre2020choice,
  title={Choice of voices: A large-scale evaluation of text-to-speech voice quality for long-form content},
  author={Cambre, Julia and Colnago, Jessica and Maddock, Jim and Tsai, Janice and Kaye, Jofish},
  booktitle={Proceedings of the 2020 CHI Conference on Human Factors in Computing Systems},
  year={2020}
}

@article{li2025long,
  title={Long-Context Speech Synthesis with Context-Aware Memory},
  author={Li, Zhipeng and Xing, Xiaofen and Xing, Jingyuan and Hu, Hangrui and Lu, Heng and Xu, Xiangmin},
  journal={arXiv preprint arXiv:2508.14713},
  year={2025}
}

@article{borsos2023soundstorm,
  title={Soundstorm: Efficient parallel audio generation},
  author={Borsos, Zal{\'a}n and Sharifi, Matt and Vincent, Damien and Kharitonov, Eugene and Zeghidour, Neil and Tagliasacchi, Marco},
  journal={arXiv preprint arXiv:2305.09636},
  year={2023}
}

@article{wang2025spark,
  title={Spark-tts: An efficient llm-based text-to-speech model with single-stream decoupled speech tokens},
  author={Wang, Xinsheng and Jiang, Mingqi and Ma, Ziyang and Zhang, Ziyu and Liu, Songxiang and Li, Linqin and Liang, Zheng and Zheng, Qixi and Wang, Rui and Feng, Xiaoqin and others},
  journal={arXiv preprint arXiv:2503.01710},
  year={2025}
}

@article{ju2025mooncast,
  title={MoonCast: High-quality zero-shot podcast generation},
  author={Ju, Zeqian and Yang, Dongchao and Yu, Jianwei and Shen, Kai and Leng, Yichong and Wang, Zhengtao and Tan, Xu and Zhou, Xinyu and Qin, Tao and Li, Xiangyang},
  journal={arXiv preprint arXiv:2503.14345},
  year={2025}
}

@article{zhao2025moss,
  title={MOSS-Speech: Towards True Speech-to-Speech Models Without Text Guidance},
  author={Zhao, Xingjian and Xu, Zhe and Cheng, Qinyuan and Fei, Zhaoye and Jin, Luozhijie and Wang, Yang and Chen, Hanfu and Jiang, Yaozhou and Gao, Qinghui and Chen, Ke and others},
  journal={arXiv preprint arXiv:2510.00499},
  year={2025}
}

@article{xie2025fireredtts,
  title={Fireredtts-2: Towards long conversational speech generation for podcast and chatbot},
  author={Xie, Kun and Shen, Feiyu and Li, Junjie and Xie, Fenglong and Tang, Xu and Hu, Yao},
  journal={arXiv preprint arXiv:2509.02020},
  year={2025}
}

@article{zhu2025zipvoice,
  title={ZipVoice: Fast and High-Quality Zero-Shot Text-to-Speech with Flow Matching},
  author={Zhu, Han and Kang, Wei and Yao, Zengwei and Guo, Liyong and Kuang, Fangjun and Li, Zhaoqing and Zhuang, Weiji and Lin, Long and Povey, Daniel},
  journal={arXiv preprint arXiv:2506.13053},
  year={2025}
}

@article{atamanenko2025inworldtts,
  title={Tts-1 technical report},
  author={Atamanenko, Oleg and Chalova, Anna and Coombes, Joseph and Cope, Nikki and Dang, Phillip and Deng, Zhifeng and Du, Jimmy and Ermolenko, Michael and Fan, Feifan and Feng, Yufei and others},
  journal={arXiv preprint arXiv:2507.21138},
  year={2025}
}

@article{xie2025soulx,
  title={SoulX-Podcast: Towards Realistic Long-form Podcasts with Dialectal and Paralinguistic Diversity},
  author={Xie, Hanke and Lin, Haopeng and Cao, Wenxiao and Guo, Dake and Tian, Wenjie and Wu, Jun and Wen, Hanlin and Shang, Ruixuan and Liu, Hongmei and Jiang, Zhiqi and others},
  journal={arXiv preprint arXiv:2510.23541},
  year={2025}
}

@article{liao2024fish,
  title={Fish-speech: Leveraging large language models for advanced multilingual text-to-speech synthesis},
  author={Liao, Shijia and Wang, Yuxuan and Li, Tianyu and Cheng, Yifan and Zhang, Ruoyi and Zhou, Rongzhi and Xing, Yijin},
  journal={arXiv preprint arXiv:2411.01156},
  year={2024}
}

@misc{gemini3,
  title        = {Gemini 3},
  author       = {{Google DeepMind}},
  howpublished = {\url{https://deepmind.google/technologies/gemini/}},
  note         = {Accessed: 2025-12-25},
  year         = {2025}
}

@misc{gpt5,
  title        = {GPT-5},
  author       = {{OpenAI}},
  howpublished = {\url{https://chagpt.com/}},
  note         = {Accessed: 2025-12-25},
  year         = {2025}
}

@article{zhang2025minimax,
  title={Minimax-speech: Intrinsic zero-shot text-to-speech with a learnable speaker encoder},
  author={Zhang, Bowen and Guo, Congchao and Yang, Geng and Yu, Hang and Zhang, Haozhe and Lei, Heidi and Mai, Jialong and Yan, Junjie and Yang, Kaiyue and Yang, Mingqi and others},
  journal={arXiv preprint arXiv:2505.07916},
  year={2025}
}

@article{zhu2025zipvoicedialog,
  title={ZipVoice-Dialog: Non-Autoregressive Spoken Dialogue Generation with Flow Matching},
  author={Zhu, Han and Kang, Wei and Guo, Liyong and Yao, Zengwei and Kuang, Fangjun and Zhuang, Weiji and Li, Zhaoqing and Han, Zhifeng and Zhang, Dong and Zhang, Xin and others},
  journal={arXiv preprint arXiv:2507.09318},
  year={2025}
}

@article{cui2025glm,
  title={GLM-TTS Technical Report},
  author={Cui, Jiayan and Yang, Zhihan and Li, Naihan and Tian, Jiankun and Ma, Xingyu and Zhang, Yi and Chen, Guangyu and Yang, Runxuan and Cheng, Yuqing and Zhou, Yizhi and others},
  journal={arXiv preprint arXiv:2512.14291},
  year={2025}
}

@article{zhou2025indextts2,
  title={IndexTTS2: A Breakthrough in Emotionally Expressive and Duration-Controlled Auto-Regressive Zero-Shot Text-to-Speech},
  author={Zhou, Siyi and Zhou, Yiquan and He, Yi and Zhou, Xun and Wang, Jinchao and Deng, Wei and Shu, Jingchen},
  journal={arXiv preprint arXiv:2506.21619},
  year={2025}
}

@article{peng2025vibevoice,
  title={Vibevoice technical report},
  author={Peng, Zhiliang and Yu, Jianwei and Wang, Wenhui and Chang, Yaoyao and Sun, Yutao and Dong, Li and Zhu, Yi and Xu, Weijiang and Bao, Hangbo and Wang, Zehua and others},
  journal={arXiv preprint arXiv:2508.19205},
  year={2025}
}

@article{du2024cosyvoice2,
  title={Cosyvoice 2: Scalable streaming speech synthesis with large language models},
  author={Du, Zhihao and Wang, Yuxuan and Chen, Qian and Shi, Xian and Lv, Xiang and Zhao, Tianyu and Gao, Zhifu and Yang, Yexin and Gao, Changfeng and Wang, Hui and others},
  journal={arXiv preprint arXiv:2412.10117},
  year={2024}
}

@article{xu2025qwen3,
  title={Qwen3-omni technical report},
  author={Xu, Jin and Guo, Zhifang and Hu, Hangrui and Chu, Yunfei and Wang, Xiong and He, Jinzheng and Wang, Yuxuan and Shi, Xian and He, Ting and Zhu, Xinfa and others},
  journal={arXiv preprint arXiv:2509.17765},
  year={2025}
}

@article{anastassiou2024seed,
  title={Seed-tts: A family of high-quality versatile speech generation models},
  author={Anastassiou, Philip and Chen, Jiawei and Chen, Jitong and Chen, Yuanzhe and Chen, Zhuo and Chen, Ziyi and Cong, Jian and Deng, Lelai and Ding, Chuang and Gao, Lu and others},
  journal={arXiv preprint arXiv:2406.02430},
  year={2024}
}

@article{manku2025emergenttts,
  title={EmergentTTS-Eval: Evaluating TTS Models on Complex Prosodic, Expressiveness, and Linguistic Challenges Using Model-as-a-Judge},
  author={Manku, Ruskin Raj and Tang, Yuzhi and Shi, Xingjian and Li, Mu and Smola, Alex},
  journal={arXiv preprint arXiv:2505.23009},
  year={2025}
}

@article{du2025cosyvoice,
  title={Cosyvoice 3: Towards in-the-wild speech generation via scaling-up and post-training},
  author={Du, Zhihao and Gao, Changfeng and Wang, Yuxuan and Yu, Fan and Zhao, Tianyu and Wang, Hao and Lv, Xiang and Wang, Hui and Ni, Chongjia and Shi, Xian and others},
  journal={arXiv preprint arXiv:2505.17589},
  year={2025}
}

@article{huang2025instructttseval,
  title={InstructTTSEval: Benchmarking Complex Natural-Language Instruction Following in Text-to-Speech Systems},
  author={Huang, Kexin and Tu, Qian and Fan, Liwei and Yang, Chenchen and Zhang, Dong and Li, Shimin and Fei, Zhaoye and Cheng, Qinyuan and Qiu, Xipeng},
  journal={arXiv preprint arXiv:2506.16381},
  year={2025}
}

@inproceedings{pan2025synthetic,
  title={Synthetic Singers: A Review of Deep-Learning-based Singing Voice Synthesis Approaches},
  author={Pan, Changhao and Yao, Dongyu and Zhang, Yu and Guo, Wenxiang and Lu, Jingyu and Zhu, Zhiyuan and Zhao, Zhou},
  booktitle={Proceedings of the 14th International Joint Conference on Natural Language Processing and the 4th Conference of the Asia-Pacific Chapter of the Association for Computational Linguistics},
  pages={396--416},
  year={2025}
}

@inproceedings{chen2025wavrag,
  title={Wavrag: Audio-integrated retrieval augmented generation for spoken dialogue models},
  author={Chen, Yifu and Ji, Shengpeng and Wang, Haoxiao and Wang, Ziqing and Chen, Siyu and He, Jinzheng and Xu, Jin and Zhao, Zhou},
  booktitle={Proceedings of the 63rd Annual Meeting of the Association for Computational Linguistics (Volume 1: Long Papers)},
  pages={12505--12523},
  year={2025}
}

@article{ji2024wavtokenizer,
  title={Wavtokenizer: an efficient acoustic discrete codec tokenizer for audio language modeling},
  author={Ji, Shengpeng and Jiang, Ziyue and Wang, Wen and Chen, Yifu and Fang, Minghui and Zuo, Jialong and Yang, Qian and Cheng, Xize and Wang, Zehan and Li, Ruiqi and others},
  journal={arXiv preprint arXiv:2408.16532},
  year={2024}
}

@article{zhang2025tcsinger,
  title={TCSinger 2: Customizable Multilingual Zero-shot Singing Voice Synthesis},
  author={Zhang, Yu and Guo, Wenxiang and Pan, Changhao and Yao, Dongyu and Zhu, Zhiyuan and Jiang, Ziyue and Wang, Yuhan and Jin, Tao and Zhao, Zhou},
  journal={arXiv preprint arXiv:2505.14910},
  year={2025}
}

@article{jiang2025megatts,
  title={Megatts 3: Sparse alignment enhanced latent diffusion transformer for zero-shot speech synthesis},
  author={Jiang, Ziyue and Ren, Yi and Li, Ruiqi and Ji, Shengpeng and Zhang, Boyang and Ye, Zhenhui and Zhang, Chen and Jionghao, Bai and Yang, Xiaoda and Zuo, Jialong and others},
  journal={arXiv preprint arXiv:2502.18924},
  year={2025}
}

@article{ji2024wavchat,
  title={WavChat: A Survey of Spoken Dialogue Models},
  author={Ji, Shengpeng and Chen, Yifu and Fang, Minghui and Zuo, Jialong and Lu, Jingyu and Wang, Hanting and Jiang, Ziyue and Zhou, Long and Liu, Shujie and Cheng, Xize and Yang, Xiaoda and Wang,  Zehan and Yang, Qian and Li, Jian and Jiang, Yidi ang He, Jingzhen and Chu, Yunfei and Xu, Jin and Zhao, Zhou},
  journal={arXiv preprint arXiv:2411.13577},
  year={2024}
}

@inproceedings{panayotov2015librispeech,
  title={Librispeech: an asr corpus based on public domain audio books},
  author={Panayotov, Vassil and Chen, Guoguo and Povey, Daniel and Khudanpur, Sanjeev},
  booktitle={2015 IEEE international conference on acoustics, speech and signal processing (ICASSP)},
  pages={5206--5210},
  year={2015},
  organization={IEEE}
}

@article{liu2024deepseek,
  title={Deepseek-v3 technical report},
  author={Liu, Aixin and Feng, Bei and Xue, Bing and Wang, Bingxuan and Wu, Bochao and Lu, Chengda and Zhao, Chenggang and Deng, Chengqi and Zhang, Chenyu and Ruan, Chong and others},
  journal={arXiv preprint arXiv:2412.19437},
  year={2024}
}

@article{koizumi2023libritts,
  title={Libritts-r: A restored multi-speaker text-to-speech corpus},
  author={Koizumi, Yuma and Zen, Heiga and Karita, Shigeki and Ding, Yifan and Yatabe, Kohei and Morioka, Nobuyuki and Bacchiani, Michiel and Zhang, Yu and Han, Wei and Bapna, Ankur},
  journal={arXiv preprint arXiv:2305.18802},
  year={2023}
}

@inproceedings{zhang2025isdrama,
  title={Isdrama: Immersive spatial drama generation through multimodal prompting},
  author={Zhang, Yu and Guo, Wenxiang and Pan, Changhao and Zhu, Zhiyuan and Jin, Tao and Zhao, Zhou},
  booktitle={Proceedings of the 33rd ACM International Conference on Multimedia},
  year={2025}
}

@article{srivastav2025open,
  title={Open ASR Leaderboard: Towards Reproducible and Transparent Multilingual and Long-Form Speech Recognition Evaluation},
  author={Srivastav, Vaibhav and Zheng, Steven and Bezzam, Eric and Bihan, Eustache Le and Koluguri, Nithin and {\.Z}elasko, Piotr and Majumdar, Somshubra and Moumen, Adel and Gandhi, Sanchit},
  journal={arXiv preprint arXiv:2510.06961},
  year={2025}
}

@inproceedings{kumar2023torchaudio,
  title={Torchaudio-squim: Reference-less speech quality and intelligibility measures in torchaudio},
  author={Kumar, Anurag and Tan, Ke and Ni, Zhaoheng and Manocha, Pranay and Zhang, Xiaohui and Henderson, Ethan and Xu, Buye},
  booktitle={ICASSP 2023-2023 IEEE International Conference on Acoustics, Speech and Signal Processing (ICASSP)},
  pages={1--5},
  year={2023},
}

@inproceedings{zhang2024stylesinger,
title={StyleSinger: Style Transfer for Out-of-Domain Singing Voice Synthesis},
author={Zhang, Yu and Huang, Rongjie and Li, Ruiqi and He, JinZheng and Xia, Yan and Chen, Feiyang and Duan, Xinyu and Huai, Baoxing and Zhao, Zhou},
booktitle={Proceedings of the AAAI Conference on Artificial Intelligence},
volume={38},
pages={19597--19605},
year={2024}
}

@article{martinez2020msp,
  title={The MSP-conversation corpus},
  author={Martinez-Lucas, Luz and Abdelwahab, Mohammed and Busso, Carlos},
  journal={Interspeech 2020},
  year={2020}
}

@inproceedings{zhou2025childmandarin,
  title={Childmandarin: A comprehensive mandarin speech dataset for young children aged 3-5},
  author={Zhou, Jiaming and Wang, Shiyao and Zhao, Shiwan and He, Jiabei and Sun, Haoqin and Wang, Hui and Liu, Cheng and Kong, Aobo and Guo, Yujie and Yang, Xi and others},
  booktitle={Proceedings of the 63rd Annual Meeting of the Association for Computational Linguistics},
  pages={12524--12537},
  year={2025}
}

@inproceedings{baba2024t05,
  title={The t05 system for the voicemos challenge 2024: Transfer learning from deep image classifier to naturalness mos prediction of high-quality synthetic speech},
  author={Baba, Kaito and Nakata, Wataru and Saito, Yuki and Saruwatari, Hiroshi},
  booktitle={2024 IEEE Spoken Language Technology Workshop (SLT)},
  pages={818--824},
  year={2024},
}

@inproceedings{liu2024generative,
  title={Generative expressive conversational speech synthesis},
  author={Liu, Rui and Hu, Yifan and Ren, Yi and Yin, Xiang and Li, Haizhou},
  booktitle={Proceedings of the 32nd ACM International Conference on Multimedia},
  pages={4187--4196},
  year={2024}
}

@article{tian2025step,
  title={Step-Audio-R1 Technical Report},
  author={Tian, Fei and Zhang, Xiangyu Tony and Zhang, Yuxin and Zhang, Haoyang and Li, Yuxin and Liu, Daijiao and Deng, Yayue and Wu, Donghang and Chen, Jun and Zhao, Liang and others},
  journal={arXiv preprint arXiv:2511.15848},
  year={2025}
}

@misc{openai2024sora,
  title = {Video generation models as world simulators},
  author = {OpenAI},
  year = {2024},
  howpublished = {\url{https://openai.com/index/video-generation-models-as-world-simulators/}},
  note = {Accessed: 2025-12-25}
}

@article{huynh2025ozspeech,
  title={OZSpeech: One-step Zero-shot Speech Synthesis with Learned-Prior-Conditioned Flow Matching},
  author={Huynh-Nguyen, Hieu-Nghia and Nguyen, Ngoc Son and Dang, Huynh Nguyen and Vo, Thieu and Hy, Truong-Son and Nguyen, Van},
  journal={arXiv preprint arXiv:2505.12800},
  year={2025}
}

@article{guo2025deepseek,
  title={Deepseek-r1: Incentivizing reasoning capability in llms via reinforcement learning},
  author={Guo, Daya and Yang, Dejian and Zhang, Haowei and Song, Junxiao and Zhang, Ruoyu and Xu, Runxin and Zhu, Qihao and Ma, Shirong and Wang, Peiyi and Bi, Xiao and others},
  journal={arXiv preprint arXiv:2501.12948},
  year={2025}
}

@inproceedings{mcauliffe2017montreal,
  title={Montreal forced aligner: Trainable text-speech alignment using kaldi.},
  author={McAuliffe, Michael and Socolof, Michaela and Mihuc, Sarah and Wagner, Michael and Sonderegger, Morgan},
  booktitle={Interspeech},
  volume={2017},
  pages={498--502},
  year={2017}
}

@inproceedings{zhang2024tcsinger,
  title={TCSinger: Zero-Shot Singing Voice Synthesis with Style Transfer and Multi-Level Style Control},
  author={Zhang, Yu and Jiang, Ziyue and Li, Ruiqi and Pan, Changhao and He, Jinzheng and Huang, Rongjie and Wang, Chuxin and Zhao, Zhou},
  booktitle={Proceedings of the 2024 Conference on Empirical Methods in Natural Language Processing},
  pages={1960--1975},
  year={2024}
}

@inproceedings{ren2020fastspeech,
  title={FastSpeech 2: Fast and High-Quality End-to-End Text to Speech},
  author={Ren, Yi and Hu, Chenxu and Tan, Xu and Qin, Tao and Zhao, Sheng and Zhao, Zhou and Liu, Tie-Yan},
  booktitle={International Conference on Learning Representations},
  year={2020}
}

@misc{kingma2013auto,
  title={Auto-encoding variational bayes},
  author={Kingma, Diederik P and Welling, Max and others},
  year={2013},
  publisher={Banff, Canada}
}

@inproceedings{shen2023naturalspeech,
  title={NaturalSpeech 2: Latent Diffusion Models are Natural and Zero-Shot Speech and Singing Synthesizers},
  author={Shen, Kai and Ju, Zeqian and Tan, Xu and Liu, Eric and Leng, Yichong and He, Lei and Qin, Tao and Zhao, Sheng and Bian, Jiang},
  booktitle={ICLR},
  year={2024}
}

@article{bain2022whisperx,
  title={WhisperX: Time-Accurate Speech Transcription of Long-Form Audio},
  author={Bain, Max and Huh, Jaesung and Han, Tengda and Zisserman, Andrew},
  journal={INTERSPEECH 2023},
  year={2023}
}

@article{ao2024sd,
  title={Sd-eval: A benchmark dataset for spoken dialogue understanding beyond words},
  author={Ao, Junyi and Wang, Yuancheng and Tian, Xiaohai and Chen, Dekun and Zhang, Jun and Lu, Lu and Wang, Yuxuan and Li, Haizhou and Wu, Zhizheng},
  journal={Advances in Neural Information Processing Systems},
  volume={37},
  pages={56898--56918},
  year={2024}
}

@inproceedings{zhang2024gtsinger,
  title={GTSinger: A Global Multi-Technique Singing Corpus with Realistic Music Scores for All Singing Tasks},
  author={Zhang, Yu and Pan, Changhao and Guo, Wenxiang and Li, Ruiqi and Zhu, Zhiyuan and Wang, Jialei and Xu, Wenhao and Lu, Jingyu and Hong, Zhiqing and Wang, Chuxin and others},
  booktitle={The Thirty-eight Conference on Neural Information Processing Systems Datasets and Benchmarks Track},
  year={2024}
}

@inproceedings{ju2024naturalspeech,
  title={NaturalSpeech 3: zero-shot speech synthesis with factorized codec and diffusion models},
  author={Ju, Zeqian and Wang, Yuancheng and Shen, Kai and Tan, Xu and Xin, Detai and Yang, Dongchao and Liu, Yanqing and Leng, Yichong and Song, Kaitao and Tang, Siliang and others},
  booktitle={Proceedings of the 41st International Conference on Machine Learning},
  pages={22605--22623},
  year={2024}
}

@article{li2024styletts,
  title={Styletts 2: Towards human-level text-to-speech through style diffusion and adversarial training with large speech language models},
  author={Li, Yinghao Aaron and Han, Cong and Raghavan, Vinay and Mischler, Gavin and Mesgarani, Nima},
  journal={Advances in Neural Information Processing Systems},
  volume={36},
  year={2024}
}

@article{chen2024f5,
  title={F5-tts: A fairytaler that fakes fluent and faithful speech with flow matching},
  author={Chen, Yushen and Niu, Zhikang and Ma, Ziyang and Deng, Keqi and Wang, Chunhui and Zhao, Jian and Yu, Kai and Chen, Xie},
  journal={arXiv preprint arXiv:2410.06885},
  year={2024}
}

@article{guo2024fireredtts,
  title={Fireredtts: A foundation text-to-speech framework for industry-level generative speech applications},
  author={Guo, Hao-Han and Liu, Kun and Shen, Fei-Yu and Wu, Yi-Chen and Xie, Feng-Long and Xie, Kun and Xu, Kai-Tuo},
  journal={arXiv preprint arXiv:2409.03283},
  year={2024}
}

@article{shi2020aishell,
  title={Aishell-3: A multi-speaker mandarin tts corpus and the baselines},
  author={Shi, Yao and Bu, Hui and Xu, Xin and Zhang, Shaoji and Li, Ming},
  journal={arXiv preprint arXiv:2010.11567},
  year={2020}
}

@article{saeki2022utmos,
  title={Utmos: Utokyo-sarulab system for voicemos challenge 2022},
  author={Saeki, Takaaki and Xin, Detai and Nakata, Wataru and Koriyama, Tomoki and Takamichi, Shinnosuke and Saruwatari, Hiroshi},
  journal={arXiv preprint arXiv:2204.02152},
  year={2022}
}
